%% file: main.tex
\documentclass[apj,twocolumn,twocolappendix,numberedappendix,appendixfloats]{openjournal}
\usepackage{amsmath}
\usepackage[dvipsnames]{xcolor} 
\usepackage[breaklinks,colorlinks,citecolor=blue,urlcolor=blue,linkcolor=blue]{hyperref}
\usepackage{orcidlink}
\usepackage{graphicx}
\usepackage{amssymb}
\usepackage[dvipsnames]{xcolor}
\usepackage{subfigure}

\input{affil}

\def\Angstrom{\textup{\AA}}
\def\Msun{{\rm M}_\odot}
\def\Zsun{{\rm Z}_\odot}

\def\figdir{fig/}

\begin{document}

\title{\bf Superlinear Type II Superluminous Supernovae 2017fck and 2019cmv: A Possible Origin from Interacting Thermonuclear Supernovae}
\shorttitle{Superlinear Type II Superluminous Supernovae 2017fck and 2019cmv}
\shortauthors{Hiramatsu et al.}

\author{Daichi~Hiramatsu$^{1,2,3}$\orcidlink{0000-0002-1125-9187}}
\author{Takashi~J.~Moriya$^{4,5}$\orcidlink{0000-0003-1169-1954}}
\author{D.~Andrew~Howell$^{2,3}$\orcidlink{0000-0003-4253-656X}}
\author{Iair~Arcavi$^{6,7}$\orcidlink{0000-0001-7090-4898}}
\author{Jamison~Burke$^{8}$\orcidlink{0000-0003-0035-6659}}
\author{Griffin~Hosseinzadeh$^{9}$\orcidlink{0000-0002-0832-2974}}
\author{Curtis~McCully$^{2}$\orcidlink{0000-0001-5807-7893}}
\author{Stefano~Valenti$^{10}$\orcidlink{0000-0001-8818-0795}}
\author{Maria~R.~Drout$^{11}$\orcidlink{0000-0001-7081-0082}}
\author{Saurabh~W.~Jha$^{12}$\orcidlink{0000-0001-8738-6011}}
\author{Youssef~Eweis$^{12,13}$\orcidlink{0009-0006-8830-7294}}
\author{Sergei~I.~Blinnikov$^{14,15}$\orcidlink{0000-0002-5726-538X}}

\UF
\LCO
\UCSB
\NAOJ
\Monash
\TAU
\CIFAR
\Shady
\UCSD
\UCD
\Toronto
\Rutgers
\RAPCorp
\NRC
\IPMU

\email{Corresponding author: dhiramatsu@ufl.edu}

\begin{abstract}
Additional power sources to traditional supernovae (SNe) are necessary to account for the extreme luminosities of superluminous SNe (SLSNe). A main power source for hydrogen-rich SLSNe (SLSNe-II) is thought to be circumstellar material (CSM) interaction. However, the nature of underlying SNe and their progenitor systems remain elusive as they are hidden below strong CSM signatures.
Here, we present optical photometry and spectroscopy of SLSNe-II 2017fck and 2019cmv. They are characterized by post-maximum ``superlinear'' light curves which we also identify in a sample of Type Ia SNe interacting with CSM (SNe~Ia-CSM), along with their light-curve correlations and spectral similarities.
Thus, we compute a numerical light-curve model grid of SNe~Ia-CSM with various SN~Ia subtypes and CSM distributions. Our model grid spans the observed parameter space of SNe~Ia-CSM in terms of their rise times, peak luminosities, and decline rates with a wide CSM mass range of $\sim2-11\,\Msun$, indicating the diversity in their progenitor systems.
For SNe~2017fck and 2019cmv, we infer high CSM masses of $\sim11$ and $6\,\Msun$, respectively, which might be produced during the common envelope evolution of a white dwarf and massive ($\geq 8\,\Msun$) or intermediate-mass ($< 8\,\Msun$) companion.
Together with the proposed connection of SLSN-II 2006gy to SNe~Ia-CSM, SNe~2017fck and 2019cmv may offer a possible thermonuclear origin for superlinear SLSNe-II with peak optical luminosities up to $10^{44}$\,erg\,s$^{-1}$.
\end{abstract}

\keywords{
\href{https://vocabs.ands.org.au/repository/api/lda/aas/the-unified-astronomy-thesaurus/current/resource.html?uri=http://astrothesaurus.org/uat/1668}{Supernovae (1668)}; 
\href{https://vocabs.ands.org.au/repository/api/lda/aas/the-unified-astronomy-thesaurus/current/resource.html?uri=http://astrothesaurus.org/uat/1728}{Type Ia supernovae (1728)}; 
\href{https://vocabs.ands.org.au/repository/api/lda/aas/the-unified-astronomy-thesaurus/current/resource.html?uri=http://astrothesaurus.org/uat/1731}{Type II supernovae (1731)}; 
\href{https://vocabs.ands.org.au/repository/api/lda/aas/the-unified-astronomy-thesaurus/current/resource.html?uri=http://astrothesaurus.org/uat/1799}{White dwarfs (1799)}; 
\href{https://vocabs.ardc.edu.au/repository/api/lda/aas/the-unified-astronomy-thesaurus/current/resource.html?uri=http://astrothesaurus.org/uat/655}{Giant stars (655)};
\href{https://vocabs.ardc.edu.au/repository/api/lda/aas/the-unified-astronomy-thesaurus/current/resource.html?uri=http://astrothesaurus.org/uat/1661}{Supergiant stars (1661)};
\href{https://vocabs.ardc.edu.au/repository/api/lda/aas/the-unified-astronomy-thesaurus/current/resource.html?uri=http://astrothesaurus.org/uat/241}{Circumstellar matter (241)};
\href{https://vocabs.ardc.edu.au/repository/api/lda/aas/the-unified-astronomy-thesaurus/current/resource.html?uri=http://astrothesaurus.org/uat/2154}{Common envelope evolution (2154)}
}

\maketitle

\section{Introduction}\label{sec:intro}

Superluminous supernovae (SLSNe) are characterized by their bright light-curve peaks ($\lesssim-21$ mag) that require additional power sources beyond those of traditional SNe (e.g., radio active decay or shock-deposited energy; \citealt{Quimby2007ApJ...668L..99Q,Barbary2009ApJ...690.1358B,Quimby2011Natur.474..487Q,Chomiuk2011ApJ...743..114C,Howell2013ApJ...779...98H}, also see e.g., \citealt{Gal-Yam2012SLSN,Gal-Yam2019SLSN,Howell2017SLSN,Moriya2018SLSN} for reviews).
Circumstellar material (CSM) interaction is thought to be the main power source for SNe with narrow Balmer-series emission lines in spectra (SNe~IIn; \citealt{Schlegel1990MNRAS.244..269S,Filippenko1997ARA&A..35..309F}), some of which become bright enough to reach into the SLSN regime (SLSNe-II; \citealt{Ofek2007,Smith2007,Gezari2008ApJ...683L.131G,Miller2009ApJ...690.1303M}), albeit with an arbitrary magnitude cut unsupported by the largest sample study to date of nearly 500 SNe~IIn/SLSNe~II \citep{Hiramatsu2026ApJ..1005...82H}. Due to the CSM contamination, however, the nature of underlying SNe and their progenitor systems remain elusive.

Type~Ia SNe (SNe~Ia) are the thermonuclear explosions of carbon-oxygen (CO) white dwarfs (WDs; \citealt{Whelan1973ApJ...186.1007W,Nomoto1982ApJ...253..798N,Nomoto1982ApJ...257..780N,Nomoto1984,Iben1984ApJS...54..335I,Webbink1984ApJ...277..355W,Woosley1986ApJ...301..601W,Nugent2011Natur.480..344N,Bloom2012ApJ...744L..17B}, see also e.g., \citealt{Howell2011Ia,Maoz2014Ia,Maguire2016Ia} for reviews). Despite their cosmological uses discovering the accelerating expansion of the Universe and revealing its energy contents \citep{Riess1998,Perlmutter1999}, open questions still remain on their progenitor systems and explosion mechanisms. The three leading theories are: the single-degenerate (SD) scenario \citep{Whelan1973ApJ...186.1007W, Nomoto1982ApJ...253..798N,Nomoto1982ApJ...257..780N,Nomoto1984,Iben1984ApJS...54..335I} in which a CO WD accretes material from a nondegerate companion; the double-degenerate (DD) scenario \citep{Iben1984ApJS...54..335I,Webbink1984ApJ...277..355W,Woosley1986ApJ...301..601W,Pakmor2010Natur.463...61P} in which a CO WD accretes from or merges with another WD; and the core-degenerate (CD) scenario \citep{Iben1984ApJS...54..335I,Terman1994} in which a CO WD merges with the stellar core of a nondegerate companion.

SNe~Ia-CSM are an intriguing intersection between SNe~IIn and SNe~Ia in that they show Balmer-series emission lines on top of a diluted SN~Ia-like continuum (see e.g., \citealt{Silverman2013_sample,Sharma2023ApJ...948...52S} for sample studies).
A few SNe~Ia-CSM show clear SN~Ia-dominated spectra at early phase (e.g., SN~2002ic; \citealt{Hamuy2003ic,Deng2004,Wood-Vasey2004} and PTF11kx; \citealt{Dilday2012,Silverman2013}, as well as SNe~2020eyj; \citealt{Kool2023Natur.617..477K} and 2020aeuh; \citealt{Tsalapatas2025A&A...704A.135T} with hydrogen-poor/helium-rich and hydrogen/helium-poor CSM, respectively), confirming their thermonuclear origin. On the other hand, most of them are dominated by CSM interaction without clean SN~Ia signatures (e.g., \citealt{Leloudas2015}), making the hydrogen/helium-poor core-collapse (SN~Ic) origin also compatible (see e.g., \citealt{Fox2015,Inserra2016} for the discussions on SN~2012ca). 
If SNe~Ia-CSM indeed come from SNe~Ia, then these are the best candidates for the CD scenario (e.g., \citealt{Taam2000,Livio2003ApJ...594L..93L,Chugai2004AstL...30...65C,Hachisu2008ApJ...679.1390H,Kashi2011MNRAS.417.1466K,Sabach2014,Soker2019,Ablimit2021}).`

SLSN-II 2006gy is another intriguing case that may bridge SLSNe-II and SNe~Ia. It was initially suggested to originate from a core-collapse or pulsational pair-instability explosion of a very massive star (as high as $\sim100\,\Msun$; \citealt{Ofek2007,Smith2007,Smith2010,Woosley2007Natur.450..390W}).
However, \cite{Jerkstrand2020} reanalyzed a nebular spectrum of SLSN-II 2006gy \citep{Kawabata2009}, identifying strong iron lines that require comparable iron yields to SNe~Ia which are higher than core-collapse and pulsational pair-instability SNe (but see also \citealt{Umeda2008ApJ...673.1014U,Andrews2022ApJ...938...19A}). They also show that their updated light curve is compatible to SNe~Ia interacting with massive CSM of $\sim13\,\Msun$.
In this context, observations of SN~IIn/SLSNe-II and numerical modeling of SNe~Ia-CSM are of great importance in revealing their true nature (see e.g., \citealt{Moriya2023MNRAS.522.6035M} for models with diluted wind-like CSM).

Here, we report optical photometry and spectroscopy of SLSNe-II 2017fck and 2019cmv with post-maximum ``superlinear'' light curves, along with a numerical light-curve model grid of SNe~Ia-CSM.
In Sections \ref{sec:disc} and \ref{sec:obs}, we summarize their discoveries, follow-up observations, and data reduction.
We then analyze their host galaxies, spectra, and light curves in Section \ref{sec:ana}, identifying some similarities to SNe~Ia-CSM, especially the linearly declining luminous light curves.
Thus in Section \ref{sec:LCmodel}, we perform numerical light-curve modeling of SNe~Ia-CSM, reproducing the diversity in SNe~Ia-CSM with various Type Ia subtypes and CSM density distributions.
Finally, we  discuss their possible progenitors systems and summarize our findings in Section \ref{sec:sd}.

\section{Discoveries} \label{sec:disc}

The \textit{Gaia} Science Alerts \citep{Hodgkin2021} discovered SN~2017fck (Gaia17bro) on 2017 July 2.54 (UT dates are used throughout in this work) at 16.48 mag in the \textit{G} band at $\text{R.A.}=05^{\text{h}}19^{\text{m}}54^{\text{s}}.370$ and $\text{Dec.}=-56^{\circ}11'08".48$, using the \textit{Gaia} Spacecraft \citep{Gaia2017fck}. The associated last \textit{G}-band non-detection limit on 2017 April 6.17 at 21.5 mag was retrieved via the \textit{Gaia} Photometric Science Alerts\footnote{\url{http://gsaweb.ast.cam.ac.uk/alerts/home}} \citep{Hodgkin2021}. The reported last non-detection on 2017 June 7.80 with the discovery \citep{Gaia2017fck} has now been marked as ``untrusted" (i.e., a detection with an unreliable flux measurement) on the \textit{Gaia} Photometric Science Alerts, and thus used only as the first detection for estimating the explosion epoch (see below).
\cite{Strader2017fck} obtained an optical spectrum of SN~2017fck on 2017 August 1.38 with the Goodman Spectrograph on the Southern Astrophysical Research Telescope (SOAR), classifying it as an SN~IIn at a redshift $z=0.09442$ from the narrow host galaxy lines (see also \citealt{Strader2017fckATel} noting a possible Ia/Ic-CSM classification).

The Zwicky Transient Facility (ZTF; \citealt{Bellm2019ZTF,Graham2019ZTF}) discovered SN~2019cmv (ZTF19aalbrgu) on 2019 March 25.51 at 17.92 mag in the \textit{g} band at $\text{R.A.}=18^{\text{h}}57^{\text{m}}52^{\text{s}}.996$ and $\text{Dec.}=+45^{\circ}35'24".00$, using the ZTF camera on the Samuel Oschin $48$-inch Schmidt Telescope at the Palomar Observatory \citep{Nordin2019cmv}. 
Upon the ZTF discovery, the \textit{Gaia} Science Alerts reported a prediscovery \textit{G}-band detection on 2019 February 24.93 at 19.51 mag. The associated last \textit{G}-band non-detection limit on 2019 February 24.85 ($\lesssim2$ hours of discovery) at 21.5 mag was retrieved via the \textit{Gaia} Photometric Science Alerts. Thus, the tighter \textit{Gaia} discovery and last non-detection are adopted in this work.
\cite{Fremling2019cmv} obtained an optical spectrum of SN~2019cmv on 2019 April 17.92 with the SPectrograph for the Rapid Acquisition of Transients (SPRAT) on the Liverpool Telescope (LT), classifying it as an SLSN-II at a redshift $z=0.097$ from the narrow Balmer-series emission lines.

We estimate an explosion epoch of each SN by simply taking the midpoint of the last non-detection and the first detection with the error being the estimated explosion epoch minus the last non-detection. This yields $\text{MJD}_0=57880\pm29$ and $58538.89\pm0.03$ for SNe~2017fck and 2019cmv, respectively, which is used as a reference epoch for all phases unless otherwise specified. 
We also obtained a host galaxy spectrum of each SN after the SN had faded (Section~\ref{sec:obs}) and determined an SN-independent redshift: $z=0.09456\pm0.00004$ and $0.0974\pm0.0004$ for SNe~2017fck and 2019cmv (Section~\ref{sec:host}), respectively, which we adopt in this work. 
We assume a standard $\Lambda$CDM cosmology with $H_0=71.0$\, km\,s$^{-1}$\,Mpc$^{-1}$, $\Omega_{\Lambda}=0.7$, and $\Omega_m=0.3$, and convert the redshift to a luminosity distance: $d_L=427.6$ Mpc ($\mu=38.16$ mag) and $441.3$ Mpc ($38.22$ mag) for SNe~2017fck and 2019cmv, respectively.

\section{Observations and Data Reduction} \label{sec:obs}

\begin{figure*}
 \centering
 \includegraphics[width=\textwidth]{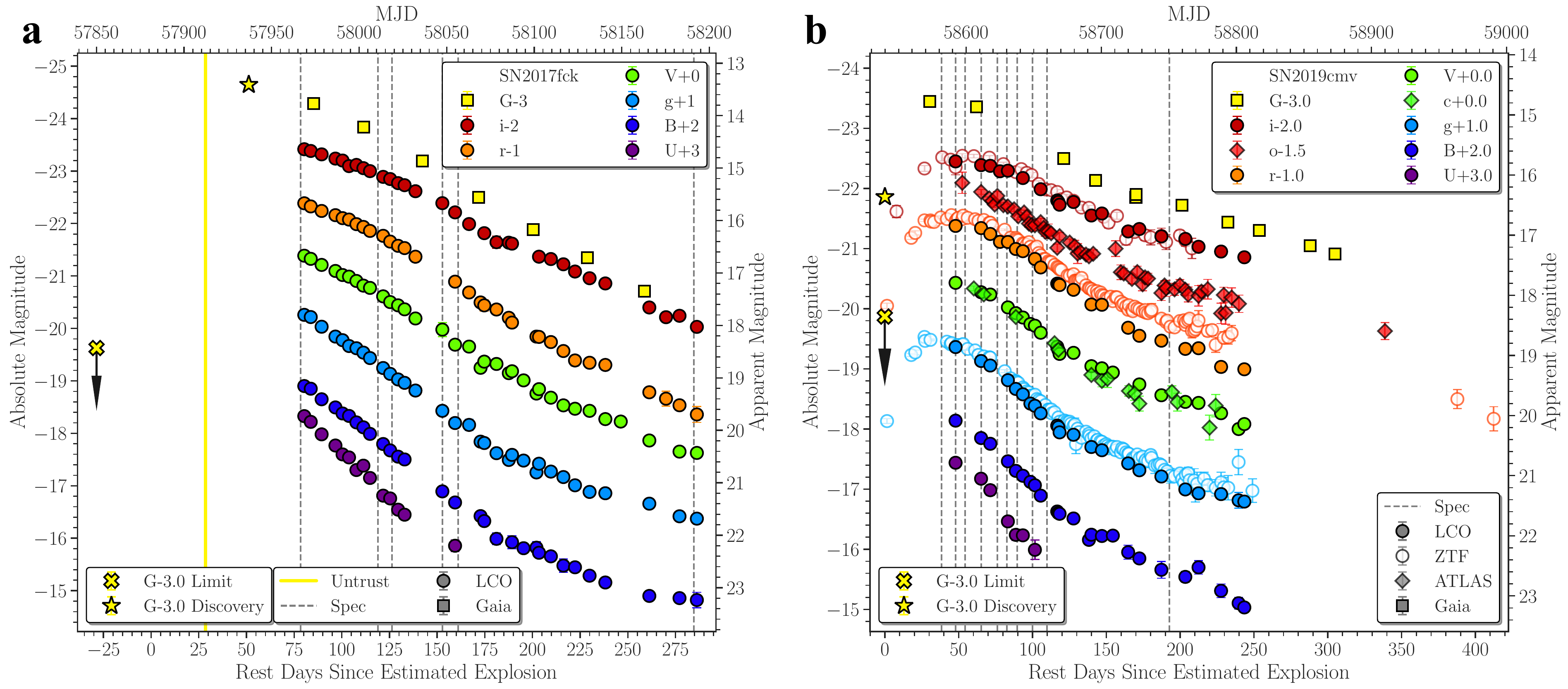}
 \caption{Extinction-corrected light curves of SNe~2017fck (\textbf{a}) and 2019cmv (\textbf{b}). Error bars denote $1\sigma$ uncertainties and are sometimes smaller than the marker size. The gray vertical dashed lines show the spectral epochs (Figure~\ref{fig:spec}). The explosion epoch of SN~2017fck is not well constrained ($\pm29$ d), while that of SN~2019cmv is extremely well constrained ($\pm0.03$ d; showing the rapid rise of at least $\sim2.0$ mag to the first \textit{Gaia} detection). ZTF \textit{r}-band light curve is systematically brighter than that of LCO since the ZTF bandpass is slightly redder, including the strong H$\alpha$ contribution at the SN redshift. Note the similar superluminous \textit{G}-band peaks ($-21.8 \leq M_G \leq -20.4$) followed by the long \textit{G}-band linear declines ($\sim0.01-0.02$ mag day$^{-1}$ for $\gtrsim200$ days). 
  } 
  \label{fig:LC}
\end{figure*}

\begin{figure*}
 \centering
 \includegraphics[width=\textwidth]{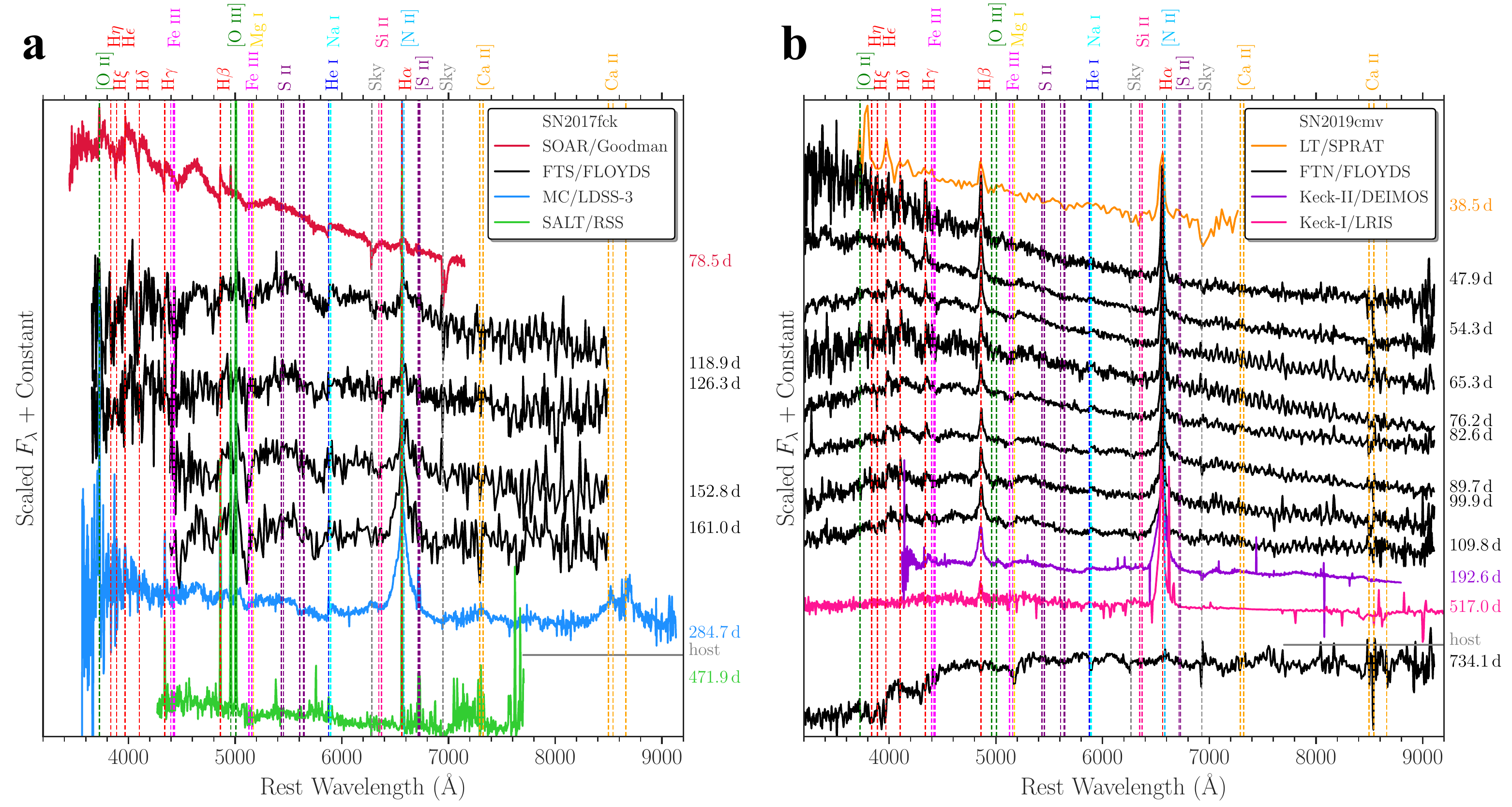}
 \caption{Extinction-corrected spectral time series of SNe~2017fck (\textbf{a}) and 2019cmv (\textbf{b}). The phase of each spectrum is given on the right. The first spectrum of each SN is the classification spectrum on TNS. The last spectra show a star-forming and elliptical host galaxies for SNe~2017fck and 2019cmv, respectively (Figure~\ref{fig:host}). Note the Balmer-series line evolution from the narrow symmetric profiles to broad asymmetric profiles and the increasing H$\alpha$/H$\beta$ flux ratios.
  } 
  \label{fig:spec}
\end{figure*}

Las Cumbres Observatory (LCO; \citealt{Brown2013}) \textit{UBgVri}-band data for SNe~2017fck and 2019cmv were obtained with the Sinistro cameras on the network of 1m telescopes at the Cerro Tololo Inter-American Observatory (District IV, Chile), Siding Spring Observatory (New South Wales, Australia), South African Astronomical Observatory (Sutherland, South Africa), and McDonald Observatory (Texas, USA), through the Global Supernova Project \citep{Howell2017GSP}. 
Using \texttt{lcogtsnpipe}\footnote{\url{https://github.com/LCOGT/lcogtsnpipe}} \citep{Valenti2016}, a PyRAF-based photometric reduction pipeline, point-spread-function fitting was performed. Reference images were obtained with a Sinistro camera after the SN had faded, and image subtraction was performed using \texttt{PyZOGY}\footnote{\url{https://github.com/dguevel/PyZOGY}} \citep{Guevel2017}, an implementation in Python of the subtraction algorithm described in \cite{Zackay2016}. 
{\it UBV}- and {\it gri}-band data were calibrated to Vega \citep{Stetson2000} and AB \citep{Albareti2017} magnitudes, respectively, using standard fields observed on the same night by the same telescope as the SN observations.

In addition, public \textit{Gaia} \textit{G}-band light curves for SNe~2017fck and 2019cmv, as well as ZTF \textit{gri}-band and the Asteroid Terrestrial-impact Last Alert System (ATLAS; \citealt{Tonry2018, Smith2020}) \textit{co}-band light curves for SN~2019cmv were retrieved respectively via the \textit{Gaia} Photometric Science Alerts, the ZTF forced-photometry service\footnote{\url{https://ztfweb.ipac.caltech.edu/cgi-bin/requestForcedPhotometry.cgi}} \citep{Masci2019,Masci2023}, and the ATLAS forced photometry server\footnote{\url{https://fallingstar-data.com/forcedphot/}} \citep{Shingles2021}, and used in the following analysis.

LCO optical spectra for SNe~2017fck and 2019cmv were taken with the FLOYDS spectrographs mounted on the 2m Faulkes Telescope North (FTN) and South (FTS) at Haleakala (USA) and Siding Spring (Australia), respectively, through the Global Supernova Project. 
A 2$\farcs$0-wide longslit was placed on the target along the parallactic angle \citep{Filippenko1982}. One-dimensional spectra were extracted, reduced, and calibrated following standard procedures using \texttt{floyds\_pipeline}\footnote{\url{https://github.com/LCOGT/floyds_pipeline}} \citep{Valenti2014}. 
A late-time optical spectrum of SN~2017fck was obtained with the Low Dispersion Survey Spectrograph 3 (LDSS-3; \citealt{Stevenson2016}) mounted on the 6.5m Magellan Clay (MC) telescope at the Las Campanas Observatory (Cerro Manqui, Chile) on 2018 March 15. The VHP-All grism coupled with a 1$\farcs$0-wide longslit for dispersion was used to obtain three $1$,$200$\,s exposures. The LDSS-3 data were reduced in a standard manner using PyRAF.
Two late-time optical spectra of SN~2019cmv were obtained with the DEep Imaging Multi-Object Spectrograph (DEIMOS; \citealt{DEIMOS2003SPIE.4841.1657F}) and Low-Resolution Imaging Spectrometer (LRIS; \citealt{LRIS1995PASP..107..375O}) mounted on the 10m Keck II and I telescopes, respectively, at the W. M. Keck Observatory (Maunakea, USA) on 2019 September 24 and 2020 September 14. The DEIMOS and LRIS observations were performed with a 1$\farcs$0-wide longslit and $900$ and $1$,$100$\,s exposures, respectively. The DEIMOS data were reduced using a custom Python pipeline, including bias subtraction, flat-field correction, cosmic-ray rejection, optimal extraction \citep{Horne1986PASP...98..609H}, and wavelength and flux calibration. The LRIS data were reduced with standard procedures in the fully automated reduction pipeline, \texttt{LPipe} \citep{Perley2019PASP..131h4503P}.
In addition, the classification spectra of SNe~2017fck \citep{Strader2017fck} and 2019cmv \citep{Fremling2019cmv} were retrieved via the Transient Name Server (TNS)\footnote{\url{https://www.wis-tns.org/}} and used in the following analysis.

The host galaxy spectrum of SN~2017fck was observed with the Robert Stobie Spectrograph (RSS; \citealt{Smith2006}) mounted on the Southern African Large Telescope (SALT) at the South African Astronomical Observatory (Sutherland, South Africa) on 2018 October 6 (through Rutgers University program 2018-1-MLT-006; PI: S. W. Jha). A 1$\farcs$5-wide longslit and the PG0900 grating in two tilt angles were used to cover the wavelength range from $470-840$\,nm. The data were reduced using a custom pipeline based on standard PyRAF spectral reduction routines and the \texttt{PySALT}\footnote{\url{https://pysalt.salt.ac.za/}} package \citep{Crawford2010}.
The host galaxy spectrum of SN~2019cmv was observed with FTN/FLOYDS on 2021 May 10 in the same setup and reduced in the same way as described above (see Section~\ref{sec:host} for the host galaxy association).

All photometry and spectroscopy of SNe~2017fck and 2019cmv are presented in Figures~\ref{fig:LC} and \ref{fig:spec}, respectively.
For both SNe~2017fck and 2019cmv, no Na~{\sc i}~D absorption is seen in the SN spectra at the host redshift (Figure~\ref{fig:spec}), indicating low host extinction at the SN position. Thus, we correct all photometry and spectroscopy only for the Milky Way (MW) extinction \citep{Schlafly2011}\footnote{Via the NASA/IPAC Infrared Science Archive (IRSA): \url{https://irsa.ipac.caltech.edu/applications/DUST/}} of $A_V=0.075$ and $0.156$ mag for SNe~2017fck and 2019cmv, respectively, assuming the \cite{Fitzpatrick1999} reddening law with $R_V=3.1$.

\section{Analysis} \label{sec:ana}

\subsection{Host Galaxies} \label{sec:host}

\begin{figure*}
 \centering
 \includegraphics[width=0.88\textwidth]{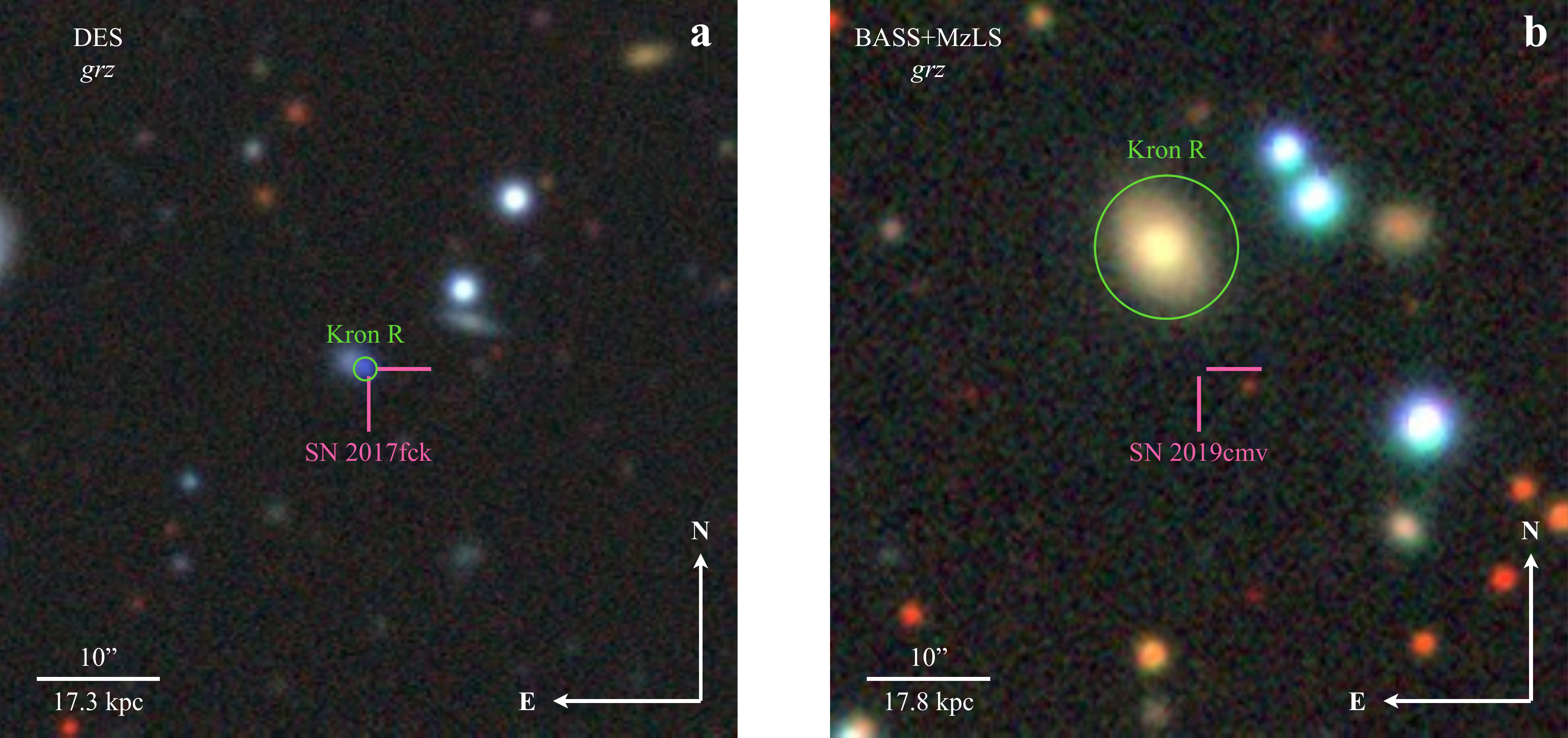}
\caption{Archival $1'\times1'$ color composites of SNe~2017fck (\textbf{a}) and 2019cmv (\textbf{b}) sites and their host galaxies, retrieved via the DESI Legacy Imaging Surveys \citep{DESILIS} Data Release 9. The each SN location and its host galaxy are marked respectively by the purple crosshairs and the green circle with a Kron radius. Note that SN~2017fck is within its host galaxy's Kron radius, while SN~2019cmv is at $1.8$ Kron radii away from its probable host galaxy.} 
  \label{fig:host}
\end{figure*}

{
\setlength{\tabcolsep}{2.6pt}
\begin{deluxetable*}{ccccccccccc}
\tablecaption{Host Galaxy Line Fluxes ($10^{-15}$\,erg\,s$^{-1}$\,cm$^{-2}$)\label{tab:hostline}}
\tablehead{
\colhead{SN} & \colhead{H$\beta$} & \colhead{[O~{\sc iii}]\,$\lambda4959$} & \colhead{[O~{\sc iii}]\,$\lambda5007$} & \colhead{Mg~{\sc i}\,$\lambda5173$} & \colhead{Na I D} & \colhead{H$\alpha$} & \colhead{[N~{\sc ii}]\,$\lambda6583$} & \colhead{[S~{\sc ii}]\,$\lambda6717$} & \colhead{[S~{\sc ii}]\,$\lambda6731$}
}
\startdata
2017fck & $1.032\pm0.009$ & $1.603\pm0.006$ & $4.765\pm0.008$ & $\cdots$ & $\cdots$ & $2.207\pm0.009$ & $0.100\pm0.009$ & $0.177\pm0.006$ & $0.217\pm0.009$ \\
2019cmv\tablenotemark{a} & $-1.7\pm0.4$ & $\cdots$ & $\cdots$ & $-3\pm1$ & $-2\pm1$ & $-0.6\pm0.4$ & $\cdots$ & $\cdots$ & $\cdots$
\enddata
\tablenotetext{a}{Negative values indicate absorption}
\end{deluxetable*}

\begin{deluxetable*}{cccccc}
\tablecaption{Host Galaxy Optical Photometry (AB mag)\label{tab:hostopt}}
\tablehead{
\colhead{SN} & \colhead{$M_g$\tablenotemark{a}} & \colhead{$M_r$\tablenotemark{a}} & \colhead{$M_i$\tablenotemark{a}} & \colhead{$M_z$\tablenotemark{a}} & \colhead{$M_{Y/y}$\tablenotemark{a}} 
}
\startdata
2017fck & $-17.385\pm0.009$ & $-17.608\pm0.009$ & $-18.07\pm0.01$ & $-17.90\pm0.02$ & $-17.7\pm0.1$\\
2019cmv & $-20.340\pm0.005$ & $-21.140\pm0.003$ & $-21.552\pm0.002$ & $-21.843\pm0.004$ & $-21.970\pm0.007$
\enddata
\tablenotetext{a}{Kron magnitudes from DES DR2 \citep{DESDR2} and PS1 DR2 \citep{PS1DR2} for the SNe~2017fck and 2019cmv hosts, respectively}
\end{deluxetable*}

\begin{deluxetable*}{ccccccc}
\tablecaption{Host Galaxy IR Photometry (Vega mag)\label{tab:hostIR}}
\tablehead{
\colhead{SN} & \colhead{$M_J$\tablenotemark{a}} & \colhead{$M_H$\tablenotemark{a}} & \colhead{$M_{K_s}$\tablenotemark{a}} & \colhead{$M_{W1}$\tablenotemark{b}} & \colhead{$M_{W2}$\tablenotemark{b}} & \colhead{$M_{W3}$\tablenotemark{b}} 
}
\startdata
2017fck & $\cdots$ & $\cdots$ & $\cdots$ & $\cdots$ & $\cdots$ & $\cdots$\\
2019cmv & $-23.4\pm0.1$ & $-23.6\pm0.2$ & $-24.4\pm0.2$ & $-24.43\pm0.01$ & $-24.45\pm0.04$ & $-25.5\pm0.5$
\enddata
\tablenotetext{a}{Extended profile-fit magnitudes from 2MASS \citep{2MASS}}
\tablenotetext{b}{$8.25"$ \textit{curve-of-growth}-corrected magnitudes from WISE \citep{WISE}}
\end{deluxetable*}

\begin{deluxetable*}{cccccccccc}
\tablecaption{Host Galaxy Properties\label{tab:hostclass}}
\tablehead{
\colhead{SN} & \colhead{Redshift} & \colhead{Offset} & \colhead{Offset\tablenotemark{a}} & \colhead{$12+ {\rm log}_{10}({\rm O/H})$\tablenotemark{b}} & \colhead{$Z({\rm O/H})$} & \colhead{${\rm SFR}(L_{\rm H\alpha})$} & \colhead{$Z({\rm SED})$} & \colhead{${\rm SFR}({\rm SED})$} & \colhead{$M_{\star}({\rm SED})$} \\ 
\colhead{} & \colhead{(z)} & \colhead{(kpc)} & \colhead{($R_{\rm Kron}$)} & \colhead{} & \colhead{(${\rm log}_{10}\,\Zsun$)} & \colhead{(${\rm log}_{10}\,\Msun \, {\rm yr}^{-1}$)} & \colhead{(${\rm log}_{10}\,\Zsun$)} & \colhead{(${\rm log}_{10}\,\Msun \, {\rm yr}^{-1}$)} & \colhead{(${\rm log}_{10}\,\Msun$)} 
}
\startdata
2017fck & $0.09456$ & $1.0\pm0.6$ & $0.6\pm0.4$ & $8.13$ ($8.09-8.24$) & $-0.64$ & $0.4\pm0.1$ & $\cdots$\tablenotemark{c} & $0.3^{+2.0}_{-0.2}$ & $8.46^{+0.06}_{-0.07}$\\
2019cmv & $0.0974$ & $18.3\pm0.6$ & $1.81\pm0.06$ & $\cdots$ ($\cdots$) & $\cdots$ & $\cdots$ & $-0.50\pm0.03$ & $-9^{+10}_{-5}$ & $10.67\pm0.03$
\enddata
\tablenotetext{a}{In terms of \textit{i}-band Kron radii}
\tablenotetext{b}{In the format of \textit{weighted average (range from various estimates)} using the line fluxes in Table~\ref{tab:hostline} and \texttt{PyMCZ} \citep{Bianco2016}}
\tablenotetext{c}{Fixed at the line flux measurements of $Z({\rm O/H})$}
\end{deluxetable*}
}

We use the archival images from the DESI Legacy Imaging Surveys \citep{DESILIS} Data Release 9 (LS DR9)\footnote{\url{https://www.legacysurvey.org/dr9/description/}}, and the catalogs from the Dark Energy Survey \citep{DES} Data Release 2 (DES DR2; \citealt{DESDR2}) and the Pan-STARRS1 \citep{PS1} Data Release 2 (PS1 DR2; \citealt{PS1DR2}), respectively, to identify the host galaxies of SNe~2017fck and 2019cmv. We obtain the Kron radii (in which $\sim90\%$ of the total luminosity should be contained; \citealt{Kron1980}) of nearby galaxy-like objects within 0$\farcm$5 of the SN location from the DES and PS1 catalogs, and associate the one with the smallest offset from the SN location in terms of Kron radius as a potential host galaxy. In Figure~\ref{fig:host}, we show the archival LS color-composite images of the SN locations with the potential host galaxies marked with their Kron radii.\footnote{The closest faint red source to the west of SN~2019cmv is a star-like object.} The host galaxy association is straightforward for SN~2017fck as it is located within the Kron radius of its compact blue host. In contrast, SN~2019cmv is located at $18.3$\,kpc ($1.81$ Kron radii) away from its potential elliptical galaxy, making the association somewhat uncertain, especially given the SLSN nature.

In order to strengthen the host galaxy identification, we obtained a spectrum of the potential host for each SN to measure its redshift.
The host galaxy spectra resemble that of typical star-forming and elliptical galaxies for SN~2017fck and 2019cmv, respectively (Figure~\ref{fig:spec}). We measure host galaxy lines by fitting a Gaussian profile to each line (Table~\ref{tab:hostline}). This yields weighted averages of $z=0.09456\pm0.00004$ and $0.0974\pm0.0004$ respectively for the potential hosts of SNe~2017fck and 2019cmv, which agree well with the SN redshifts. Thus, we associate these as the host galaxies in this work.
However, we also note that SLSNe are preferentially hosted in faint galaxies ($-13.5 \gtrsim M_g \gtrsim -16.5$; \citealt{Perley2016}), and an ultra-faint host galaxy for SN~2019cmv (below the detection limit of $M_g\gtrsim-13.3$ at the SN redshift) cannot be ruled out.


For SN~2017fck, we place the host galaxies in the Baldwin-Phillips-Terlevich (BPT) diagrams \citep{Baldwin1981} based on the line ratios of [O~{\sc iii}] $\lambda5007$/H$\beta$, [N~{\sc ii}] $\lambda6583$/H$\alpha$, and [S~{\sc ii}] $\lambda6717$/H$\alpha$ (Table~\ref{tab:hostline}). According to the \cite{Kewley2006} classification scheme, the host galaxy lies in the star-forming region in the BPT diagrams. 
Thus, we estimate a star-formation rate (SFR) from the H$\alpha$ luminosity using the \cite{Kennicutt1998} calibration, giving $0.4\pm0.1\, \Msun \, {\rm yr}^{-1}$. 
We also estimate a host galaxy metallicity from the measured line ratios and various estimates using \texttt{PyMCZ}\footnote{\url{https://github.com/nyusngroup/pyMCZ}} \citep{Bianco2016}, which yields a weighted average of $12+ {\rm log}_{10}(\rm O/H) = 8.13$, roughly corresponding to $0.2\,\Zsun$ \citep{Asplund2009}.

\begin{figure}
    \centering
    \includegraphics[width=0.48\textwidth]{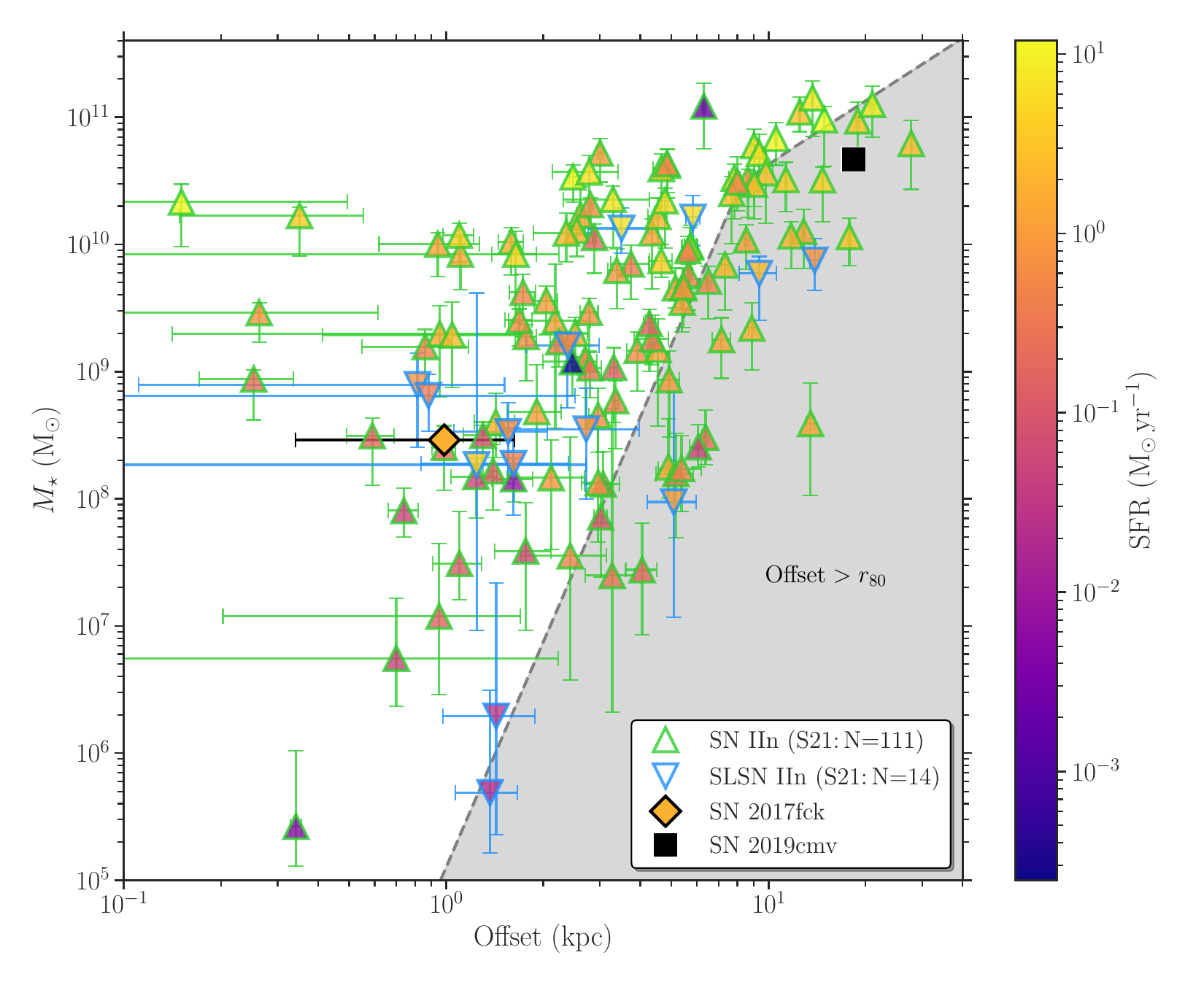}
    \caption{Comparison of the host stellar masses ($M_\star$) and offsets of SNe~2017fck and 2019cmv with the PTF sample from \citet[S21]{Schulze2021}, color coded by the star formation rate (SFR). SN~2019cmv is colored black since its SFR is well outside the sample range (${\rm log}_{10}[{\rm SFR}/\Msun\,{\rm yr}^{-1}]\simeq-9$; Table~\ref{tab:hostclass}). Error bars denote $1\sigma$ uncertainties. The gray shaded region shows where the offset is larger than the 80\% light radius ($r_{80}$) of the host galaxy given its stellar mass \citep{Miller2019,Mowla2019}. Note that SN~2017fck appears consistent with the SN~IIn and SLSN~IIn samples, while SN~2019cmv seems an outlier due to its low SFR.
    }
    \label{fig:MOffSFR}
\end{figure}

In order to extract the host galaxy properties of both SNe~2017fck and 2019cmv, we perform galaxy spectral energy distribution (SED) modeling with \texttt{Prospector}\footnote{\url{https://github.com/bd-j/prospector}} \citep{Johnson2021}. 
We follow the procedures described in \cite{Schulze2021}, assuming a linear-exponential star formation history (${\rm SFH} \propto t \times e^{-t/\tau}$, where $t$ is the age and $\tau$ is the $e$-folding timescale of SFH), the \cite{Chabrier2003} initial mass function, the \cite{Calzetti2000} extinction model, and the \cite{Byler2017} ionized gas model.
Then the galaxy parameters\footnote{For SN~2017fck, the metallicity is fixed at the measured value from the host emission lines.}: mass, age, $\tau$, extinction, and metallicity were inferred in a Bayesian way with \texttt{emcee}\footnote{\url{https://github.com/dfm/emcee}} \citep{emcee2013}.
To construct SED, we retrieve the host galaxy optical photometry from DES DR2\footnote{\url{https://des.ncsa.illinois.edu/releases/dr2/dr2-access}} and PS1 DR2\footnote{\url{https://catalogs.mast.stsci.edu/panstarrs/}} (Table~\ref{tab:hostopt}), and infrared (IR) photometry from the Two Micron All Sky Survey (2MASS; \citealt{2MASS})\footnote{\url{https://irsa.ipac.caltech.edu/cgi-bin/Gator/nph-scan?submit=Select&projshort=2MASS}} and the Wide-field Infrared Survey Explorer (WISE; \citealt{WISE})\footnote{\url{https://irsa.ipac.caltech.edu/cgi-bin/Gator/nph-scan?mission=irsa&submit=Select&projshort=WISE}} where available (Table~\ref{tab:hostIR}).

The measured and inferred host galaxy parameters are summarized in Table~{\ref{tab:hostclass}}. 
The host galaxy of SN~2017fck is a low-mass, star-forming galaxy at a subsolar metallicy, typical of SLSN host galaxies \citep{Perley2016}, while that of SN~2019cmv is a massive elliptical galaxy at a subsolar metallicity.
In Figure~{\ref{fig:MOffSFR}}, we show the comparison of the stellar masses ($M_\star$), SN offsets, and SFR of the host galaxies of SNe~2017fck and 2019cmv with the Palomar Transient Factory (PTF) sample\footnote{\url{https://github.com/steveschulze/PTF}} from \cite{Schulze2021}. The host galaxy of SN~2017fck lies in the typical regions covered by SNe~IIn and SLSNe~IIn. Although the large host offset of SN~2019cmv alone does not stand out, the host galaxy is still an outlier given its low SFR.
In this context, we consider the Ia-CSM origin as a viable scenario. We also note that SNe~Ia-CSM are also generally hosted in spiral or dwarf irregular galaxies \citep{Silverman2013_sample,Sharma2023ApJ...948...52S}, but longer delay times are expected for the SN Ia-CSM progenitor systems than for the typical SLSN-II massive progenitors (e.g., \citealt{Ablimit2021}; see Section~\ref{sec:progenitor} for further discussion).

\subsection{Spectral Evolution} \label{sec:spec}

\begin{figure*}
    \centering
    \includegraphics[width=\textwidth]{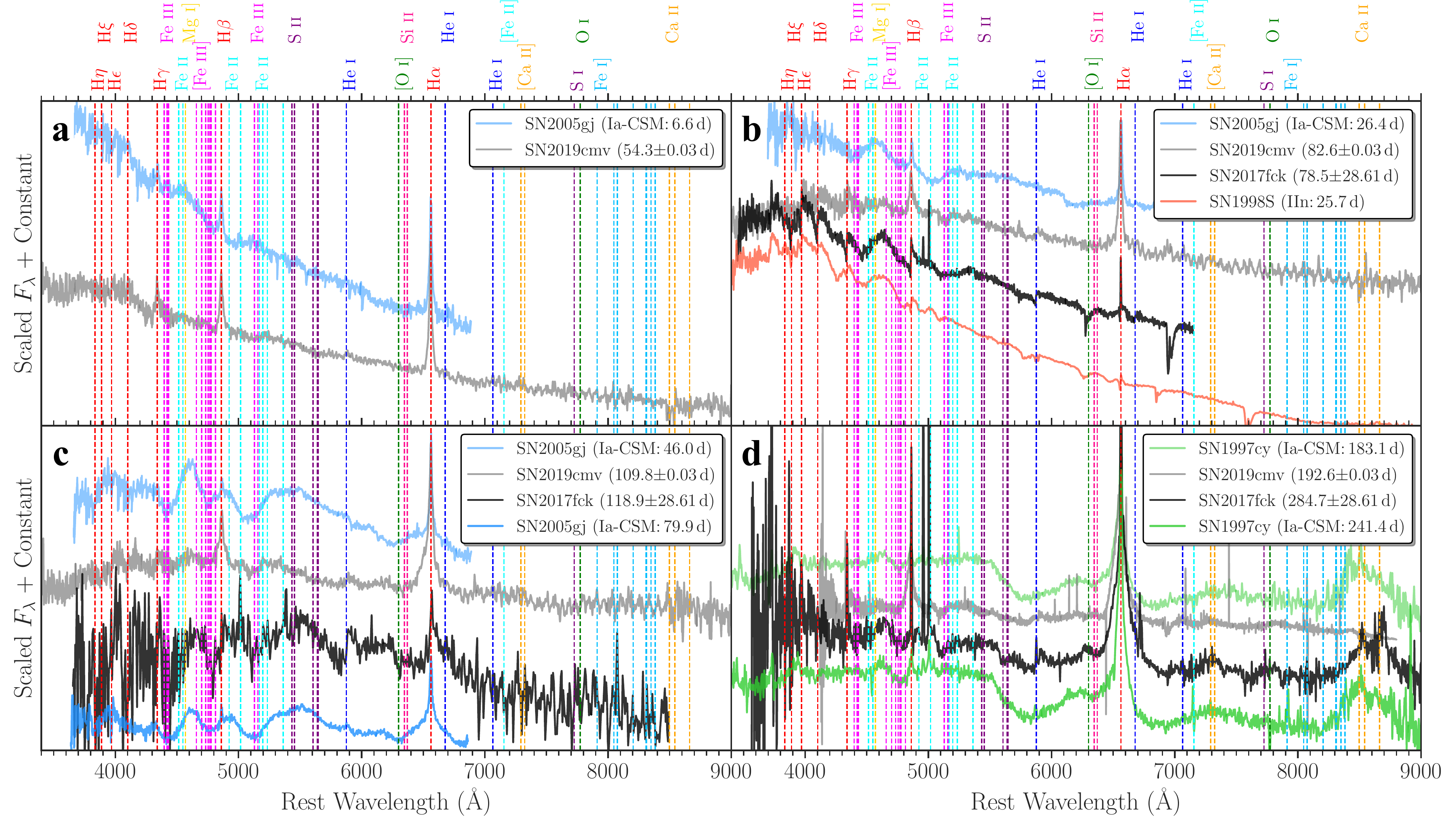}
    \caption{Spectral comparisons of SNe~2017fck and 2019cmv at four different phases (\textbf{a}--\textbf{d}) with interacting SN~IIn 2008S \citep{Fassia2001} and SNe~Ia-CSM 1997cy \citep{Germany2000,Turatto2000} and 2005gj \citep{Prieto2007}, retrieved via the Weizmann Interactive Supernova Data Repository \citep{Yaron2012}.
    The spectral matches for SNe~2019cmv and 2017fck were identified by GELATO and plotted at the top and bottom in each panel, respectively.
    The spectra are characterized by: (\textbf{a}, \textbf{b}) the narrow Balmer-series emission lines on top of the blue continuum; (\textbf{c}) the strong Fe~{\sc ii/iii} absorption and broad asymmetric H$\alpha$ emission lines with the suppressed red wing and increasing H$\alpha$/H$\beta$ flux ratios; and then (\textbf{d}) the broad H$\alpha$ and Ca~{\sc ii} NIR triplet emission as well as weak Fe~{\sc i}] emission and Si~{\sc ii} and S~{\sc ii} absorption lines on top of the persistent continuum; note that the spectral evolution of SN~2019cmv is slower compared to SNe~2017fck and 2005gj in terms of the continuum and emission lines.
    }
    \label{fig:speccomp}
\end{figure*}

\begin{figure}
    \centering
    \includegraphics[width=0.48\textwidth]{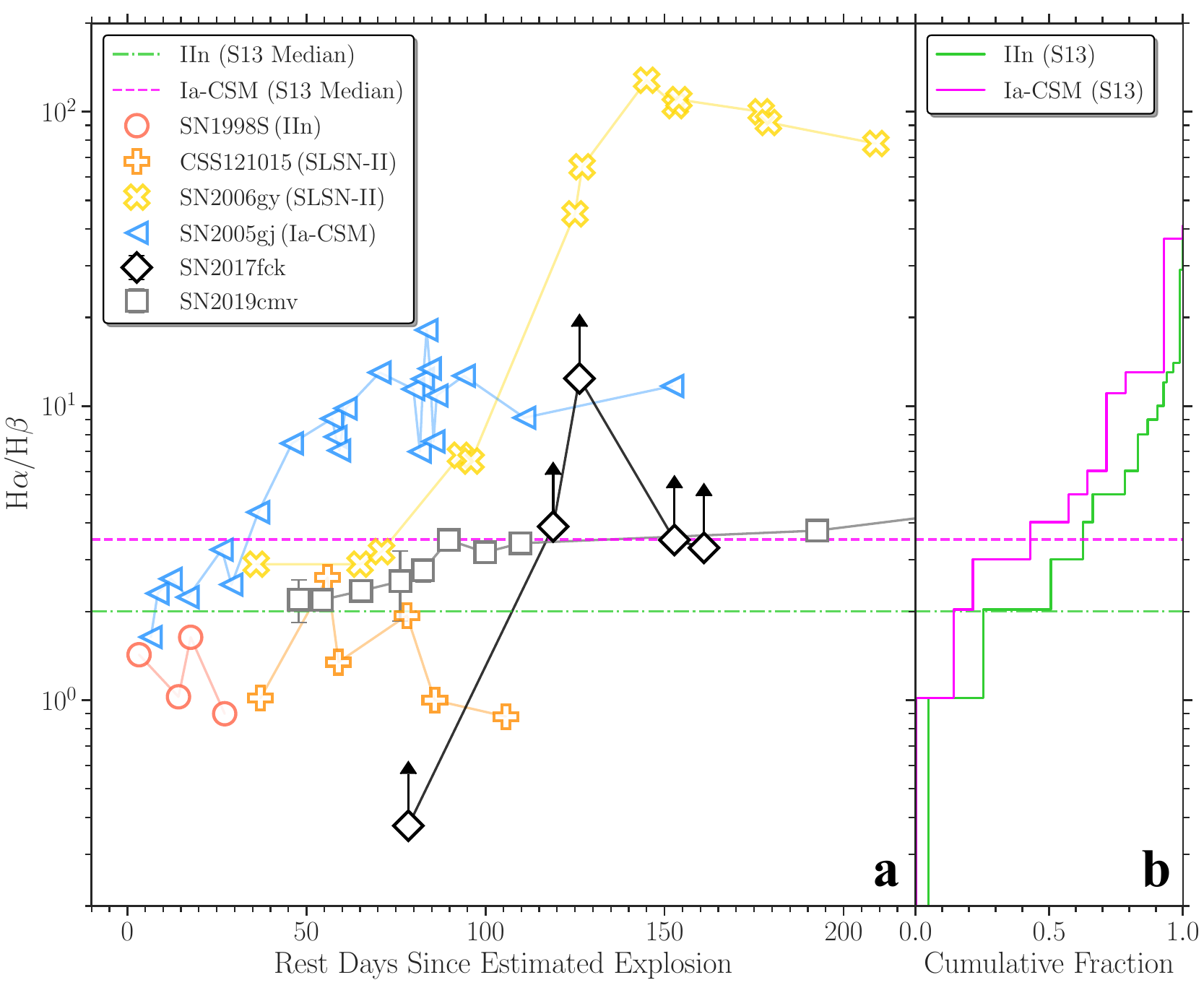}
    \caption{Comparison of the H$\alpha$/H$\beta$ flux-ratio evolution (\textbf{a}) of SNe~2017fck and 2019cmv with SN~IIn 2008S \citep{Fassia2001}, SLSNe-II~CSS121015 \citep{Benetti2014} and 2006gy \citep{Smith2010}, and SN~Ia-CSM 2005gj \citep{Prieto2007}, as well as the SN IIn and Ia-CSM cumulative fractions (\textbf{b}) from the \citet[S13]{Silverman2013_sample} sample.
    Error bars and arrows respectively denote $1\sigma$ uncertainties and limits.
    The median of H$\alpha$/H$\beta$ flux ratio over all epochs is smaller for SNe~IIn (e.g., SN~1998S) than for SNe~Ia-CSM (e.g., SN 2005gj).
    Note the increasing H$\alpha$/H$\beta$ flux ratios (and limits) of SNe~2017fck and 2019cmv that are above the SN~IIn median.
    }
    \label{fig:HaHb}
\end{figure}

We use the spectral library fitting code GEneric cLAssification TOol (GELATO\footnote{\url{https://gelato.tng.iac.es/}}; \citealt{Harutyunyan2008}) to cross-correlate the spectra of SNe~2017fck and 2019cmv to other well-observed interacting SNe.
In Figure~\ref{fig:speccomp}, we show good spectral matches to SNe~2017fck and 2019cmv, namely SN~IIn 1998S \citep{Fassia2001} and SNe~Ia-CSM 1997cy \citep{Germany2000,Turatto2000} and 2005gj \citep{Prieto2007}, at four different phases: $\sim50$, $80$, $110$, and $\gtrsim200$ days after explosion.\footnote{The first and last spectra of SN~2019cmv (Figure~\ref{fig:spec}) are excluded in the comparison due to the low signal-to-noise ratios.} 

Around the light-curve maximum ($\lesssim80$ days; Figure~\ref{fig:LC}), the spectra of SNe~2017fck and 2019cmv are characterized by the narrow Balmer-series emission lines on top of the blue continuum. For SN~2017fck, weak Fe~{\sc ii/iii} absorption lines, similar to SN~2005gj, and narrow Balmer-series and He~{\sc i} P Cygni features, similar to SN~1998S, are also seen; if the narrow absorption originates from unshocked CSM, this gives its velocity of $\sim400$\,km\,s$^{-1}$.
During the light-curve decline ($\gtrsim80$ days), strong Fe~{\sc ii/iii} absorption and weak He~{\sc i} emission lines appear, H$\alpha$ emission lines become broader and asymmetric with the suppressed red wing, and the H$\alpha$/H$\beta$ flux ratio increases, typical of SNe~Ia-CSM \citep{Silverman2013_sample,Sharma2023ApJ...948...52S}. However, the spectral evolution of SN~2019cmv is slower compared to SNe~2017fck and 2005gj in that the continuum and H$\beta$ emission line stay bluer and stronger, respectively (i.e., more IIn-like evolution), which might suggest a different level of CSM contamination on the underlying SN (\citealt{Leloudas2015}; see Section~\ref{sec:LCmodel} for further discussion).
At $\sim200$ days, the prominent spectral features remain the same for SN~2019cmv with the decreasing continuum. Even at $\gtrsim280$ days, the spectrum of SN~2017fck is not yet fully nebular, similar to SN~1997cy, showing the broad H$\alpha$ and Ca~{\sc ii} near-IR (NIR) triplet as well as weak Fe~{\sc i}] emission and Si~{\sc ii} and S~{\sc ii} absorption lines on top of the persistent continuum below $5500\,\Angstrom$ due to CSM interaction. Unlike typical core-collapse SNe, SNe~2017fck and 2019cmv lack strong [O~{\sc i}] $\lambda\lambda6300,6364$, O~{\sc i} $\lambda7774$, and Mg~{\sc i}] $\lambda4570$ emission lines \citep{Fox2015}. To diagnose the underlying SN composition, fully nebular spectra at later epochs ($>500$ days after explosion; see Section~\ref{sec:bolo} for further discussion) would be necessary, as was done for SLSN-II~2006gy, suggesting a possible Ia-CSM origin (\citealt{Jerkstrand2020}; but see also \citealt{Andrews2022ApJ...938...19A} for an opposing interpretation).

To quantify the spectral evolution, we measure H$\alpha$ and H$\beta$ line fluxes of SNe~2017fck and 2019cmv by fitting a multi-Gaussian profile (to account for the asymmetry) to each line. Since no obvious H$\beta$ line is seen in the spectra of SN~2017fck, the upper limit at each epoch is estimated assuming the same full width at half maximum as H$\alpha$.
In Figure~\ref{fig:HaHb}, we show the comparison of the H$\alpha$/H$\beta$ flux-ratio evolution of SNe~2017fck and 2019cmv with well-observed interacting SNe (SLSN-II CSS121015 is included in the comparison because of its overall spectral similarities to SN~Ia-CSM 2005gj; \citealt{Benetti2014}).
As seen in the IIn and Ia-CSM cumulative fractions from the \cite{Silverman2013_sample} sample, H$\alpha$/H$\beta$ flux ratios are generally smaller for SNe~IIn (e.g., SN~1998S) than for SNe~Ia-CSM (e.g., SN 2005gj). 
SNe~2017fck and 2019cmv show the increasing H$\alpha$/H$\beta$ flux ratios (and limits), which appears more consistent with those of SNe~Ia-CSM than IIn, although not conclusive given the wide overlapping SN~Ia-CSM and IIn distributions. 
It is also interesting to note the large increasing H$\alpha$/H$\beta$ flux ratios of SN~2006gy, which might also support its possible Ia-CSM origin.

\subsection{Bolometric Light Curves} \label{sec:bolo}

\begin{figure}
    \centering
    \includegraphics[width=0.48\textwidth]{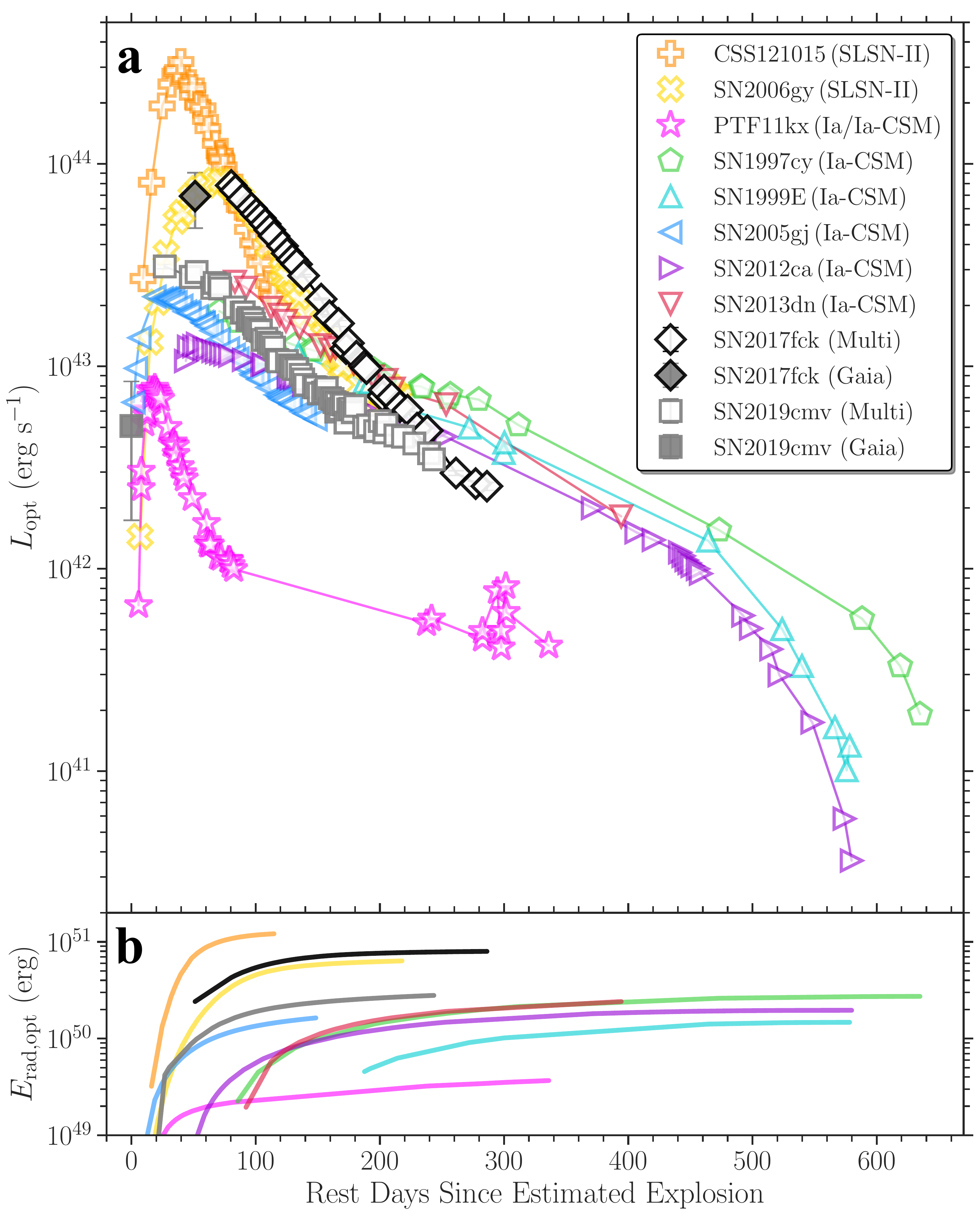}
    \caption{Comparison of the optical light curves ($L_{\rm opt}$; \textbf{a}) and cumulative radiated energies ($E_{\rm rad,opt}$; \textbf{b}) of SNe~2017fck and 2019cmv with SLSNe-II~CSS121015 \citep{Benetti2014} and 2006gy \citep{Jerkstrand2020}, as well as SNe~Ia-CSM PTF11kx \citep{Dilday2012}, 1997cy \citep{Germany2000,Turatto2000}, 1999E \citep{Rigon2003}, 2005gj \citep{Prieto2007}, 2012ca \citep{Inserra2016}, and 2013dn \citep{Fox2015}; the possible Ia-CSM origin has also been discussed for CSS121015 and SN~2006gy.
    For SNe~2017fck and 2019cmv, the pre-maximum and post-maximum luminosity is calculated from the \textit{Gaia} \textit{G}-band photometry and the LCO/ZTF/ATLAS multi-band photometry, respectively. Error bars denote $1\sigma$ uncertainties. 
    Note that except the transitional SN~Ia/Ia-CSM PTF11kx, the other SLSNe-II and SNe~Ia-CSM, including SNe~2017fck and 2019cmv, are characterized by the bright maximum ($\gtrsim10^{43}$\,erg\,s$^{-1}$) and ``superlinear" post-maximum declines (for $\gtrsim100$ days) with the high radiated energies ($\gtrsim10^{50}$\,erg).
    }
    \label{fig:LCsamp}
\end{figure}

\begin{figure*}
    \centering
    \includegraphics[width=\textwidth]{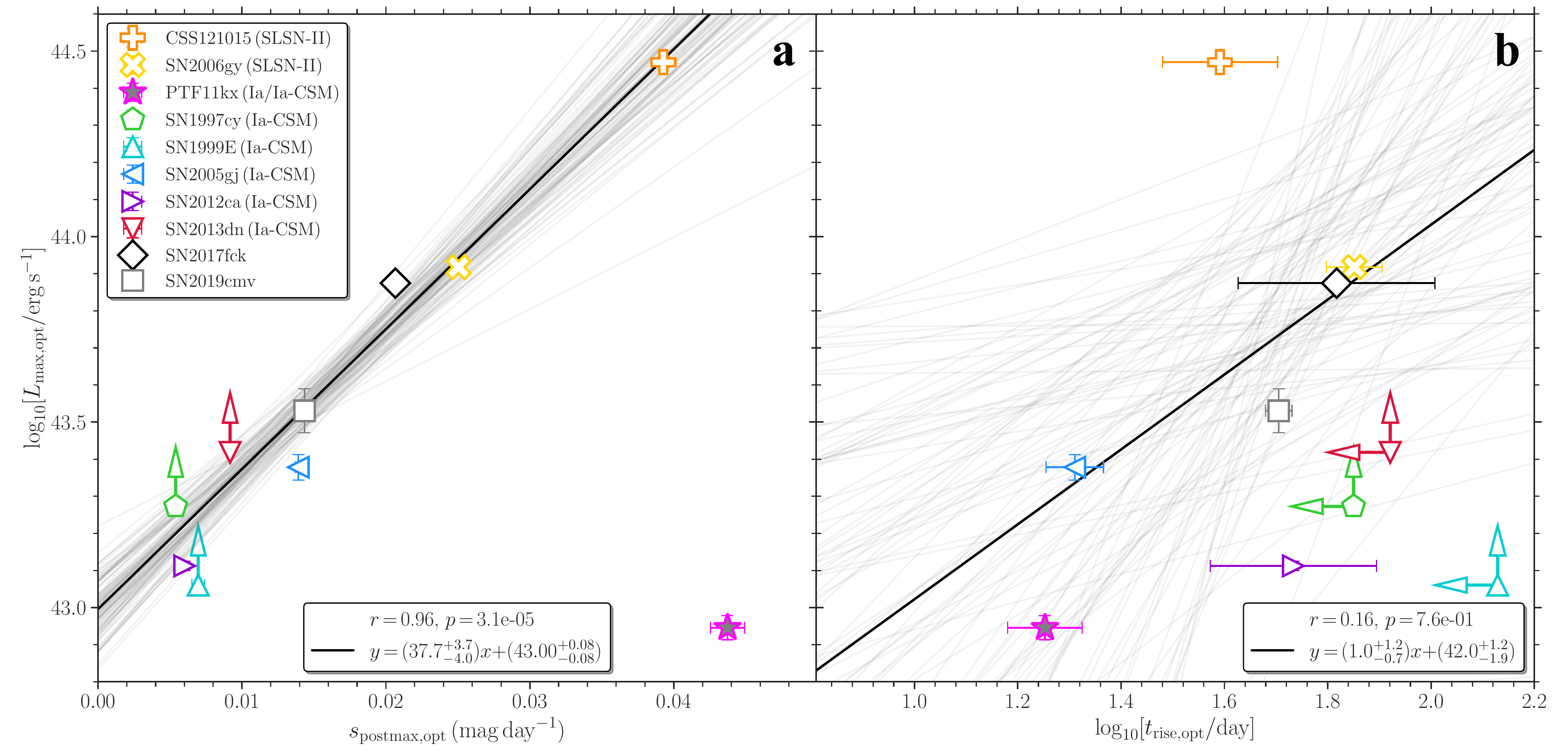}
    \caption{Pearson correlation tests (coefficient $r$ and $p$-value) and liner fits between the maximum optical luminosity ($L_{\rm max,opt}$), post-maximum decline slope ($s_{\rm postmax,opt}$; \textbf{a}), and rise time to the maximum ($t_{\rm rise,opt}$; \textbf{b}) of SNe~2017fck and 2019cmv with the same sample as in Figure~\ref{fig:LCsamp}, except PTF11kx (the grayed star) given its transitional nature.
    Error bars and arrows respectively denote $1\sigma$ uncertainties and limits.
    The gray transparent and black solid lines represent randomly drawn 100 fits and the best fits, respectively, from \texttt{emcee}.
    Note the significant correlation between $L_{\rm max,opt}$ and $s_{\rm postmax,opt}$, and the negligible correlation between $L_{\rm max,opt}$ and $t_{\rm rise,opt}$.
    }
    \label{fig:LCcorr}
\end{figure*}

We fit a blackbody SED to every epoch of photometry containing at least three LCO, ZTF, and/or ATLAS filters obtained within $1$ day of each other to estimate blackbody temperature and radius. Then we integrate the fitted blackbody SED over the optical wavelength range ($3250-8900\Angstrom$) to obtain  pseudobolometric (i.e., optical) luminosity at each epoch. For the early epochs where only \textit{Gaia} \textit{G}-band photometry is available, we convert its magnitude to optical luminosity assuming a zero pseudobolometric correction, given the comparably wide optical wavelength coverage.

In Figure~\ref{fig:LCsamp}, we show the comparison of the optical light curves and radiated energies of SNe~2017fck and 2019cmv with well-observed SLSNe-II and SNe~Ia-CSM.
PTF11kx is a transitional SN~Ia/Ia-CSM\footnote{SN~2002ic is another transional SN~Ia/Ia-CSM \citep{Hamuy2003ic,Deng2004,Wood-Vasey2004}, but not included in the comparison due to the sparse multi-band light-curve sampling.} in the sense that it initially showed Ia-dominated spectra and then transitioned to CSM-dominated spectra \citep{Dilday2012,Silverman2013}, which is different from the other SNe~Ia-CSM in the comparison with CSM-dominated spectra throughout their evolution.
Except PTF11kx, the light-curve characteristics of SNe~2017fck and 2019cmv are similar to those of the other SLSNe-II and SNe~Ia-CSM in that they show bright maxima ($\gtrsim10^{43}$\,erg\,s$^{-1}$) and linear post-maximum declines for $\gtrsim100$ days, which we refer as ``superlinear" declines, with high radiated energies ($\gtrsim10^{50}$\,erg) powered by CSM interaction. For SNe~Ia-CSM~1997cy, 1999E, and 2012ca, the late-time observations reveal the light-curve drops from the superlinear decline phase around $500-600$ days, which may indicate the end of CSM-dominated phase and the beginning of nebular phase.

In order to quantify their light-curve characteristics, we measure the maximum optical luminosity ($L_{\rm max,opt}$) and rise time ($t_{\rm rise,opt}$) from the estimated explosion of each SN by fitting a parabola around the peak when available. Then we fit a line from the peak to decline up to $100$ days post maximum to measure the decline slope ($s_{\rm postmax,opt}$). When the peak is not observed, we use the first data point as the lower and upper limits of $L_{\rm max,opt}$ and $t_{\rm rise,opt}$, respectively, and measure $s_{\rm postmax,opt}$ from the first data point assuming a constant linear decline from the peak.

In Figure~\ref{fig:LCcorr}, we show the comparisons of the measured $L_{\rm max,opt}$, $t_{\rm rise,opt}$, and $s_{\rm postmax,opt}$. 
Except PTF11kx, a positive trend in $L_{\rm max,opt}$\,vs.\,$s_{\rm postmax,opt}$ can be seen, i.e., brighter SNe decline faster. The points are more scattered in $L_{\rm max,opt}$\,vs.\,$t_{\rm rise,opt}$, likely due to poorly constrained explosion epochs of some sample SNe.
We then perform Pearson correlation tests for $L_{\rm max,opt}$\,vs.\,$s_{\rm postmax,opt}$ and $L_{\rm max,opt}$\,vs.\,$t_{\rm rise,opt}$, finding a significant and negligible correlations, respectively.
We also perform linear fitting in a Bayesian way with \texttt{emcee}, assuming flat priors to all parameters. The tightness of the linear fits also suggests a significant and negligible correlations for $L_{\rm max,opt}$\,vs.\,$s_{\rm postmax,opt}$ and $L_{\rm max,opt}$\,vs.\,$t_{\rm rise,opt}$, respectively.
These correlations may indicate some similarities in the progenitor systems and CSM configurations for the sample SLSNe-II and SNe~Ia-CSM, which we try to reproduce with numerical Ia-CSM light-curve modeling in Section~\ref{sec:LCmodel}.

\section{SN~Ia-CSM Light-Curve Modeling} \label{sec:LCmodel}

\begin{figure}
    \centering
    \includegraphics[width=0.48\textwidth]{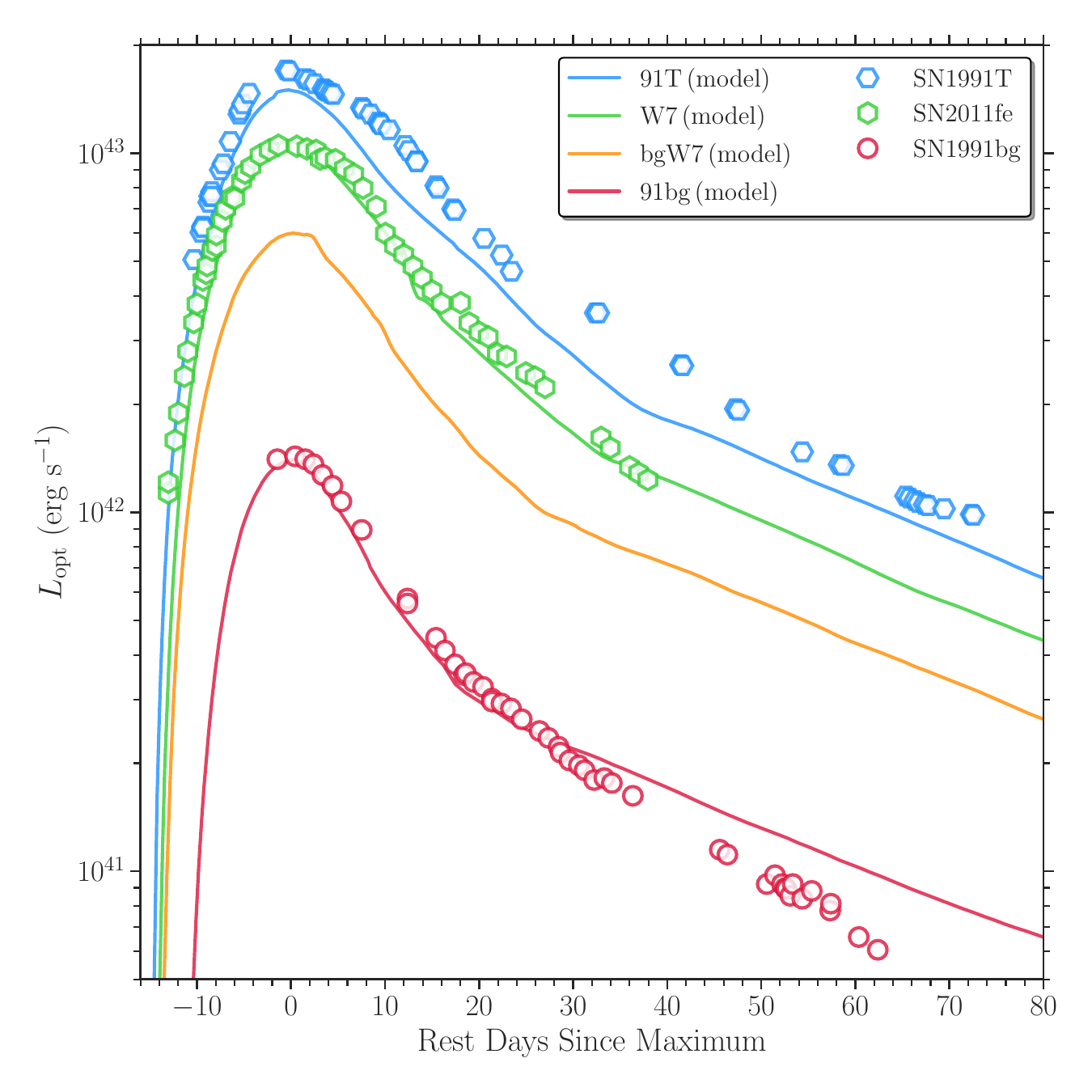}
    \caption{SN~Ia light-curve models: 91T ($E_{\rm exp}=1.3\times10^{51}$\,erg, $M_{\rm Ni}=0.80\,\Msun$); W7 ($E_{\rm exp}=1.3\times10^{51}$\,erg, $M_{\rm Ni}=0.58\,\Msun$); bgW7 ($E_{\rm exp}=0.8\times10^{51}$\,erg, $M_{\rm Ni}=0.30\,\Msun$); and 91bg ($E_{\rm exp}=0.6\times10^{51}$\,erg, $M_{\rm Ni}=0.06\,\Msun$), compared with well-observed prototypical events: SNe~1991T \citep{Lira1998AJ....115..234L}; 2011fe \citep{Pereira2013A&A...554A..27P}; and 1991bg \citep{Taubenberger2008MNRAS.385...75T}.
    }
    \label{fig:Iasub}
\end{figure}

\begin{deluxetable}{cccc}
\tablecaption{
CSM Model Grid Properties
 \label{tab:modellist}
}
\tablehead{
\colhead{Density} & \colhead{Radius ($R$)} & \colhead{Mass ($M_{\rm CSM}$)} & \colhead{Mass Loss ($\dot{M}$)\tablenotemark{a}}\\
\colhead{$\rho_{\rm CSM}(r)$} & \colhead{($10^{16}\,\mathrm{cm}$)} & \colhead{($\Msun$)} & \colhead{($\Msun$\,yr$^{-1}$)} 
}
\startdata
$r^{-2}$  & $1.6$ & $0.5$ & $9\times10^{-3}$\\ 
 & $1.0$ & $0.9$ & $3\times10^{-2}$\\ 
 & $2.0$ & $1.8$ & $3\times10^{-2}$\\ 
 & $1.0$ & $1.9$ & $6\times10^{-2}$\\ 
 & $1.2$ & $7.5$ & $2\times10^{-1}$\\ 
 & $1.4$ & $8.8$ & $2\times10^{-1}$\\ 
 & $1.0$ & $9.4$ & $3\times10^{-1}$\\ 
 & $1.6$ & $10.0$ & $2\times10^{-1}$\\ 
 & $1.2$ & $11.3$ & $3\times10^{-1}$\\ 
 & $1.4$ & $13.1$ & $3\times10^{-1}$\\ 
 & $1.6$ & $15.0$ & $3\times10^{-1}$\\ 
 & $1.8$ & $16.9$ & $3\times10^{-1}$\\ 
 & $2.0$ & $18.8$ & $3\times10^{-1}$\\ 
\hline
$r^{-1}$ & $1.0$ & $1.6$ & $5\times10^{-2}$\\ 
 & $1.5$ & $3.6$ & $7\times10^{-2}$\\ 
 & $0.5$ & $3.8$ & $2\times10^{-1}$\\ 
 & $1.0$ & $4.7$ & $1\times10^{-1}$\\ 
 & $2.0$ & $6.3$ & $1\times10^{-1}$\\ 
 & $1.5$ & $10.7$ & $2\times10^{-1}$\\ 
 & $3.0$ & $14.3$ & $1\times10^{-1}$\\ 
 & $1.0$ & $15.7$ & $5\times10^{-1}$\\ 
 & $2.0$ & $19.0$ & $3\times10^{-1}$\\ 
 & $1.1$ & $19.0$ & $5\times10^{-1}$\\ 
 & $3.5$ & $19.4$ & $2\times10^{-1}$\\ 
\hline
$r^0$ & $1.0$ & $0.6$ & $2\times10^{-2}$\\ 
 & $1.0$ & $2.1$ & $7\times10^{-2}$\\ 
 & $2.0$ & $5.1$ & $8\times10^{-2}$\\ 
 & $1.0$ & $6.3$ & $2\times10^{-1}$\\ 
 & $1.5$ & $7.1$ & $1\times10^{-1}$\\ 
 & $2.0$ & $16.8$ & $3\times10^{-1}$\\ 
 & $3.0$ & $17.0$ & $2\times10^{-1}$\\ 
 & $1.5$ & $21.3$ & $4\times10^{-1}$ 
\enddata
\tablenotetext{a}{Average values assuming a steady wind-like mass loss, $\dot M=M_{\rm CSM}/(R/v)$, with a velocity, $v=100$\,km\,s$^{-1}$}
\end{deluxetable}

\begin{figure*}
    \centering
    \includegraphics[width=\textwidth]{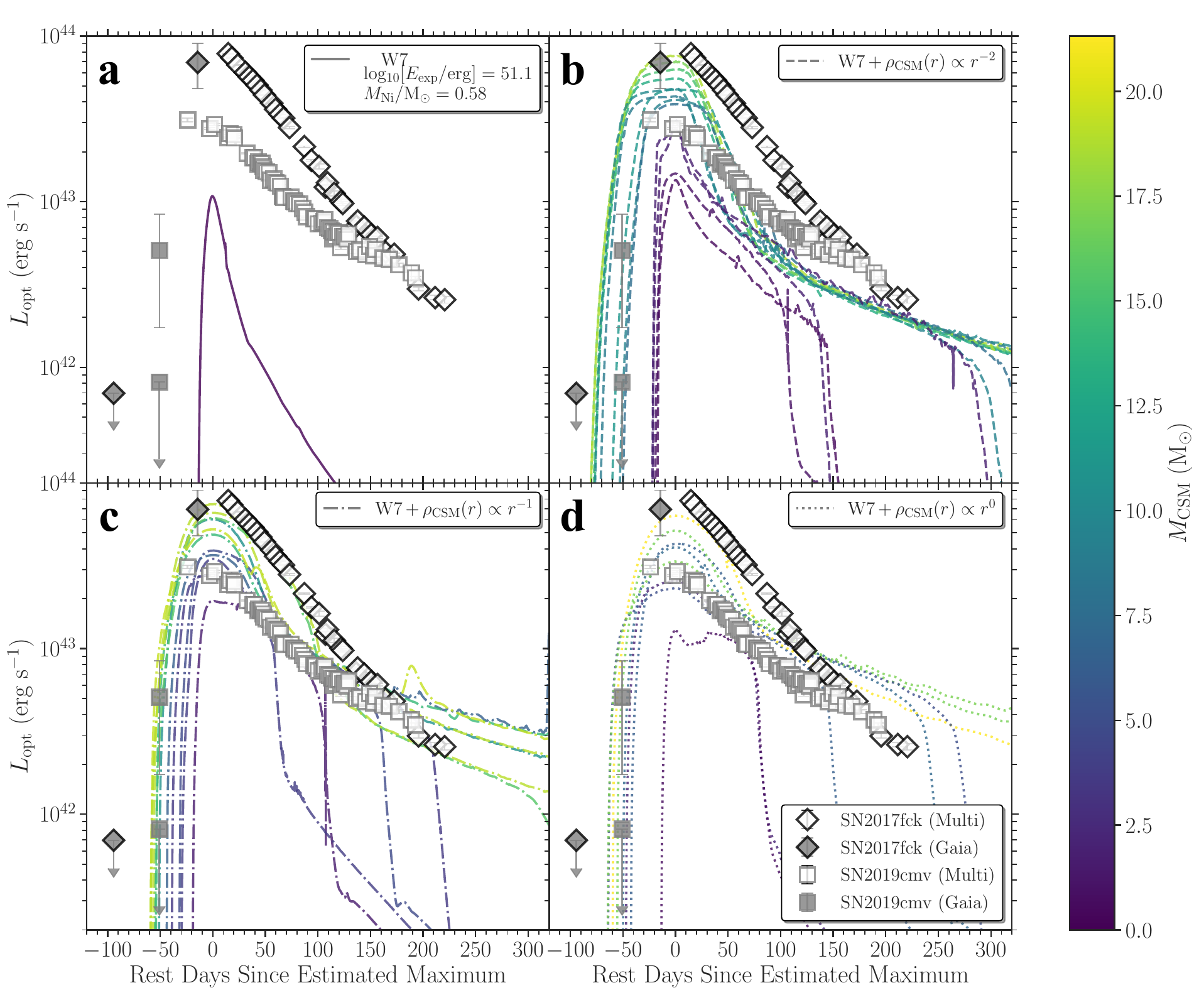}
    \caption{Comparisons of the optical light curves of SNe~2017fck and 2019cmv with the W7 (\textbf{a}) and W7+CSM models with three different CSM density distributions: $\rho_{\rm CSM}(r) \propto r^{-2}$, $r^{-1}$, and $r^{0}$ (\textbf{b}--\textbf{d}; Table~\ref{tab:modellist}), color coded by the CSM mass ($M_{\rm CSM}$). The downward arrows denote the last non-detection limits from \textit{Gaia}. With increasing $M_{\rm CSM}$, the peak luminosity and timescale 
    The CSM models with the $r^{-2}$ (and $r^0$) distributions decline too rapidly (and slowly) after the peak compared to SNe~2017fck and 2019cmv. Note that the peak luminosities and decline rates of SNe~2017fck and 2019cmv are best reproduced by the CSM models with the $r^{-1}$ distribution.
    }
    \label{fig:LCmod}
\end{figure*}

\begin{figure*}
    \centering
    \includegraphics[width=0.96\textwidth]{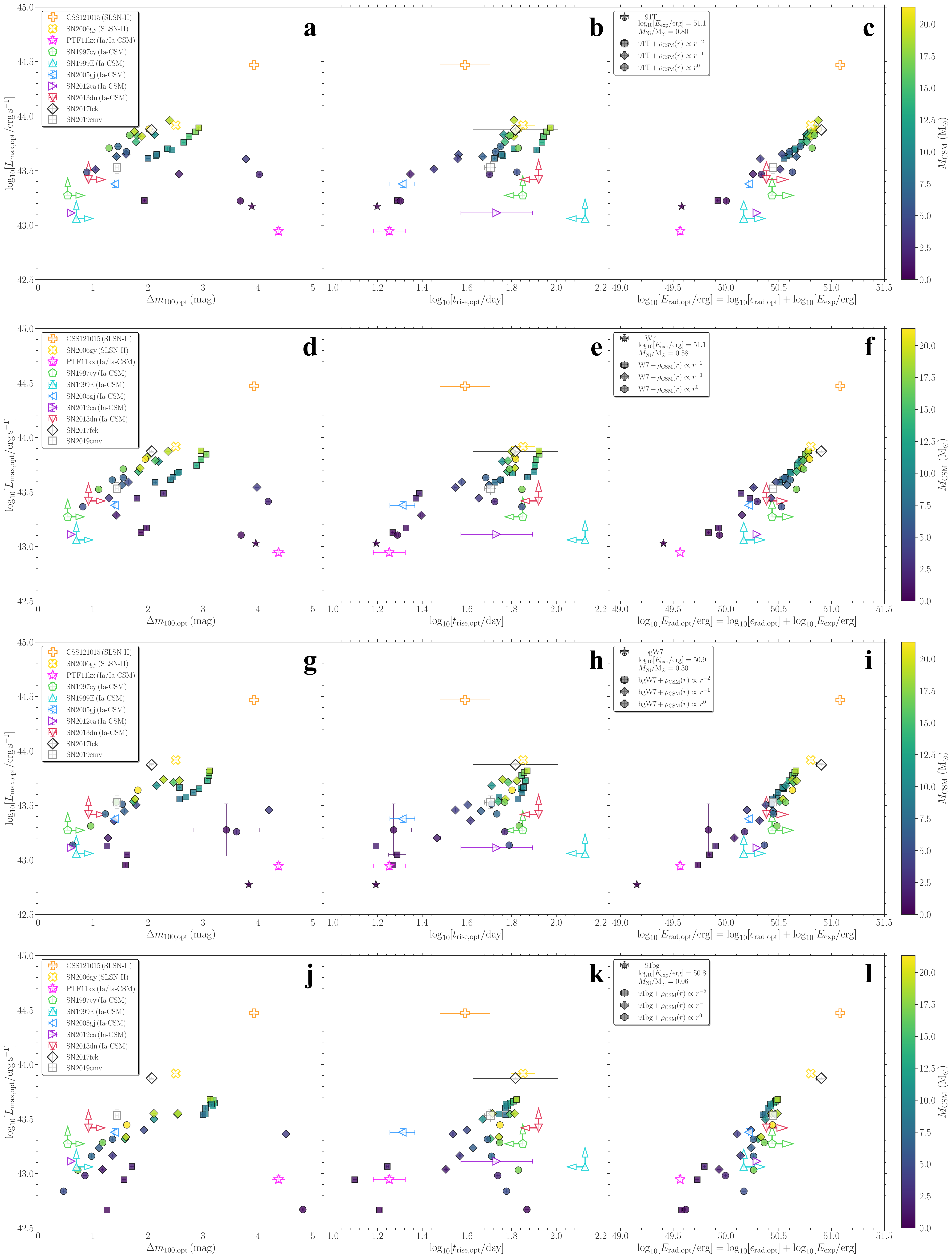}
    \caption{Comparisons of $L_{\rm max,opt}$, post-maximum decline for $100$ days ($\Delta m_{\rm 100,opt}$), $t_{\rm rise,opt}$, and $E_{\rm rad,opt}$ ($=\epsilon_{\rm rad,opt}E_{\rm exp}$) of SNe~2017fck and 2019cmv with the same sample as in Figure~\ref{fig:LCsamp} as well as the 91T+CSM (\textbf{a--c}; Figure~\ref{fig:LCmodT}), W7+CSM (\textbf{d--f}; Figure~\ref{fig:LCmod}), bgW7+CSM (\textbf{g--i}; Figure~\ref{fig:LCmodbt}), and 91bg+W7 (\textbf{j--l}; Figure~\ref{fig:LCmodbg}) models, color coded by $M_{\rm CSM}$. With increasing $M_{\rm CSM}$, $\epsilon_{\rm rad,opt}$ ranges from $0.05$ to $0.5$.
    The pure W7 model is the closest match to PTF11kx given its early normal Ia-dominated evolution. The Ia-CSM models span a wide $L_{\rm max,opt}$ range, covering most of the sample except CSS121015. 
    Note that the Ia-CSM models are able to reproduce the observed correlation between $L_{\rm max,opt}$ and $\Delta m_{\rm 100,opt}$, and also predict a correlation between $L_{\rm max,opt}$ and $t_{\rm rise,opt}$.
    }
    \label{fig:LCmodcorr}
\end{figure*}

\begin{figure}
    \centering
    \includegraphics[width=0.48\textwidth]{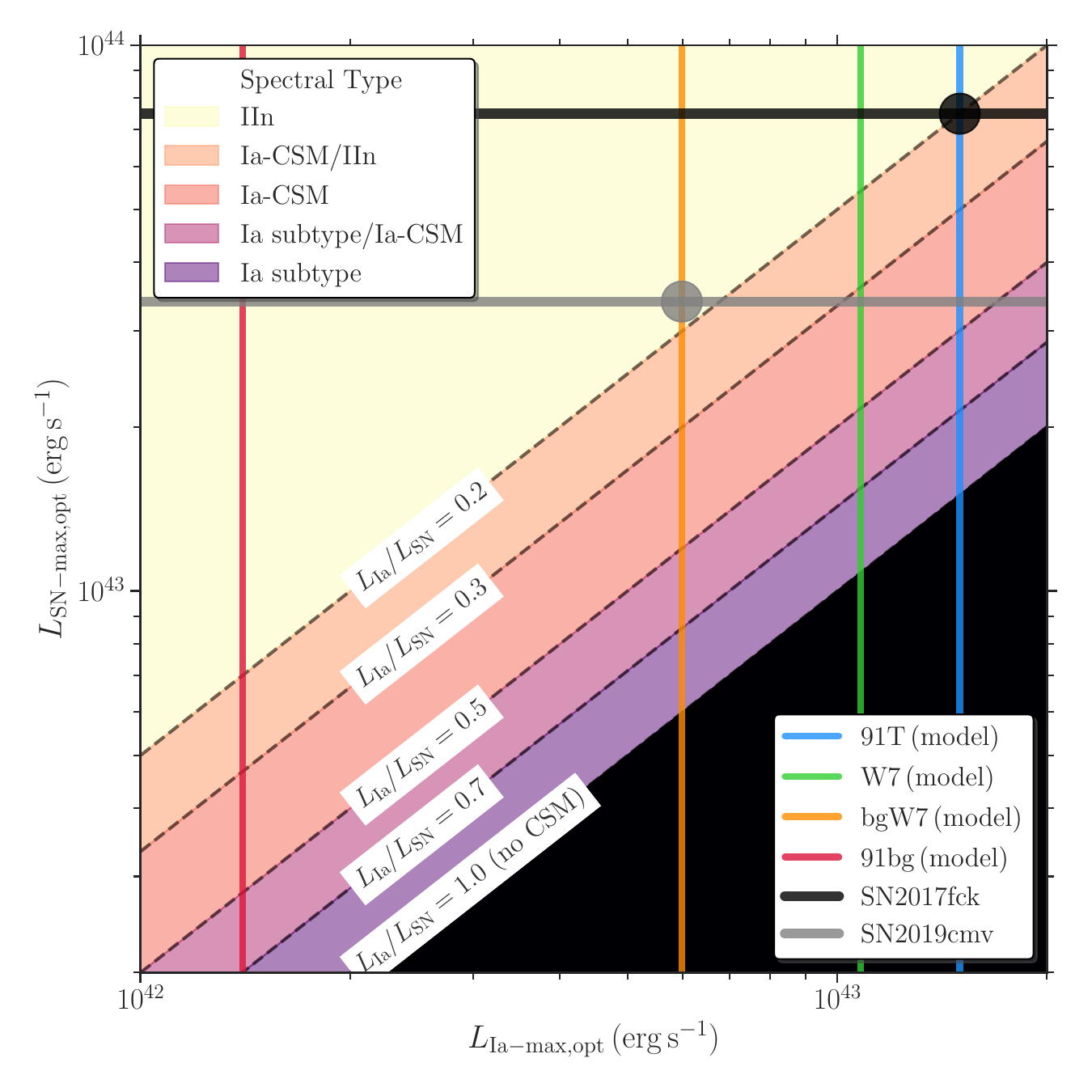}
    \caption{SN spectral types with varying CSM interaction, parameterized by the fractional optical luminosity at maximum of the underlying SN~Ia ($L_{\rm Ia-max,opt}$) to total SN ($L_{\rm SN-max,opt}$), as adopted from \cite{Leloudas2015}. The SN~Ia models (91T, W7, bgW7, 91T; Figure~\ref{fig:Iasub}) are shown as the vertical lines, while SNe~2017fck and 2019cmv are shown as the horizontal lines. With decreasing $L_{\rm Ia}/L_{\rm SN}$, i.e., increasing CSM interaction, SN spectral types change from Ia subtype (e.g., 1991T and 1991bg-like) to Ia subtype/Ia-CSM (i.e., Ia subtype with interaction signatures), Ia-CSM, Ia-CSM/IIn (i.e., mixture of both), and interaction-dominated IIn. Note that the appropriate underlying SN Ia models are 91T and bgW7 for SNe~2017fck and 2019cmv (the black and gray circles at the intersections), respectively, given their Ia-CSM/IIn and IIn spectral types (Figures~\ref{fig:speccomp} and \ref{fig:HaHb}). 
    }
    \label{fig:LsnLmod}
\end{figure}

We construct a numerical light-curve model grid of SNe~Ia that interact with dense CSM to investigate their expected properties. For this purpose, we use the one-dimensional multi-group radiation hydrodynamics code \texttt{STELLA} \citep{Blinnikov1998,Blinnikov2000,Blinnikov2006} which has been used for numerical modeling of SNe powered by CSM interaction. This allows us to calculate the optical luminosity to compare with observations (e.g., \citealt{Moriya2013,Noebauer2016,Hiramatsu2021SP}).

As a fiducial model, we take the W7 ejecta model of SN~Ia \citep{Nomoto1984} which has an ejecta mass of $1.38\,\Msun$, nickel mass ($M_{\rm Ni}$) of $0.58\,\Msun$, and kinetic energy ($E_{\rm exp}$) of $1.3\times10^{51}\,\mathrm{erg}$. In order to span the luminosity range of SN~Ia subtypes from subluminous 1991bg-like to luminous 1991T-like, we vary $M_{\rm Ni}$ and $E_{\rm exp}$ of the W7 ejecta model to produce three additional models: 91bg ($M_{\rm Ni}=0.06\,\Msun$, $E_{\rm exp}=0.6\times10^{51}$\,erg), 91T ($M_{\rm Ni}=0.80\,\Msun$, $E_{\rm exp}=1.3\times10^{51}$\,erg), and bgW7 with intermediate luminosity between 91bg and W7 ($M_{\rm Ni}=0.30\,\Msun$, $E_{\rm exp}=0.8\times10^{51}$\,erg). The SN~Ia light-curve models are shown in Figure~\ref{fig:Iasub}.

On top of each SN~Ia model, we attach solar-metallicity CSM with three different density distributions: $\rho_\mathrm{CSM}(r)\propto r^{-2}$, $r^{-1}$, and $r^0$ (i.e., constant). We change the CSM density and radius to have different CSM masses ($M_{\rm CSM}$). The CSM parameters are chosen so that $M_{\rm CSM}$ does not become excessively large ($\lesssim20\,\Msun$; see Section~\ref{sec:sd} and \citealt{Ablimit2021} for binary population synthesis calculations). We assume large CSM radii ($R=(0.5-3.5)\times 10^{16}\,\mathrm{cm}$) because the CSM interaction signatures persist for a long time in the SNe of interest (Figure~\ref{fig:LCsamp}). Average mass-loss rates ($\dot M$) can be estimated by assuming a steady wind-like mass loss with a velocity, $v$; $\dot M=M_{\rm CSM}/(R/v)$.
The CSM properties used in our numerical light-curve model grid are summarized in Table~\ref{tab:modellist}.

In Figure~\ref{fig:LCmod}, we show the comparisons of the optical light curves of SNe~2017fck and 2019cmv with the W7+CSM models. In these comparisons, we shift the reference epoch to the estimated maximum for clarity.
The light curves of SNe~2017fck and 2019cmv are obviously brighter and broader than the pure W7 model. 
The W7+CSM models with all density distributions cover comparable maximum luminosities to SNe~2017fck and 2019cmv, with sudden drops occurring when the SN ejecta reach outer CSM extent. In general, the W7+CSM models with the $r^{-2}$ density distribution decline faster from the maximum due to the less dense outer CSM than the other distributions. On the other hand, the W7+CSM models with the constant density distribution decline slower due to the denser outer CSM region.
The W7+CSM models with the $r^{-1}$ density distribution best reproduces the maximum luminosities and declines of SN~2017fck and 2019cmv, suggesting their similar CSM density distributions responsible for their superlinear declines. The same light-curve morphology can also be seen in the 91bg+CSM, 91T+CSM, and bgW7+CSM models as shown in Figures~\ref{fig:LCmodbg}, \ref{fig:LCmodT}, and \ref{fig:LCmodbt}, respectively, with different luminosity ranges from the underlying SN~Ia models.

In Figure~\ref{fig:LCmodcorr}, we show the comparisons of $L_{\rm max,opt}$, $t_{\rm rise,opt}$, post-maximum decline for $100$ days ($\Delta m_{\rm 100,opt}$)\footnote{The choice of $\Delta m_{\rm 100,opt}$, instead of $s_{\rm postmax,opt}$, is made due to numerical noise in some models, making it harder to measure the linear slope.}, and $E_{\rm rad,opt}$ ($=\epsilon_{\rm rad,opt}E_{\rm exp}$ where $\epsilon_{\rm rad,opt}$ is the optical radiation conversion efficiency) of the 91T+CSM, W7+CSM, bgW7+CSM, and 91bg+CSM models with the observed SLSN-II and SN~Ia-CSM sample (Figure~\ref{fig:LCsamp}). With increasing $M_{\rm CSM}$, $\epsilon_{\rm rad,opt}$ ranges from $0.05$ to $0.5$ (i.e., $5-50\%$ efficiency).
The transitional SN~Ia/Ia-CSM PTF11kx lies in the similar parameter space to the pure W7 model around maximum given its early normal Ia-dominated evolution.
All Ia-CSM models with the reasonable $M_{\rm CSM}$ range ($\lesssim20\,\Msun$) cannot occupy the parameter space of SLSN-II CSS121015 due to the kinetic energy budgets of $\lesssim10^{51}$\,erg (i.e., the maximum $E_{\rm rad,opt}$ with $\epsilon_{\rm rad,opt}=1$; Fig.~\ref{fig:LCsamp}), supporting the massive star origin despite the spectral similarities to SN~Ia-CSM 2005gj, as also suggested by \cite{Benetti2014}.
In contrast, the 91T+CSM and W7+CSM models with $M_{\rm CSM}\approx15-19\,\Msun$ can reproduce the parameter space of SLSN-II 2006gy reasonably well, indicating the possible Ia-CSM origin, which is in broad agreement with \cite{Jerkstrand2020} who find $M_{\rm CSM}\sim13\,\Msun$. 
On the faint end, the W7+CSM and bgW7+CSM models with $M_{\rm CSM}\approx1.6$ and $5.1\,\Msun$, respectively, can reproduce the parameter space of SNe~Ia-CSM 2005gj and 2012ca. 

In addition, all Ia-CSM models reproduce a positive correlation in $L_{\rm max,opt}$\,vs.\,$\Delta m_{\rm 100,opt}$ observed in the SN~Ia-CSM sample, suggesting that the light-curve diversity in SNe~Ia-CSM ($L_{\rm max,opt}\approx10^{43}-10^{44}$\,erg\,s$^{-1}$ from SNe~2012ca and 2005gj to 2006gy) can be explained by the diversity in CSM distributions (i.e., $M_{\rm CSM}\approx1.6-15\,\Msun$). 
The models also predict a positive correlation in $L_{\rm max,opt}$\,vs.\,$t_{\rm rise,opt}$, which needs to be tested with more SNe~Ia-CSM with better-constrained explosion epochs. Another constraining observable would be a drop from the linear decline phase, as observed in SNe~1997cy, 1999E, and 2012ca, which should correlate with the outer CSM extent (Figure~\ref{fig:LCsamp}).

Together with the SN~Ia-CSM light-curve modeling, we utilize the spectral evolution of SNe~2017fck and 2019cmv to characterize their possible underlying SN~Ia and CSM properties. In Figure~\ref{fig:LsnLmod}, we show the comparison of the peak optical luminosities of total SN ($L_{\rm SN}$) and underlying SN~Ia ($L_{\rm Ia}$), where their fractional luminosity ($L_{\rm Ia}/L_{\rm SN}$) defines the parameter space of SN spectral types deduced from the Monte Carlo simulation of \cite{Leloudas2015}. With increasing CSM interaction, i.e., decreasing $L_{\rm Ia}/L_{\rm SN}$, the resultant SN spectral types continuously change from Ia subtype ($L_{\rm Ia}/L_{\rm SN}>0.7$), to Ia subtype/Ia-CSM ($0.5<L_{\rm Ia}/L_{\rm SN}<0.7$), Ia-CSM ($0.3<L_{\rm Ia}/L_{\rm SN}<0.5$), Ia-CSM/IIn ($0.2<L_{\rm Ia}/L_{\rm SN}<0.3$), and interaction-dominated IIn ($L_{\rm Ia}/L_{\rm SN}<0.2$). 
As discussed in Section~\ref{sec:spec} (Figures~\ref{fig:speccomp} and \ref{fig:HaHb}), the spectral evolution of SN~2017fck starts with IIn-like around light-curve peak and changes to Ia-CSM-like, while that of SN~2019cmv stays IIn-like throughout with a hint of Ia-CSM. Thus in this context, we consider the spectral types of SNe~2017fck and 2019cmv as Ia-CSM/IIn and IIn, respectively, which determines their probable underlying SN~Ia models to be 91T and bgW7,\footnote{We prefer bgW7 over 91bg for SN~2019cmv due to its subtle spectral similarity to Ia-CSM.} given their spectral parameter space at the intersections of $L_{\rm SN}$ and $L_{\rm Ia}$. Consequently, as in Figure~\ref{fig:LCmodcorr}, the 91T+CSM (Figure~\ref{fig:LCmodT}) and bgW7+CSM (Figure~\ref{fig:LCmodbt}) models provide good matches to SNe~2017fck and 2019cmv with $M_{\rm CSM}\approx10.7$ and $6.3\,\Msun$, respectively, both with the $\rho_{\rm CSM}(r)\propto r^{-1}$ distribution.

\section{Discussion and Summary} \label{sec:sd}

\subsection{SN~Ia-CSM Progenitor System} \label{sec:progenitor}

The SN~Ia-CSM light-curve models can cover the observed SN~Ia-CSM parameter space with various CSM distributions (i.e., $M_{\rm CSM}\sim1.6-10.7$ or $15\,\Msun$, excluding or including SLSN-II~2006gy, respectively); however, a remaining question is whether SN~Ia progenitor systems are able to produce such high-mass CSM.
One possible scenario is an interacting binary system consisting of a CO WD and intermediate-mass ($<8\,\Msun$) or massive ($\geq8\,\Msun$) companion star that undergoes common envelope (CE) evolution and explodes by a core merger inside the CE, i.e., the CD scenario (e.g., \citealt{Iben1984ApJS...54..335I,Terman1994,Taam2000,Livio2003ApJ...594L..93L,Chugai2004AstL...30...65C,Hachisu2008ApJ...679.1390H,Kashi2011MNRAS.417.1466K,Sabach2014,Soker2019,Ablimit2021}). In such a scenario, the initially massive primary turns into a CO WD after stable Roche‑lobe overflow, from which the secondary receives mass and grows into an O/B star. When this O/B companion evolves to the Hertzsprung‐gap or core‑helium‑burning phase, it overflows its Roche lobe where the resulting mass ratio (WD/companion $\ll 1$) makes the transfer unstable, forming a massive CE that engulfs the CO WD.


From the binary population synthesis calculations by \cite{Ablimit2021}, CE masses of $\sim2-7$ and $6-12\,\Msun$ and delay times of $\sim0.1-10$ and $0.03-0.12$\,Gyr are expected with an intermediate-mass and massive companion star, respectively.
Therefore, CE evolution could in principle span the inferred CSM mass range of SNe~Ia-CSM (except SN~2006gy), although detailed simulations are necessary to confirm whether the final product of CE evolution is indeed a core merger inducing an SN~Ia explosion.
In this scenario, the progenitor system of SN~2017fck ($M_{\rm CSM}\approx10.7\,\Msun$) would be a CO WD + massive companion, while that of SN~2019cmv ($M_{\rm CSM}\approx6.3\,\Msun$) would be a CO WD + intermediate-mass companion whose longer delay time would then allow the large offset from the potential elliptical host galaxy. Additionally, the inferred $\rho_{\rm CSM}(r)\propto r^{-1}$ distribution for SNe~2017fck and 2019cmv may correspond to the shallow density distribution expected in homologously expanding CE \citep{Valsan2023MNRAS.526.5365V}. 
In contrast, the progenitor system of SN~2006gy ($M_{\rm CSM}\approx15-19\,\Msun$) would merit further numerical investigation.

\subsection{SN~Ia-CSM Signatures} \label{sec:criteria}

{
\setlength{\tabcolsep}{2.2pt}
\begin{deluxetable*}{cccccccc}[t]
\tablecaption{\cite{Silverman2013_sample} and \cite{Sharma2023ApJ...948...52S} SN~Ia-CSM Signatures\tablenotemark{a} \label{tab:S13}}
\tablehead{
\colhead{SN} & \colhead{Peak Luminosity} & \colhead{Spectral Homogeneity:} & \colhead{H$\alpha$ Profile:} & \colhead{Line Flux:} & \colhead{Early UV} & \colhead{Mass Loss} & \colhead{Host Galaxy:} \\
\colhead{} & \colhead{$-21.3 \leq M_{R/r} \leq -19$} & \colhead{H$\alpha$} & \colhead{narrow P-Cygni} & \colhead{weak He~{\sc i}} & \colhead{\& late MIR} & \colhead{$\sim10^{-3}-10^{-1}$} & \colhead{spiral}  \\
\colhead{} & \colhead{\& Rise Time} & \colhead{Ca~{\sc ii} NIR triplet} & \colhead{strong fluctuations} & \colhead{weak H$\beta$} & \colhead{no radio} & \colhead{$\Msun\,{\rm yr}^{-1}$} & \colhead{or}  \\
\colhead{} & \colhead{$\sim20-40$\,days} & \colhead{diluted SN~Ia} & \colhead{suppressed red wing} & \colhead{large H$\alpha$/H$\beta$} & \colhead{no X-ray} & \colhead{(assuming wind)} & \colhead{dwarf}
}
\startdata
2017fck & $\checkmark?$ & $\checkmark$ & $\checkmark$ & $\checkmark$ & -- & $\checkmark$ & $\checkmark$\\
2019cmv & $\checkmark?$ & $?$ & $\checkmark?$ & $\checkmark$ & -- & $\checkmark$ & $\times$
\enddata
\tablenotetext{a}{Check marks, check+question marks, and cross marks (respectively) indicate observations consistent, perhaps consistent, and inconsistent with the signatures. Dashed lines indicate the lack of observational constraints, and lone question marks indicate unknowns.}
\end{deluxetable*}
}

As the summary of this work, we check the observed and modeled properties of SNe~2017fck and 2019cmv against the SN~Ia-CSM signatures from the PTF \citep{Silverman2013_sample} and ZTF \citep{Sharma2023ApJ...948...52S} samples.
The checklist is shown in Table~\ref{tab:S13} and discussed in the following.

\begin{itemize}

\item Peak absolute magnitudes of $-21.3 \leq M_{R/r} \leq -19$ and rise times of $\sim20-40$ days are observed for SNe~Ia-CSM.
In the \textit{G} band (with the similar effective wavelength to the \textit{R/r} bands, SNe~2017fck and 2019cmv show the peak magnitudes of $-21.8 \leq M_G \leq -20.4$ and rise times of $\sim50-66$ days (Figures~\ref{fig:LC} and \ref{fig:LCsamp}) that are slightly brighter and substantially longer, respectively, but still comparable to the modeled SN~Ia-CSM population.
These are reaching into the SLSN regime.

\item Spectral homogeneity is seen for SNe~Ia-CSM in that they show strong and broad H$\alpha$ ($\sim2$,$000$\,km\,s$^{-1}$) and Ca~{\sc ii} NIR triplet ($\sim10$,$000$\,km\,s$^{-1}$) on top of a diluted SN~Ia-like ``quasi-continuum'' with Fe~{\sc ii/iii} absorption forest.
The post-maximum spectra of SN~2017fck ($\sim120-280$ d) show all the above features (Figure~\ref{fig:speccomp}), while the spectra of SN~2019cmv (observed up to $\sim520$ d) are dominated by H$\alpha$, not showing a clear quasi-continuum or Ca~{\sc ii} NIR triplet, which likely suggests different levels of CSM interaction (Figure~\ref{fig:LsnLmod}).
On the other hand, if SLSN-II 2006gy is indeed an SN~Ia-CSM, some spectral heterogeneity within the SN~Ia-CSM population is expected.

\item For the H$\alpha$ profiles of SNe~Ia-CSM, narrow P-Cygni profiles (with absorption at $50-100$\,km\,s$^{-1}$), fluctuating emission profiles (until $\sim100-150$ days after maximum), and suppressed red wings (after $\sim75-100$ days after maximum) are observed.
SNe~2017fck and 2019cmv show H$\alpha$ fluctuations and suppressed red wings (Figure~\ref{fig:spec}). The early-time ($\sim79$ d) spectrum of SN~2017fck shows a narrow H$\alpha$ P-Cygni profile ($\sim400$\,km\,s$^{-1}$) as well, while the spectral resolutions of the other SN~2017fck and 2019cmv spectra are not high enough to resolve it (Figure~\ref{fig:speccomp}). 

\item Weak He~{\sc i} and H$\beta$ emission lines and large H$\alpha$/H$\beta$ flux ratios are observed for SNe~Ia-CSM.
SN~2017fck and 2019cmv show all the above features (Figures~\ref{fig:speccomp} and \ref{fig:HaHb}). However, we note that the observed H$\alpha$/H$\beta$ flux ratios of SNe~IIn and Ia-CSM cover a wide range, making the diagnosis somewhat inconclusive. 
Interestingly, SN~2006gy also shows large H$\alpha$/H$\beta$ flux ratios.

\item Multi-wavelength observations of SNe~Ia-CSM have revealed early ultraviolet (UV) emission without radio or X-ray counterparts (within the first few months) and late Mid-IR (MIR) emission ($\sim0.5-2$ years).
Unfortunately, we do not have any of the above constraints for SNe~2017fck and 2019cmv.

\item Using rise times, H$\alpha$ luminosities, and X-ray upper limits with several analytical assumptions, wind mass-loss rates of $\sim10^{-3}-10^{-1}\,\Msun\,{\rm yr}^{-1}$ can be inferred for SNe~Ia-CSM.
With the numerical CSM light-curve modeling (Figures~\ref{fig:LCmod} and \ref{fig:LCmodcorr}), we infer large CSM masses of $\sim6.3-10.7\,\Msun$ with the $\rho_{\rm CSM}(r)\propto r^{-1}$ distribution and average mass-loss rates of $(1-2)\times10^{-1}\,\Msun\,{\rm yr}^{-1}$, respectively, for SNe~2017fck and 2019cmv (Table~\ref{tab:modellist}). Additionally, the models can cover the SN~Ia-CSM parameter space with a CSM mass range of $\sim1.6-10.7\,\Msun$, or $15\,\Msun$ if including SN~2006gy.

\item Spiral galaxies (with MW-like luminosities and solar metallicities) and dwarf irregulars (with Magellanic Clouds-like luminosities and subsolar metallicities) are identified for SNe~Ia-CSM.
The host galaxy of SN~2017fck is a low-mass, star-forming galaxy at a subsolar metallicy (Figures~\ref{fig:host} and \ref{fig:MOffSFR} and Table~\ref{tab:hostclass}). In contrast, SN~2019cmv has a large offset from the potential massive elliptical galaxy at a subsolar metallicity, although a ultra-faint host galaxy ($M_g\gtrsim-13.3$) cannot be ruled out.
Given the expected longer delay times of SN Ia-CSM progenitor systems than typical SLSN-II massive progenitors, however, we consider the Ia-CSM origin more likely.

\end{itemize}

In addition to the above SN~Ia-CSM signatures, we suggest the superlinear light curves and correlations (Figures~\ref{fig:LCsamp} and \ref{fig:LCcorr}) as possible signatures, which can also be reproduced by the numerical SN~Ia-CSM models with various underlying Ia subtypes and CSM density distributions (Figures~\ref{fig:LCmod}, \ref{fig:LCmodcorr}, and \ref{fig:LsnLmod}). 
With more well-observed samples anticipated from the Vera C. Rubin
Observatory’s Legacy Survey of Space and Time (limiting magnitude $\approx 25$; \citealt{LSST2019ApJ...873..111I}), the late-time light-curve drops and nebular spectra would be another set of constraining signatures, probing the extent of CSM and underlying SN composition.

SN~2017fck satisfies all the above SN~Ia-CSM signatures where observations are available, while SN~2019cmv may be missing some of the spectral signatures given its slower evolution.
With the possible addition of SNe~2017fck and 2019cmv to the SN~Ia-CSM population, SNe~Ia-CSM would cover wider light-curve parameter space, even into the SLSN regime (up to $L_{\rm max,opt}\approx10^{44}$\,erg\,s$^{-1}$), requiring the diversity in CSM distributions (i.e., $M_{\rm CSM}\sim1.6-10.7\,\Msun$) which might be produced during the CE phase of a WD and intermediate-mass or massive companion stars. Further numerical simulations of the CD scenario inside the CE is required to test its explosive outcomes and CSM environments.

\section*{Acknowledgments}

We are grateful to Lars Bildsten, Timothy D. Brandt, Jared A. Goldberg, Morgan MacLeod, Keiichi Maeda, Yuan Qi Ni, Akihiro Suzuki, Tomoya Takiwaki, Nozomu Tominaga, and Daichi Tsuna for their comments and discussions.

D.H. is supported by STScI grants HST-GO-17770.002, JWST-GO-12468.001, and JWST-GO-09964.001.
D.H., D.A.H., G.H., C.M., and J.B. were supported by National Science Foundation (NSF) grants AST-1313484, AST-1911225, and AST-1911151, as well as by National Aeronautics and Space Administration (NASA) grant 80NSSC19kf1639.
D.H. is thankful for support and hospitality by the National Astronomical Observatory of Japan where many discussions of this work took place.
I.A. is a CIFAR Azrieli Global Scholar in the Gravity and the Extreme Universe Program and acknowledges support from that program, from the European Research Council (ERC) under the European Union's Horizon 2020 research and innovation program (grant agreement number 852097), from the Israel Science Foundation (grant numbers 2108/18 and 2752/19), from the United States - Israel Binational Science Foundation, and from the Israeli Council for Higher Education Alon Fellowship.
This work at Rutgers University (S.W.J. and Y.E.) was supported by NSF awards AST-1615455 and AST-2407567. S.W.J. also gratefully acknowledges support from a Guggenheim Fellowship.
S.V. and the UC Davis time-domain research team acknowledge support by NSF grants AST-2407565.
S.B.'s work was carried out within the framework of the state assignment of NRC ``Kurchatov Institute".

This work makes use of observations from the Las Cumbres
Observatory global telescope network.
The authors wish to recognize and acknowledge the very significant cultural role and reverence that the summit of Haleakal$\bar{\text{a}}$ has always had within the indigenous Hawaiian community. We are most fortunate to have the opportunity to conduct observations from the mountain.
This paper includes data gathered
with the 6.5m Magellan Telescopes located at the Las
Campanas Observatory, Chile.
Some of the data presented herein were obtained at the W. M. Keck
Observatory, which is operated as a scientific partnership among the
California Institute of Technology, the University of California, and
NASA; the observatory was made possible by the generous financial
support of the W. M. Keck Foundation.
Some of the observations reported in this paper were obtained with the Southern African Large Telescope (SALT).

Numerical computations were in part carried out on PC cluster at the Center for Computational Astrophysics, National Astronomical Observatory of Japan.

This research has made use of NASA’s Astrophysics Data System (ADS), the NASA/IPAC Extragalactic Database (NED) and NASA/IPAC Infrared Science Archive (IRSA) (which is funded by NASA and operated by the California Institute of Technology), and IRAF (which is distributed by the National Optical Astronomy Observatory (NOAO), operated by the Association of Universities for Research in Astronomy (AURA), Inc., under cooperative agreement with NSF).

TNS is supported by funding from the Weizmann Institute of Science, as well as grants from the Israeli Institute for Advanced Studies and the European Union via ERC grant No. 725161.

We acknowledge ESA \textit{Gaia}, DPAC and the Photometric Science Alerts Team (\url{http://gsaweb.ast.cam.ac.uk/alerts}).


ZTF is supported by NSF grant AST-1440341 and a collaboration including Caltech, IPAC, the Weizmann Institute for Science, the Oskar Klein Center at Stockholm University, the University of Maryland, the University of Washington, Deutsches Elektronen-Synchrotron and Humboldt University, Los Alamos National Laboratories, the TANGO Consortium of Taiwan, the University of Wisconsin at Milwaukee, and Lawrence Berkeley National Laboratories. Operations are conducted by COO, IPAC, and UW.
The ZTF forced-photometry service was funded under the Heising-Simons Foundation grant
\#12540303 (PI: Graham).

This work has made use of data from the Asteroid Terrestrial-impact Last Alert System (ATLAS) project. ATLAS is primarily funded to search for near earth asteroids through NASA grants NN12AR55G, 80NSSC18K0284, and 80NSSC18K1575; byproducts of the NEO search include images and catalogs from the survey area. This work was partially funded by Kepler/K2 grant J1944/80NSSC19K0112 and HST GO-15889, and STFC grants ST/T000198/1 and ST/S006109/1. The ATLAS science products have been made possible through the contributions of the University of Hawaii Institute for Astronomy, the Queen’s University Belfast, the Space Telescope Science Institute, the South African Astronomical Observatory, and The Millennium Institute of Astrophysics (MAS), Chile.

The PS1 and the PS1 public science archive have been made possible through contributions by the Institute for Astronomy, the University of Hawaii, the Pan-STARRS Project Office, the Max-Planck Society and its participating institutes, the Max Planck Institute for Astronomy, Heidelberg and the Max Planck Institute for Extraterrestrial Physics, Garching, The Johns Hopkins University, Durham University, the University of Edinburgh, the Queen's University Belfast, the Harvard-Smithsonian Center for Astrophysics, the Las Cumbres Observatory Global Telescope Network Incorporated, the National Central University of Taiwan, the Space Telescope Science Institute, NASA under Grant No. NNX08AR22G issued through the Planetary Science Division of the NASA Science Mission Directorate, NSF Grant No. AST-1238877, the University of Maryland, Eotvos Lorand University, the Los Alamos National Laboratory, and the Gordon and Betty Moore Foundation.

The Legacy Surveys consist of three individual and complementary projects: the Dark Energy Camera Legacy Survey (DECaLS; Proposal ID \#2014B-0404; PIs: David Schlegel and Arjun Dey), the Beijing-Arizona Sky Survey (BASS; NOAO Prop. ID \#2015A-0801; PIs: Zhou Xu and Xiaohui Fan), and the Mayall z-band Legacy Survey (MzLS; Prop. ID \#2016A-0453; PI: Arjun Dey). DECaLS, BASS and MzLS together include data obtained, respectively, at the Blanco telescope, Cerro Tololo Inter-American Observatory, NSF’s NOIRLab; the Bok telescope, Steward Observatory, University of Arizona; and the Mayall telescope, Kitt Peak National Observatory, NOIRLab. The Legacy Surveys project is honored to be permitted to conduct astronomical research on Iolkam Du’ag (Kitt Peak), a mountain with particular significance to the Tohono O’odham Nation.

The Legacy Survey team makes use of data products from the Near-Earth Object Wide-field Infrared Survey Explorer (NEOWISE), which is a project of the Jet Propulsion Laboratory/California Institute of Technology. NEOWISE is funded by the National Aeronautics and Space Administration.

The Legacy Surveys imaging of the DESI footprint is supported by the Director, Office of Science, Office of High Energy Physics of the U.S. Department of Energy under Contract No. DE-AC02-05CH1123, by the National Energy Research Scientific Computing Center, a DOE Office of Science User Facility under the same contract; and by the U.S. National Science Foundation, Division of Astronomical Sciences under Contract No. AST-0950945 to NOAO.


\textit{Facilities:} 
ADS, ATLAS, FTN (FLOYDS), FTS (FLOYDS), Gaia, IRSA, Keck:I (LRIS), Keck:II (DEIMOS), LCOGT (Sinistro), Liverpool:2m (SPRAT), Magellan:Clay (LDSS-3), NED, PO:1.2m (ZTF), PS1, SALT (RSS), SOAR (Goodman), ZTF.


\textit{Software:}
\texttt{Astropy} \citep{AstropyCollaboration2013,AstropyCollaboration2018,AstropyCollaboration2022}, 
\texttt{emcee} \citep{emcee2013}, 
\texttt{floyds\_pipeline} \citep{Valenti2014}, \texttt{fundamentals} \citep{Young2023zndo...8037510Y},  \texttt{FSPS} \citep{Conroy2009,Conroy2010}, 
\texttt{lcogtsnpipe} \citep{Valenti2016}, 
\texttt{LPipe} \citep{Perley2019PASP..131h4503P},
\texttt{Matplotlib} \citep{Hunter2007}, 
\texttt{NumPy} \citep{Oliphant2006}, 
\texttt{PyMCZ} \citep{Bianco2016}, 
\texttt{PyRAF} \citep{PyRAF2012}, 
\texttt{PySALT} \citep{Crawford2010}, 
\texttt{PyZOGY} \citep{Guevel2017}, 
\texttt{Prospector} \citep{Johnson2021}, 
\texttt{SciPy} \citep{SciPy2020}, 
\texttt{seaborn} \citep{Waskom2020}, 
\texttt{sedpy} \citep{sedpy}, 
\texttt{SExtractor} \citep{Bertin1996}, 
\texttt{STELLA} \citep{Blinnikov1998,Blinnikov2000,Blinnikov2006,Blinnikov2004,Baklanov2005}.

\appendix
\section{Extra SN Ia-CSM Light-Curve Models}\label{sec:extra}

\begin{figure*}
    \centering
    \includegraphics[width=0.98\textwidth]{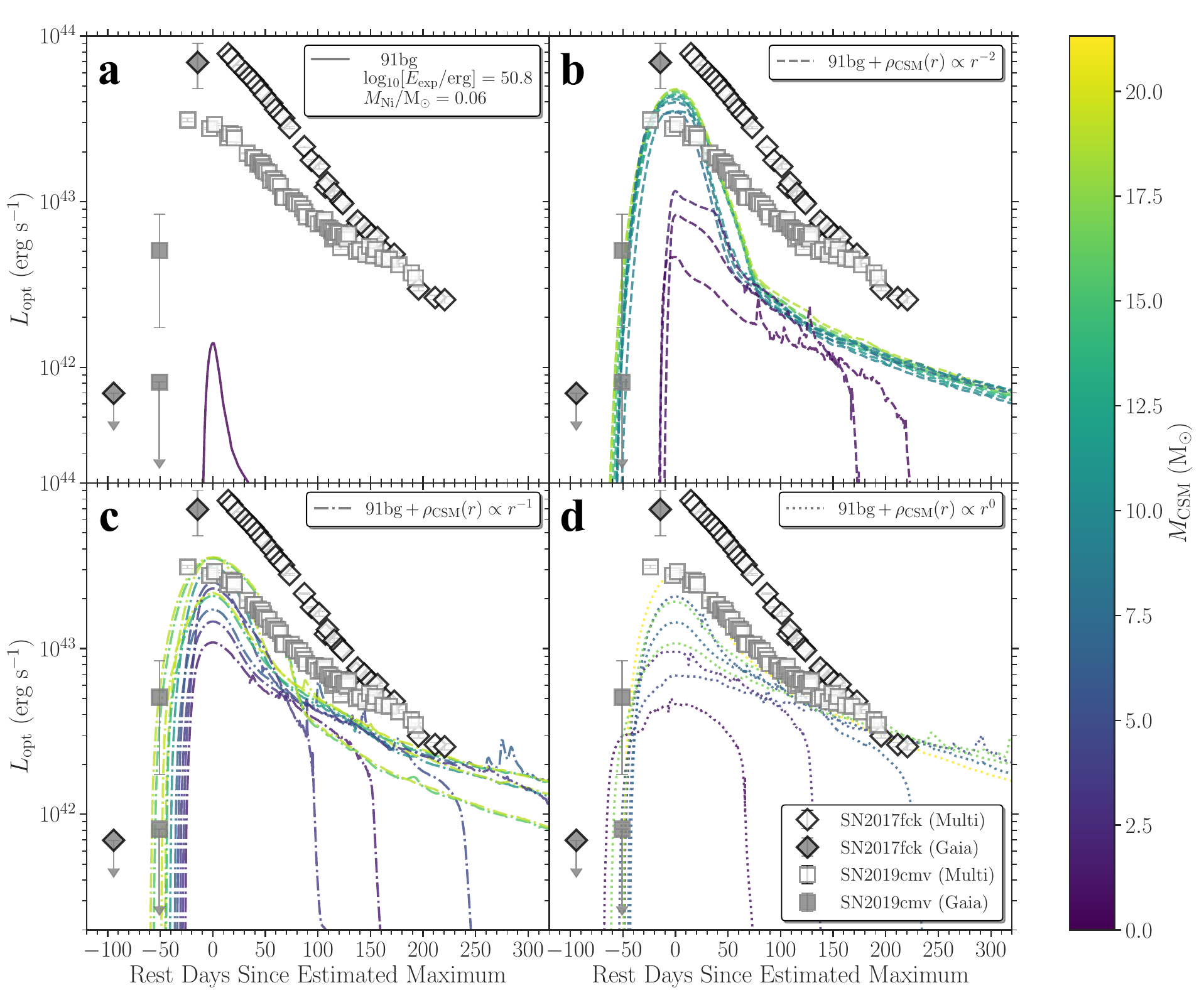}
    \caption{Same as Figure~\ref{fig:LCmod}, but with 91bg model.
    }
    \label{fig:LCmodbg}
\end{figure*}

\begin{figure*}
    \centering
    \includegraphics[width=0.98\textwidth]{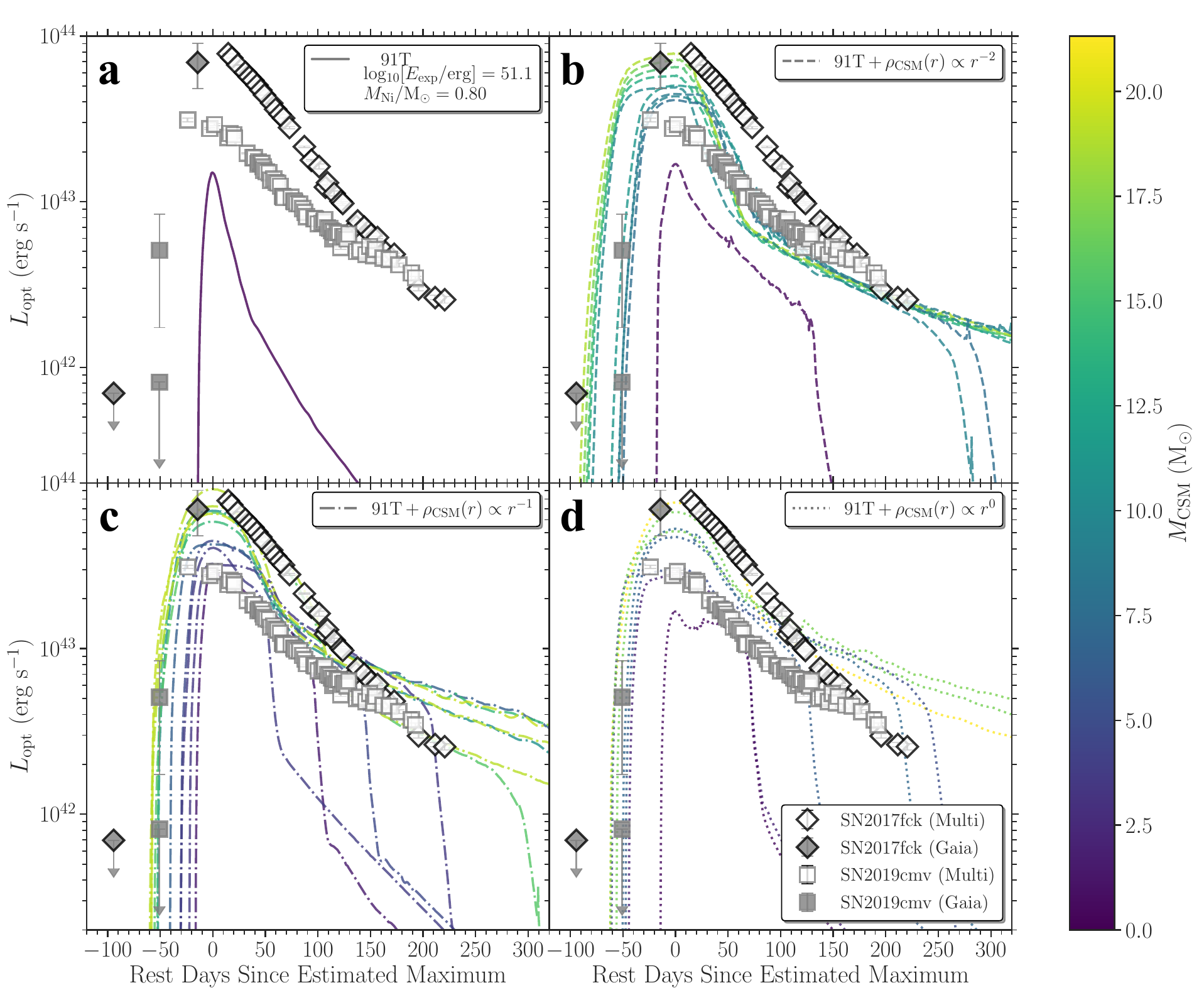}
    \caption{Same as Figure~\ref{fig:LCmod}, but with 91T model.
    }
    \label{fig:LCmodT}
\end{figure*}

\begin{figure*}
    \centering
    \includegraphics[width=0.98\textwidth]{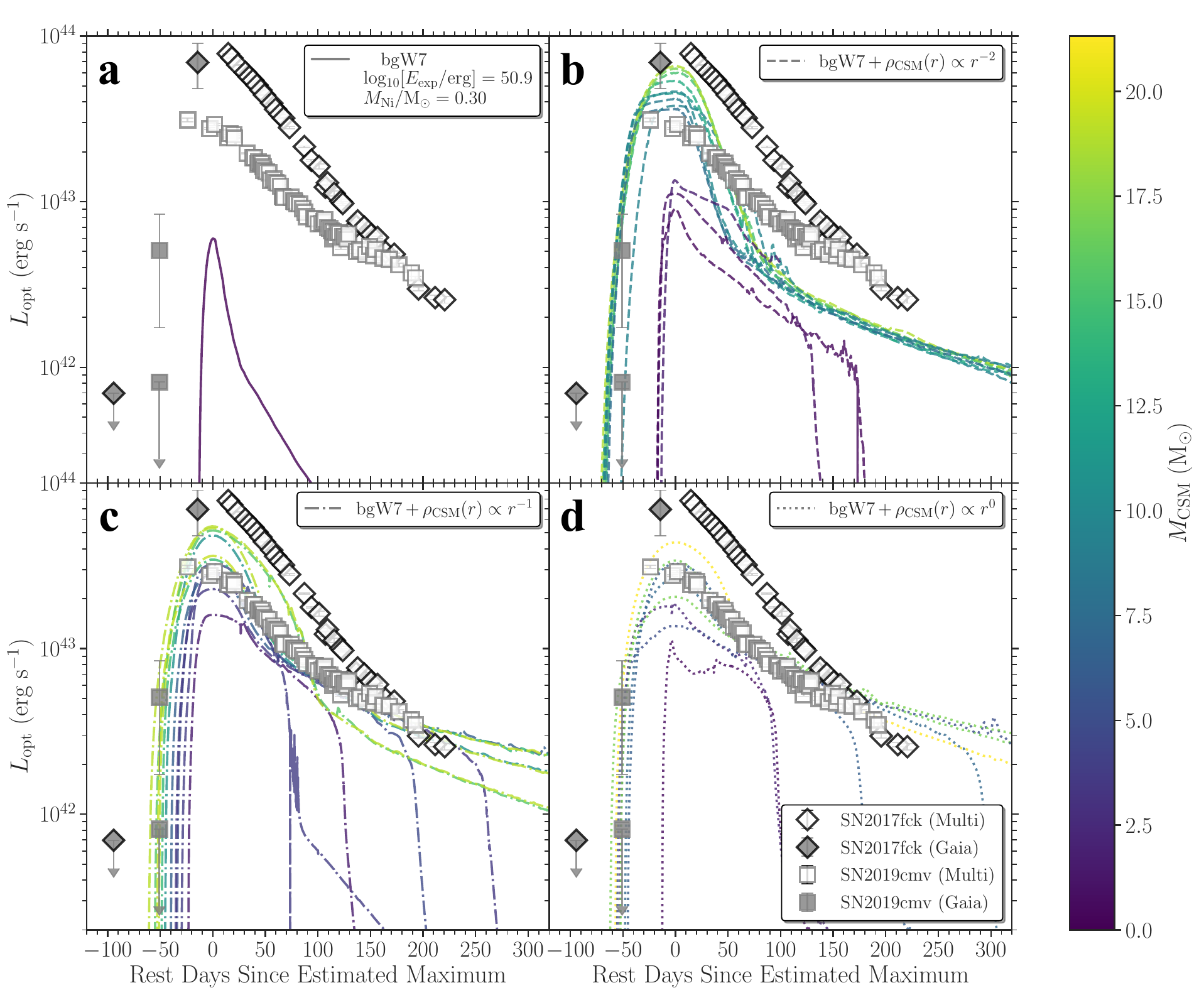}
    \caption{Same as Figure~\ref{fig:LCmod}, but with bgW7 model.
    }
    \label{fig:LCmodbt}
\end{figure*}

The CSM light-curve grids for 91T, 91bg, and bgW7 models are shown in Figures~\ref{fig:LCmodbg}, \ref{fig:LCmodT}, and \ref{fig:LCmodbt}.

\bibliographystyle{aasjournal}
\bibliography{main}

\end{document}

%% file: affil.tex
\newcommand{\UF}{\affiliation{$^{1}$Department of Astronomy, University of Florida, Bryant Space Science Center, Gainesville, FL 32611-2055, USA}}
\newcommand{\LCO}{\affiliation{$^{2}$Las Cumbres Observatory, 6740 Cortona Drive, Suite 102, Goleta, CA 93117-5575, USA}}
\newcommand{\UCSB}{\affiliation{$^{3}$Department of Physics, University of California, Santa Barbara, CA 93106-9530, USA}}
\newcommand{\NAOJ}{\affiliation{$^{4}$National Astronomical Observatory of Japan, National Institutes of Natural Sciences, 2-21-1 Osawa, Mitaka, Tokyo 181-8588, Japan}}
\newcommand{\Monash}{\affiliation{$^{5}$School of Physics and Astronomy, Faculty of Science, Monash University, Clayton, VIC 3800, Australia}}
\newcommand{\TAU}{\affiliation{$^{6}$School of Physics and Astronomy, Tel Aviv University, Tel Aviv 69978, Israel}}
\newcommand{\CIFAR}{\affiliation{$^{7}$CIFAR Azrieli Global Scholars program, CIFAR, Toronto, Canada}}
\newcommand{\Shady}{\affiliation{$^{8}$Shady Side Academy, 423 Fox Chapel Rd, Pittsburgh, PA 15238, USA}}
\newcommand{\UCSD}{\affiliation{$^{9}$Department of Astronomy \& Astrophysics, University of California, San Diego, 9500 Gilman Drive, MC 0424, La Jolla, CA 92093-0424, USA}}
\newcommand{\UCD}{\affiliation{$^{10}$Department of Physics, University of California, 1 Shields Avenue, Davis, CA 95616-5270, USA}}
\newcommand{\Toronto}{\affiliation{$^{11}$Department of Astronomy and Astrophysics, University of Toronto, Toronto, ON M5S 3H4, Canada}}
\newcommand{\Rutgers}{\affiliation{$^{12}$Department of Physics and Astronomy, Rutgers, The State University of New Jersey, 136 Frelinghuysen Road, Piscataway, NJ 08854, USA}}
\newcommand{\RAPCorp}{\affiliation{$^{13}$Renaissance of American Physics and Astronomy, 310 Adams Street, Gravity, IA 50848, USA}}
\newcommand{\NRC}{\affiliation{$^{14}$NRC `Kurchatov Institute', Acad. Kurchatov Sq. 1, Moscow 123182, Russia}}
\newcommand{\IPMU}{\affiliation{$^{15}$Kavli Institute for the Physics and Mathematics of the Universe (WPI), The University of Tokyo Institutes for Advanced Study, The University of Tokyo, 5-1-5 Kashiwanoha, Kashiwa, Chiba 277-8583, Japan}}


%% file: main.bib
@ARTICLE{Valenti2016,
       author = {{Valenti}, S. and {Howell}, D.~A. and {Stritzinger}, M.~D. and
         {Graham}, M.~L. and {Hosseinzadeh}, G. and {Arcavi}, I. and
         {Bildsten}, L. and {Jerkstrand}, A. and {McCully}, C. and
         {Pastorello}, A. and {Piro}, A.~L. and {Sand}, D. and {Smartt}, S.~J. and
         {Terreran}, G. and {Baltay}, C. and {Benetti}, S. and {Brown}, P. and
         {Filippenko}, A.~V. and {Fraser}, M. and {Rabinowitz}, D. and
         {Sullivan}, M. and {Yuan}, F.},
        title = "{The diversity of Type II supernova versus the similarity in their progenitors}",
      journal = {\mnras},
         year = 2016,
        month = jul,
       volume = {459},
       number = {4},
        pages = {3939-3962},
          doi = {10.1093/mnras/stw870},
archivePrefix = {arXiv},
       eprint = {1603.08953},
 primaryClass = {astro-ph.SR},
       adsurl = {https://ui.adsabs.harvard.edu/abs/2016MNRAS.459.3939V}
}

@ARTICLE{Brown2013,
       author = {{Brown}, T.~M. and {Baliber}, N. and {Bianco}, F.~B. and {Bowman}, M. and
         {Burleson}, B. and {Conway}, P. and {Crellin}, M. and
         {Depagne}, {\'E}. and {De Vera}, J. and {Dilday}, B. and
         {Dragomir}, D. and {Dubberley}, M. and {Eastman}, J.~D. and
         {Elphick}, M. and {Falarski}, M. and {Foale}, S. and {Ford}, M. and
         {Fulton}, B.~J. and {Garza}, J. and {Gomez}, E.~L. and {Graham}, M. and
         {Greene}, R. and {Haldeman}, B. and {Hawkins}, E. and {Haworth}, B. and
         {Haynes}, R. and {Hidas}, M. and {Hjelstrom}, A.~E. and
         {Howell}, D.~A. and {Hygelund}, J. and {Lister}, T.~A. and
         {Lobdill}, R. and {Martinez}, J. and {Mullins}, D.~S. and
         {Norbury}, M. and {Parrent}, J. and {Paulson}, R. and {Petry}, D.~L. and
         {Pickles}, A. and {Posner}, V. and {Rosing}, W.~E. and {Ross}, R. and
         {Sand}, D.~J. and {Saunders}, E.~S. and {Shobbrook}, J. and
         {Shporer}, A. and {Street}, R.~A. and {Thomas}, D. and {Tsapras}, Y. and
         {Tufts}, J.~R. and {Valenti}, S. and {Vander Horst}, K. and
         {Walker}, Z. and {White}, G. and {Willis}, M.},
        title = "{Las Cumbres Observatory Global Telescope Network}",
      journal = {\pasp},
         year = "2013",
        month = "Sep",
       volume = {125},
       number = {931},
        pages = {1031},
          doi = {10.1086/673168},
archivePrefix = {arXiv},
       eprint = {1305.2437},
 primaryClass = {astro-ph.IM},
       adsurl = {https://ui.adsabs.harvard.edu/abs/2013PASP..125.1031B}
}

@MISC{Guevel2017,
       author = {{Guevel}, David and {Hosseinzadeh}, Griffin},
        title = "{Dguevel/Pyzogy: Initial Release}",
         year = 2017,
        month = nov,
          eid = {10.5281/zenodo.1043973},
          doi = {10.5281/zenodo.1043973},
      version = {v0.0.1},
    publisher = {Zenodo},
       adsurl = {https://ui.adsabs.harvard.edu/abs/2017zndo...1043973G}
}

@ARTICLE{Zackay2016,
       author = {{Zackay}, Barak and {Ofek}, Eran O. and {Gal-Yam}, Avishay},
        title = "{Proper Image Subtraction{\textemdash}Optimal Transient Detection, Photometry, and Hypothesis Testing}",
      journal = {\apj},
         year = 2016,
        month = oct,
       volume = {830},
       number = {1},
          eid = {27},
        pages = {27},
          doi = {10.3847/0004-637X/830/1/27},
archivePrefix = {arXiv},
       eprint = {1601.02655},
 primaryClass = {astro-ph.IM},
       adsurl = {https://ui.adsabs.harvard.edu/abs/2016ApJ...830...27Z}
}

@MISC{Waskom2020,
       author = {{Waskom}, Michael and {Botvinnik}, Olga and {Ostblom}, Joel and
         {Lukauskas}, Saulius and {Hobson}, Paul and {MaozGelbart} and
         {Gemperline}, David C and {Augspurger}, Tom and {Halchenko}, Yaroslav and
         {Cole}, John B. and {Warmenhoven}, Jordi and {de Ruiter}, Julian and
         {Pye}, Cameron and {Hoyer}, Stephan and {Vanderplas}, Jake and
         {Villalba}, Santi and {Kunter}, Gero and {Quintero}, Eric and
         {Bachant}, Pete and {Martin}, Marcel and {Meyer}, Kyle and
         {Swain}, Corban and {Miles}, Alistair and {Brunner}, Thomas and
         {O'Kane}, Drew and {Yarkoni}, Tal and {Williams}, Mike Lee and
         {Evans}, Constantine},
        title = "{mwaskom/seaborn: v0.10.0 (January 2020)}",
         year = 2020,
        month = jan,
          eid = {10.5281/zenodo.3629446},
          doi = {10.5281/zenodo.3629446},
      version = {v0.10.0},
    publisher = {Zenodo},
       adsurl = {https://ui.adsabs.harvard.edu/abs/2020zndo...3629446W}
}

@ARTICLE{Stetson2000,
       author = {{Stetson}, Peter B.},
        title = "{Homogeneous Photometry for Star Clusters and Resolved Galaxies. II. Photometric Standard Stars}",
      journal = {\pasp},
         year = "2000",
        month = "Jul",
       volume = {112},
       number = {773},
        pages = {925-931},
          doi = {10.1086/316595},
archivePrefix = {arXiv},
       eprint = {astro-ph/0004144},
 primaryClass = {astro-ph},
       adsurl = {https://ui.adsabs.harvard.edu/abs/2000PASP..112..925S}
}

@ARTICLE{Albareti2017,
       author = {{Albareti}, Franco D. and {Allende Prieto}, Carlos and
         {Almeida}, Andres and {Anders}, Friedrich and {Anderson}, Scott and
         {Andrews}, Brett H. and {Arag{\'o}n-Salamanca}, Alfonso and
         {Argudo-Fern{\'a}ndez}, Maria and {Armengaud}, Eric and
         {Aubourg}, Eric and {Avila-Reese}, Vladimir and {Badenes}, Carles and
         {Bailey}, Stephen and {Barbuy}, Beatriz and {Barger}, Kat and
         {Barrera-Ballesteros}, Jorge and {Bartosz}, Curtis and {Basu}, Sarbani and
         {Bates}, Dominic and {Battaglia}, Giuseppina and {Baumgarten}, Falk and
         {Baur}, Julien and {Bautista}, Julian and {Beers}, Timothy C. and
         {Belfiore}, Francesco and {Bershady}, Matthew and
         {Bertran de Lis}, Sara and {Bird}, Jonathan C. and {Bizyaev}, Dmitry and
         {Blanc}, Guillermo A. and {Blanton}, Michael and {Blomqvist}, Michael and
         {Bolton}, Adam S. and {Borissova}, J. and {Bovy}, Jo and {Brand
        t}, William Nielsen and {Brinkmann}, Jonathan and {Brownstein}, Joel R. and
         {Bundy}, Kevin and {Burtin}, Etienne and {Busca}, Nicol{\'a}s G. and
         {Orlando Camacho Chavez}, Hugo and {Cano D{\'\i}az}, M. and
         {Cappellari}, Michele and {Carrera}, Ricardo and {Chen}, Yanping and
         {Cherinka}, Brian and {Cheung}, Edmond and {Chiappini}, Cristina and
         {Chojnowski}, Drew and {Chuang}, Chia-Hsun and {Chung}, Haeun and
         {Cirolini}, Rafael Fernando and {Clerc}, Nicolas and {Cohen}, Roger E. and
         {Comerford}, Julia M. and {Comparat}, Johan and
         {Correa do Nascimento}, Janaina and {Cousinou}, Marie-Claude and
         {Covey}, Kevin and {Crane}, Jeffrey D. and {Croft}, Rupert and
         {Cunha}, Katia and {Darling}, Jeremy and {Davidson}, James W., Jr. and
         {Dawson}, Kyle and {Da Costa}, Luiz and {Da Silva Ilha}, Gabriele and
         {Deconto Machado}, Alice and {Delubac}, Timoth{\'e}e and
         {De Lee}, Nathan and {De la Macorra}, Axel and {De la Torre}, Sylvain and
         {Diamond-Stanic}, Aleksandar M. and {Donor}, John and
         {Downes}, Juan Jose and {Drory}, Niv and {Du}, Cheng and
         {Du Mas des Bourboux}, H{\'e}lion and {Dwelly}, Tom and
         {Ebelke}, Garrett and {Eigenbrot}, Arthur and {Eisenstein}, Daniel J. and
         {Elsworth}, Yvonne P. and {Emsellem}, Eric and {Eracleous}, Michael and
         {Escoffier}, Stephanie and {Evans}, Michael L. and
         {Falc{\'o}n-Barroso}, Jes{\'u}s and {Fan}, Xiaohui and
         {Favole}, Ginevra and {Fernandez-Alvar}, Emma and {Fernand
        ez-Trincado}, J.~G. and {Feuillet}, Diane and {Fleming}, Scott W. and
         {Font-Ribera}, Andreu and {Freischlad}, Gordon and {Frinchaboy}, Peter and
         {Fu}, Hai and {Gao}, Yang and {Garcia}, Rafael A. and
         {Garcia-Dias}, R. and {Garcia-Hern{\'a}ndez}, D.~A. and
         {Garcia P{\'e}rez}, Ana E. and {Gaulme}, Patrick and {Ge}, Junqiang and
         {Geisler}, Douglas and {Gillespie}, Bruce and {Gil Marin}, Hector and
         {Girardi}, L{\'e}o and {Goddard}, Daniel and
         {Gomez Maqueo Chew}, Yilen and {Gonzalez-Perez}, Violeta and
         {Grabowski}, Kathleen and {Green}, Paul and {Grier}, Catherine J. and
         {Grier}, Thomas and {Guo}, Hong and {Guy}, Julien and {Hagen}, Alex and
         {Hall}, Matt and {Harding}, Paul and {Harley}, R.~E. and
         {Hasselquist}, Sten and {Hawley}, Suzanne and {Hayes}, Christian R. and
         {Hearty}, Fred and {Hekker}, Saskia and {Hernandez Toledo}, Hector and
         {Ho}, Shirley and {Hogg}, David W. and {Holley-Bockelmann}, Kelly and
         {Holtzman}, Jon A. and {Holzer}, Parker H. and {Hu}, Jian and
         {Huber}, Daniel and {Hutchinson}, Timothy Alan and {Hwang}, Ho Seong and
         {Ibarra-Medel}, H{\'e}ctor J. and {Ivans}, Inese I. and
         {Ivory}, KeShawn and {Jaehnig}, Kurt and {Jensen}, Trey W. and
         {Johnson}, Jennifer A. and {Jones}, Amy and {Jullo}, Eric and
         {Kallinger}, T. and {Kinemuchi}, Karen and {Kirkby}, David and
         {Klaene}, Mark and {Kneib}, Jean-Paul and {Kollmeier}, Juna A. and
         {Lacerna}, Ivan and {Lane}, Richard R. and {Lang}, Dustin and
         {Laurent}, Pierre and {Law}, David R. and {Leauthaud}, Alexie and
         {Le Goff}, Jean-Marc and {Li}, Chen and {Li}, Cheng and {Li}, Niu and
         {Li}, Ran and {Liang}, Fu-Heng and {Liang}, Yu and {Lima}, Marcos and
         {Lin}, Lihwai and {Lin}, Lin and {Lin}, Yen-Ting and {Liu}, Chao and
         {Long}, Dan and {Lucatello}, Sara and {MacDonald}, Nicholas and
         {MacLeod}, Chelsea L. and {Mackereth}, J. Ted and {Mahadevan}, Suvrath and
         {Maia}, Marcio Antonio Geimba and {Maiolino}, Roberto and
         {Majewski}, Steven R. and {Malanushenko}, Olena and
         {Malanushenko}, Viktor and {Mallmann}, N{\'\i}colas Dullius and
         {Manchado}, Arturo and {Maraston}, Claudia and {Marques-Chaves}, Rui and
         {Martinez Valpuesta}, Inma and {Masters}, Karen L. and
         {Mathur}, Savita and {McGreer}, Ian D. and {Merloni}, Andrea and
         {Merrifield}, Michael R. and {M{\'e}sz{\'a}ros}, Szabolcs and
         {Meza}, Andres and {Miglio}, Andrea and {Minchev}, Ivan and
         {Molaverdikhani}, Karan and {Montero-Dorta}, Antonio D. and
         {Mosser}, Benoit and {Muna}, Demitri and {Myers}, Adam and
         {Nair}, Preethi and {Nandra}, Kirpal and {Ness}, Melissa and
         {Newman}, Jeffrey A. and {Nichol}, Robert C. and {Nidever}, David L. and
         {Nitschelm}, Christian and {O'Connell}, Julia and {Oravetz}, Audrey and
         {Oravetz}, Daniel J. and {Pace}, Zachary and {Padilla}, Nelson and
         {Palanque-Delabrouille}, Nathalie and {Pan}, Kaike and {Parejko}, John and
         {Paris}, Isabelle and {Park}, Changbom and {Peacock}, John A. and
         {Peirani}, Sebastien and {Pellejero-Ibanez}, Marcos and
         {Penny}, Samantha and {Percival}, Will J. and {Percival}, Jeffrey W. and
         {Perez-Fournon}, Ismael and {Petitjean}, Patrick and {Pieri}, Matthew and
         {Pinsonneault}, Marc H. and {Pisani}, Alice and {Prada}, Francisco and
         {Prakash}, Abhishek and {Price-Jones}, Natalie and
         {Raddick}, M. Jordan and {Rahman}, Mubdi and {Raichoor}, Anand and
         {Barboza Rembold}, Sandro and {Reyna}, A.~M. and {Rich}, James and
         {Richstein}, Hannah and {Ridl}, Jethro and {Riffel}, Rogemar A. and
         {Riffel}, Rog{\'e}rio and {Rix}, Hans-Walter and {Robin}, Annie C. and
         {Rockosi}, Constance M. and {Rodr{\'\i}guez-Torres}, Sergio and
         {Rodrigues}, Tha{\'\i}se S. and {Roe}, Natalie and {Roman Lopes}, A. and
         {Rom{\'a}n-Z{\'u}{\~n}iga}, Carlos and {Ross}, Ashley J. and
         {Rossi}, Graziano and {Ruan}, John and {Ruggeri}, Rossana and
         {Runnoe}, Jessie C. and {Salazar-Albornoz}, Salvador and
         {Salvato}, Mara and {Sanchez}, Sebastian F. and {Sanchez}, Ariel G. and
         {Sanchez-Gallego}, Jos{\'e} R. and {Santiago}, Bas{\'\i}lio Xavier and
         {Schiavon}, Ricardo and {Schimoia}, Jaderson S. and {Schlafly}, Eddie and
         {Schlegel}, David J. and {Schneider}, Donald P. and
         {Sch{\"o}nrich}, Ralph and {Schultheis}, Mathias and {Schwope}, Axel and
         {Seo}, Hee-Jong and {Serenelli}, Aldo and {Sesar}, Branimir and
         {Shao}, Zhengyi and {Shetrone}, Matthew and {Shull}, Michael and
         {Silva Aguirre}, Victor and {Skrutskie}, M.~F. and
         {Slosar}, An{\v{z}}e and {Smith}, Michael and {Smith}, Verne V. and
         {Sobeck}, Jennifer and {Somers}, Garrett and {Souto}, Diogo and
         {Stark}, David V. and {Stassun}, Keivan G. and {Steinmetz}, Matthias and
         {Stello}, Dennis and {Storchi Bergmann}, Thaisa and
         {Strauss}, Michael A. and {Streblyanska}, Alina and
         {Stringfellow}, Guy S. and {Suarez}, Genaro and {Sun}, Jing and
         {Taghizadeh-Popp}, Manuchehr and {Tang}, Baitian and {Tao}, Charling and
         {Tayar}, Jamie and {Tembe}, Mita and {Thomas}, Daniel and
         {Tinker}, Jeremy and {Tojeiro}, Rita and {Tremonti}, Christy and
         {Troup}, Nicholas and {Trump}, Jonathan R. and {Unda-Sanzana}, Eduardo and
         {Valenzuela}, O. and {Van den Bosch}, Remco and
         {Vargas-Maga{\~n}a}, Mariana and {Vazquez}, Jose Alberto and
         {Villanova}, Sandro and {Vivek}, M. and {Vogt}, Nicole and
         {Wake}, David and {Walterbos}, Rene and {Wang}, Yuting and
         {Wang}, Enci and {Weaver}, Benjamin Alan and {Weijmans}, Anne-Marie and
         {Weinberg}, David H. and {Westfall}, Kyle B. and {Whelan}, David G. and
         {Wilcots}, Eric and {Wild}, Vivienne and {Williams}, Rob A. and
         {Wilson}, John and {Wood-Vasey}, W.~M. and {Wylezalek}, Dominika and
         {Xiao}, Ting and {Yan}, Renbin and {Yang}, Meng and {Ybarra}, Jason E. and
         {Yeche}, Christophe and {Yuan}, Fang-Ting and {Zakamska}, Nadia and
         {Zamora}, Olga and {Zasowski}, Gail and {Zhang}, Kai and {Zhao}, Cheng and
         {Zhao}, Gong-Bo and {Zheng}, Zheng and {Zheng}, Zheng and
         {Zhou}, Zhi-Min and {Zhu}, Guangtun and {Zinn}, Joel C. and {Zou}, Hu},
        title = "{The 13th Data Release of the Sloan Digital Sky Survey: First Spectroscopic Data from the SDSS-IV Survey Mapping Nearby Galaxies at Apache Point Observatory}",
      journal = {\apjs},
         year = "2017",
        month = "Dec",
       volume = {233},
       number = {2},
          eid = {25},
        pages = {25},
          doi = {10.3847/1538-4365/aa8992},
archivePrefix = {arXiv},
       eprint = {1608.02013},
 primaryClass = {astro-ph.GA},
       adsurl = {https://ui.adsabs.harvard.edu/abs/2017ApJS..233...25A}
}

@ARTICLE{Valenti2014,
       author = {{Valenti}, S. and {Sand}, D. and {Pastorello}, A. and {Graham}, M.~L. and
         {Howell}, D.~A. and {Parrent}, J.~T. and {Tomasella}, L. and
         {Ochner}, P. and {Fraser}, M. and {Benetti}, S. and {Yuan}, F. and
         {Smartt}, S.~J. and {Maund}, J.~R. and {Arcavi}, I. and {Gal-Yam}, A. and
         {Inserra}, C. and {Young}, D.},
        title = "{The first month of evolution of the slow-rising Type IIP SN 2013ej in M74$^{★}$}",
      journal = {\mnras},
         year = "2014",
        month = "Feb",
       volume = {438},
       number = {1},
        pages = {L101-L105},
          doi = {10.1093/mnrasl/slt171},
archivePrefix = {arXiv},
       eprint = {1309.4269},
 primaryClass = {astro-ph.CO},
       adsurl = {https://ui.adsabs.harvard.edu/abs/2014MNRAS.438L.101V}
}

@ARTICLE{Yaron2012,
       author = {{Yaron}, Ofer and {Gal-Yam}, Avishay},
        title = "{WISeREP{\textemdash}An Interactive Supernova Data Repository}",
      journal = {\pasp},
         year = "2012",
        month = "Jul",
       volume = {124},
       number = {917},
        pages = {668},
          doi = {10.1086/666656},
archivePrefix = {arXiv},
       eprint = {1204.1891},
 primaryClass = {astro-ph.IM},
       adsurl = {https://ui.adsabs.harvard.edu/abs/2012PASP..124..668Y}
}

@ARTICLE{Schlafly2011,
       author = {{Schlafly}, Edward F. and {Finkbeiner}, Douglas P.},
        title = "{Measuring Reddening with Sloan Digital Sky Survey Stellar Spectra and Recalibrating SFD}",
      journal = {\apj},
         year = "2011",
        month = "Aug",
       volume = {737},
       number = {2},
          eid = {103},
        pages = {103},
          doi = {10.1088/0004-637X/737/2/103},
archivePrefix = {arXiv},
       eprint = {1012.4804},
 primaryClass = {astro-ph.GA},
       adsurl = {https://ui.adsabs.harvard.edu/abs/2011ApJ...737..103S}
}

@ARTICLE{Kewley2006,
       author = {{Kewley}, Lisa J. and {Groves}, Brent and {Kauffmann}, Guinevere and
         {Heckman}, Tim},
        title = "{The host galaxies and classification of active galactic nuclei}",
      journal = {\mnras},
         year = 2006,
        month = nov,
       volume = {372},
       number = {3},
        pages = {961-976},
          doi = {10.1111/j.1365-2966.2006.10859.x},
archivePrefix = {arXiv},
       eprint = {astro-ph/0605681},
 primaryClass = {astro-ph},
       adsurl = {https://ui.adsabs.harvard.edu/abs/2006MNRAS.372..961K}
}

@ARTICLE{AstropyCollaboration2013,
       author = {{Astropy Collaboration} and {Robitaille}, Thomas P. and {Tollerud}, Erik J. and {Greenfield}, Perry and {Droettboom}, Michael and {Bray}, Erik and {Aldcroft}, Tom and {Davis}, Matt and {Ginsburg}, Adam and {Price-Whelan}, Adrian M. and {Kerzendorf}, Wolfgang E. and {Conley}, Alexander and {Crighton}, Neil and {Barbary}, Kyle and {Muna}, Demitri and {Ferguson}, Henry and {Grollier}, Fr{\'e}d{\'e}ric and {Parikh}, Madhura M. and {Nair}, Prasanth H. and {Unther}, Hans M. and {Deil}, Christoph and {Woillez}, Julien and {Conseil}, Simon and {Kramer}, Roban and {Turner}, James E.~H. and {Singer}, Leo and {Fox}, Ryan and {Weaver}, Benjamin A. and {Zabalza}, Victor and {Edwards}, Zachary I. and {Azalee Bostroem}, K. and {Burke}, D.~J. and {Casey}, Andrew R. and {Crawford}, Steven M. and {Dencheva}, Nadia and {Ely}, Justin and {Jenness}, Tim and {Labrie}, Kathleen and {Lim}, Pey Lian and {Pierfederici}, Francesco and {Pontzen}, Andrew and {Ptak}, Andy and {Refsdal}, Brian and {Servillat}, Mathieu and {Streicher}, Ole},
        title = "{Astropy: A community Python package for astronomy}",
      journal = {\aap},
         year = 2013,
        month = oct,
       volume = {558},
          eid = {A33},
        pages = {A33},
          doi = {10.1051/0004-6361/201322068},
archivePrefix = {arXiv},
       eprint = {1307.6212},
 primaryClass = {astro-ph.IM},
       adsurl = {https://ui.adsabs.harvard.edu/abs/2013A&A...558A..33A}
}

@ARTICLE{AstropyCollaboration2018,
       author = {{Astropy Collaboration} and {Price-Whelan}, A.~M. and
         {Sip{\H{o}}cz}, B.~M. and {G{\"u}nther}, H.~M. and {Lim}, P.~L. and
         {Crawford}, S.~M. and {Conseil}, S. and {Shupe}, D.~L. and
         {Craig}, M.~W. and {Dencheva}, N. and {Ginsburg}, A. and {Vand
        erPlas}, J.~T. and {Bradley}, L.~D. and {P{\'e}rez-Su{\'a}rez}, D. and
         {de Val-Borro}, M. and {Aldcroft}, T.~L. and {Cruz}, K.~L. and
         {Robitaille}, T.~P. and {Tollerud}, E.~J. and {Ardelean}, C. and
         {Babej}, T. and {Bach}, Y.~P. and {Bachetti}, M. and {Bakanov}, A.~V. and
         {Bamford}, S.~P. and {Barentsen}, G. and {Barmby}, P. and
         {Baumbach}, A. and {Berry}, K.~L. and {Biscani}, F. and {Boquien}, M. and
         {Bostroem}, K.~A. and {Bouma}, L.~G. and {Brammer}, G.~B. and
         {Bray}, E.~M. and {Breytenbach}, H. and {Buddelmeijer}, H. and
         {Burke}, D.~J. and {Calderone}, G. and {Cano Rodr{\'\i}guez}, J.~L. and
         {Cara}, M. and {Cardoso}, J.~V.~M. and {Cheedella}, S. and {Copin}, Y. and
         {Corrales}, L. and {Crichton}, D. and {D'Avella}, D. and {Deil}, C. and
         {Depagne}, {\'E}. and {Dietrich}, J.~P. and {Donath}, A. and
         {Droettboom}, M. and {Earl}, N. and {Erben}, T. and {Fabbro}, S. and
         {Ferreira}, L.~A. and {Finethy}, T. and {Fox}, R.~T. and
         {Garrison}, L.~H. and {Gibbons}, S.~L.~J. and {Goldstein}, D.~A. and
         {Gommers}, R. and {Greco}, J.~P. and {Greenfield}, P. and
         {Groener}, A.~M. and {Grollier}, F. and {Hagen}, A. and {Hirst}, P. and
         {Homeier}, D. and {Horton}, A.~J. and {Hosseinzadeh}, G. and {Hu}, L. and
         {Hunkeler}, J.~S. and {Ivezi{\'c}}, {\v{Z}}. and {Jain}, A. and
         {Jenness}, T. and {Kanarek}, G. and {Kendrew}, S. and {Kern}, N.~S. and
         {Kerzendorf}, W.~E. and {Khvalko}, A. and {King}, J. and {Kirkby}, D. and
         {Kulkarni}, A.~M. and {Kumar}, A. and {Lee}, A. and {Lenz}, D. and
         {Littlefair}, S.~P. and {Ma}, Z. and {Macleod}, D.~M. and
         {Mastropietro}, M. and {McCully}, C. and {Montagnac}, S. and
         {Morris}, B.~M. and {Mueller}, M. and {Mumford}, S.~J. and {Muna}, D. and
         {Murphy}, N.~A. and {Nelson}, S. and {Nguyen}, G.~H. and
         {Ninan}, J.~P. and {N{\"o}the}, M. and {Ogaz}, S. and {Oh}, S. and
         {Parejko}, J.~K. and {Parley}, N. and {Pascual}, S. and {Patil}, R. and
         {Patil}, A.~A. and {Plunkett}, A.~L. and {Prochaska}, J.~X. and
         {Rastogi}, T. and {Reddy Janga}, V. and {Sabater}, J. and
         {Sakurikar}, P. and {Seifert}, M. and {Sherbert}, L.~E. and
         {Sherwood-Taylor}, H. and {Shih}, A.~Y. and {Sick}, J. and
         {Silbiger}, M.~T. and {Singanamalla}, S. and {Singer}, L.~P. and
         {Sladen}, P.~H. and {Sooley}, K.~A. and {Sornarajah}, S. and
         {Streicher}, O. and {Teuben}, P. and {Thomas}, S.~W. and
         {Tremblay}, G.~R. and {Turner}, J.~E.~H. and {Terr{\'o}n}, V. and
         {van Kerkwijk}, M.~H. and {de la Vega}, A. and {Watkins}, L.~L. and
         {Weaver}, B.~A. and {Whitmore}, J.~B. and {Woillez}, J. and
         {Zabalza}, V. and {Astropy Contributors}},
        title = "{The Astropy Project: Building an Open-science Project and Status of the v2.0 Core Package}",
      journal = {\aj},
         year = 2018,
        month = sep,
       volume = {156},
       number = {3},
          eid = {123},
        pages = {123},
          doi = {10.3847/1538-3881/aabc4f},
archivePrefix = {arXiv},
       eprint = {1801.02634},
 primaryClass = {astro-ph.IM},
       adsurl = {https://ui.adsabs.harvard.edu/abs/2018AJ....156..123A}
}

@ARTICLE{AstropyCollaboration2022,
       author = {{Astropy Collaboration} and {Price-Whelan}, Adrian M. and {Lim}, Pey Lian and {Earl}, Nicholas and {Starkman}, Nathaniel and {Bradley}, Larry and {Shupe}, David L. and {Patil}, Aarya A. and {Corrales}, Lia and {Brasseur}, C.~E. and {N{\"o}the}, Maximilian and {Donath}, Axel and {Tollerud}, Erik and {Morris}, Brett M. and {Ginsburg}, Adam and {Vaher}, Eero and {Weaver}, Benjamin A. and {Tocknell}, James and {Jamieson}, William and {van Kerkwijk}, Marten H. and {Robitaille}, Thomas P. and {Merry}, Bruce and {Bachetti}, Matteo and {G{\"u}nther}, H. Moritz and {Aldcroft}, Thomas L. and {Alvarado-Montes}, Jaime A. and {Archibald}, Anne M. and {B{\'o}di}, Attila and {Bapat}, Shreyas and {Barentsen}, Geert and {Baz{\'a}n}, Juanjo and {Biswas}, Manish and {Boquien}, M{\'e}d{\'e}ric and {Burke}, D.~J. and {Cara}, Daria and {Cara}, Mihai and {Conroy}, Kyle E. and {Conseil}, Simon and {Craig}, Matthew W. and {Cross}, Robert M. and {Cruz}, Kelle L. and {D'Eugenio}, Francesco and {Dencheva}, Nadia and {Devillepoix}, Hadrien A.~R. and {Dietrich}, J{\"o}rg P. and {Eigenbrot}, Arthur Davis and {Erben}, Thomas and {Ferreira}, Leonardo and {Foreman-Mackey}, Daniel and {Fox}, Ryan and {Freij}, Nabil and {Garg}, Suyog and {Geda}, Robel and {Glattly}, Lauren and {Gondhalekar}, Yash and {Gordon}, Karl D. and {Grant}, David and {Greenfield}, Perry and {Groener}, Austen M. and {Guest}, Steve and {Gurovich}, Sebastian and {Handberg}, Rasmus and {Hart}, Akeem and {Hatfield-Dodds}, Zac and {Homeier}, Derek and {Hosseinzadeh}, Griffin and {Jenness}, Tim and {Jones}, Craig K. and {Joseph}, Prajwel and {Kalmbach}, J. Bryce and {Karamehmetoglu}, Emir and {Ka{\l}uszy{\'n}ski}, Miko{\l}aj and {Kelley}, Michael S.~P. and {Kern}, Nicholas and {Kerzendorf}, Wolfgang E. and {Koch}, Eric W. and {Kulumani}, Shankar and {Lee}, Antony and {Ly}, Chun and {Ma}, Zhiyuan and {MacBride}, Conor and {Maljaars}, Jakob M. and {Muna}, Demitri and {Murphy}, N.~A. and {Norman}, Henrik and {O'Steen}, Richard and {Oman}, Kyle A. and {Pacifici}, Camilla and {Pascual}, Sergio and {Pascual-Granado}, J. and {Patil}, Rohit R. and {Perren}, Gabriel I. and {Pickering}, Timothy E. and {Rastogi}, Tanuj and {Roulston}, Benjamin R. and {Ryan}, Daniel F. and {Rykoff}, Eli S. and {Sabater}, Jose and {Sakurikar}, Parikshit and {Salgado}, Jes{\'u}s and {Sanghi}, Aniket and {Saunders}, Nicholas and {Savchenko}, Volodymyr and {Schwardt}, Ludwig and {Seifert-Eckert}, Michael and {Shih}, Albert Y. and {Jain}, Anany Shrey and {Shukla}, Gyanendra and {Sick}, Jonathan and {Simpson}, Chris and {Singanamalla}, Sudheesh and {Singer}, Leo P. and {Singhal}, Jaladh and {Sinha}, Manodeep and {Sip{\H{o}}cz}, Brigitta M. and {Spitler}, Lee R. and {Stansby}, David and {Streicher}, Ole and {{\v{S}}umak}, Jani and {Swinbank}, John D. and {Taranu}, Dan S. and {Tewary}, Nikita and {Tremblay}, Grant R. and {de Val-Borro}, Miguel and {Van Kooten}, Samuel J. and {Vasovi{\'c}}, Zlatan and {Verma}, Shresth and {de Miranda Cardoso}, Jos{\'e} Vin{\'\i}cius and {Williams}, Peter K.~G. and {Wilson}, Tom J. and {Winkel}, Benjamin and {Wood-Vasey}, W.~M. and {Xue}, Rui and {Yoachim}, Peter and {Zhang}, Chen and {Zonca}, Andrea and {Astropy Project Contributors}},
        title = "{The Astropy Project: Sustaining and Growing a Community-oriented Open-source Project and the Latest Major Release (v5.0) of the Core Package}",
      journal = {\apj},
         year = 2022,
        month = aug,
       volume = {935},
       number = {2},
          eid = {167},
        pages = {167},
          doi = {10.3847/1538-4357/ac7c74},
archivePrefix = {arXiv},
       eprint = {2206.14220},
 primaryClass = {astro-ph.IM},
       adsurl = {https://ui.adsabs.harvard.edu/abs/2022ApJ...935..167A}
}

@ARTICLE{Hunter2007,
       author = {{Hunter}, John D.},
        title = "{Matplotlib: A 2D Graphics Environment}",
      journal = {Computing in Science and Engineering},
         year = 2007,
        month = may,
       volume = {9},
       number = {3},
        pages = {90-95},
          doi = {10.1109/MCSE.2007.55},
       adsurl = {https://ui.adsabs.harvard.edu/abs/2007CSE.....9...90H}
}

@book{Oliphant2006,
author="Oliphant, Travis E.",
title="A Guide to NumPy",
year="2006",
publisher="USA: Trelgol Publishing",
url={https://web.mit.edu/dvp/Public/numpybook.pdf}
}

@MISC{PyRAF2012,
       author = {{Science Software Branch at STScI}},
        title = "{PyRAF: Python alternative for IRAF}",
         year = 2012,
        month = jul,
          eid = {ascl:1207.011},
        pages = {ascl:1207.011},
archivePrefix = {ascl},
       eprint = {1207.011},
       adsurl = {https://ui.adsabs.harvard.edu/abs/2012ascl.soft07011S}
}

@ARTICLE{Bertin1996,
       author = {{Bertin}, E. and {Arnouts}, S.},
        title = "{SExtractor: Software for source extraction.}",
      journal = {\aaps},
         year = 1996,
        month = jun,
       volume = {117},
        pages = {393-404},
          doi = {10.1051/aas:1996164},
       adsurl = {https://ui.adsabs.harvard.edu/abs/1996A&AS..117..393B}
}

@ARTICLE{SciPy2020,
       author = {{Virtanen}, Pauli and {Gommers}, Ralf and {Oliphant}, Travis E. and {Haberland}, Matt and {Reddy}, Tyler and {Cournapeau}, David and {Burovski}, Evgeni and {Peterson}, Pearu and {Weckesser}, Warren and {Bright}, Jonathan and {van der Walt}, St{\'e}fan J. and {Brett}, Matthew and {Wilson}, Joshua and {Millman}, K. Jarrod and {Mayorov}, Nikolay and {Nelson}, Andrew R.~J. and {Jones}, Eric and {Kern}, Robert and {Larson}, Eric and {Carey}, C.~J. and {Polat}, {\.I}lhan and {Feng}, Yu and {Moore}, Eric W. and {VanderPlas}, Jake and {Laxalde}, Denis and {Perktold}, Josef and {Cimrman}, Robert and {Henriksen}, Ian and {Quintero}, E.~A. and {Harris}, Charles R. and {Archibald}, Anne M. and {Ribeiro}, Ant{\^o}nio H. and {Pedregosa}, Fabian and {van Mulbregt}, Paul and {SciPy 1. 0 Contributors}},
        title = "{SciPy 1.0: fundamental algorithms for scientific computing in Python}",
      journal = {Nature Methods},
         year = 2020,
        month = feb,
       volume = {17},
        pages = {261-272},
          doi = {10.1038/s41592-019-0686-2},
archivePrefix = {arXiv},
       eprint = {1907.10121},
 primaryClass = {cs.MS},
       adsurl = {https://ui.adsabs.harvard.edu/abs/2020NatMe..17..261V}
}

@ARTICLE{Blinnikov1998,
   author = {{Blinnikov}, S.~I. and {Eastman}, R. and {Bartunov}, O.~S. and 
	{Popolitov}, V.~A. and {Woosley}, S.~E.},
    title = "{A Comparative Modeling of Supernova 1993J}",
  journal = {\apj},
   eprint = {astro-ph/9711055},
     year = 1998,
    month = mar,
   volume = 496,
    pages = {454-472},
      doi = {10.1086/305375},
   adsurl = {http://adsabs.harvard.edu/abs/1998ApJ...496..454B}
}

@ARTICLE{Blinnikov2000,
       author = {{Blinnikov}, Sergei and {Lundqvist}, Peter and {Bartunov}, Oleg and
         {Nomoto}, Ken'ichi and {Iwamoto}, Koichi},
        title = "{Radiation Hydrodynamics of SN 1987A. I. Global Analysis of the Light Curve for the First 4 Months}",
      journal = {\apj},
         year = "2000",
        month = "Apr",
       volume = {532},
        pages = {1132-1149},
          doi = {10.1086/308588},
archivePrefix = {arXiv},
       eprint = {astro-ph/9911205},
 primaryClass = {astro-ph},
       adsurl = {https://ui.adsabs.harvard.edu/\#abs/2000ApJ...532.1132B}
}

@ARTICLE{Blinnikov2004,
       author = {{Blinnikov}, S. and {Sorokina}, E.},
        title = "{Type Ia Supernova models: Latest developments}",
      journal = {\apss},
         year = 2004,
        month = feb,
       volume = {290},
       number = {1},
        pages = {13-28},
          doi = {10.1023/B:ASTR.0000022161.03559.42},
archivePrefix = {arXiv},
       eprint = {astro-ph/0212530},
 primaryClass = {astro-ph},
       adsurl = {https://ui.adsabs.harvard.edu/abs/2004Ap&SS.290...13B}
}

@ARTICLE{Baklanov2005,
       author = {{Baklanov}, P.~V. and {Blinnikov}, S.~I. and {Pavlyuk}, N.~N.},
        title = "{Parameters of the classical type-IIP supernova SN 1999em}",
      journal = {Astronomy Letters},
         year = 2005,
        month = jul,
       volume = {31},
       number = {7},
        pages = {429-441},
          doi = {10.1134/1.1958107},
       adsurl = {https://ui.adsabs.harvard.edu/abs/2005AstL...31..429B}
}

@ARTICLE{Blinnikov2006,
       author = {{Blinnikov}, S.~I. and {R{\"o}pke}, F.~K. and {Sorokina}, E.~I. and
         {Gieseler}, M. and {Reinecke}, M. and {Travaglio}, C. and {Hillebrand
        t}, W. and {Stritzinger}, M.},
        title = "{Theoretical light curves for deflagration models of type Ia supernova}",
      journal = {\aap},
         year = "2006",
        month = "Jul",
       volume = {453},
        pages = {229-240},
          doi = {10.1051/0004-6361:20054594},
archivePrefix = {arXiv},
       eprint = {astro-ph/0603036},
 primaryClass = {astro-ph},
       adsurl = {https://ui.adsabs.harvard.edu/\#abs/2006A&A...453..229B}
}

@ARTICLE{Filippenko1982,
       author = {{Filippenko}, A.~V.},
        title = "{The importance of atmospheric differential refraction in spectrophotometry.}",
      journal = {\pasp},
         year = 1982,
        month = aug,
       volume = {94},
        pages = {715-721},
          doi = {10.1086/131052},
       adsurl = {https://ui.adsabs.harvard.edu/abs/1982PASP...94..715F}
}

@ARTICLE{Bianco2016,
       author = {{Bianco}, F.~B. and {Modjaz}, M. and {Oh}, S.~M. and {Fierroz}, D. and
         {Liu}, Y.~Q. and {Kewley}, L. and {Graur}, O.},
        title = "{Monte Carlo method for calculating oxygen abundances and their uncertainties from strong-line flux measurements}",
      journal = {Astronomy and Computing},
         year = 2016,
        month = jul,
       volume = {16},
        pages = {54-66},
          doi = {10.1016/j.ascom.2016.03.002},
archivePrefix = {arXiv},
       eprint = {1505.06213},
 primaryClass = {astro-ph.IM},
       adsurl = {https://ui.adsabs.harvard.edu/abs/2016A&C....16...54B}
}

@ARTICLE{Asplund2009,
       author = {{Asplund}, Martin and {Grevesse}, Nicolas and {Sauval}, A. Jacques and
         {Scott}, Pat},
        title = "{The Chemical Composition of the Sun}",
      journal = {\araa},
         year = 2009,
        month = sep,
       volume = {47},
       number = {1},
        pages = {481-522},
          doi = {10.1146/annurev.astro.46.060407.145222},
archivePrefix = {arXiv},
       eprint = {0909.0948},
 primaryClass = {astro-ph.SR},
       adsurl = {https://ui.adsabs.harvard.edu/abs/2009ARA&A..47..481A}
}

@ARTICLE{Kennicutt1998,
       author = {{Kennicutt}, Robert C., Jr.},
        title = "{Star Formation in Galaxies Along the Hubble Sequence}",
      journal = {\araa},
         year = 1998,
        month = jan,
       volume = {36},
        pages = {189-232},
          doi = {10.1146/annurev.astro.36.1.189},
archivePrefix = {arXiv},
       eprint = {astro-ph/9807187},
 primaryClass = {astro-ph},
       adsurl = {https://ui.adsabs.harvard.edu/abs/1998ARA&A..36..189K}
}

@ARTICLE{Baldwin1981,
       author = {{Baldwin}, J.~A. and {Phillips}, M.~M. and {Terlevich}, R.},
        title = "{Classification parameters for the emission-line spectra of extragalactic objects.}",
      journal = {\pasp},
         year = 1981,
        month = feb,
       volume = {93},
        pages = {5-19},
          doi = {10.1086/130766},
       adsurl = {https://ui.adsabs.harvard.edu/abs/1981PASP...93....5B}
}

@ARTICLE{Fitzpatrick1999,
       author = {{Fitzpatrick}, Edward L.},
        title = "{Correcting for the Effects of Interstellar Extinction}",
      journal = {\pasp},
         year = 1999,
        month = jan,
       volume = {111},
       number = {755},
        pages = {63-75},
          doi = {10.1086/316293},
archivePrefix = {arXiv},
       eprint = {astro-ph/9809387},
 primaryClass = {astro-ph},
       adsurl = {https://ui.adsabs.harvard.edu/abs/1999PASP..111...63F}
}

@ARTICLE{Schulze2021,
       author = {{Schulze}, Steve and {Yaron}, Ofer and {Sollerman}, Jesper and {Leloudas}, Giorgos and {Gal}, Amit and {Wright}, Angus H. and {Lunnan}, Ragnhild and {Gal-Yam}, Avishay and {Ofek}, Eran O. and {Perley}, Daniel A. and {Filippenko}, Alexei V. and {Kasliwal}, Mansi M. and {Kulkarni}, Shrinivas R. and {Neill}, James D. and {Nugent}, Peter E. and {Quimby}, Robert M. and {Sullivan}, Mark and {Strotjohann}, Nora Linn and {Arcavi}, Iair and {Ben-Ami}, Sagi and {Bianco}, Federica and {Bloom}, Joshua S. and {De}, Kishalay and {Fraser}, Morgan and {Fremling}, Christoffer U. and {Horesh}, Assaf and {Johansson}, Joel and {Kelly}, Patrick L. and {Kne{\v{z}}evi{\'c}}, Nikola and {Kne{\v{z}}evi{\'c}}, Sladjana and {Maguire}, Kate and {Nyholm}, Anders and {Papadogiannakis}, Sem{\'e}li and {Petrushevska}, Tanja and {Rubin}, Adam and {Yan}, Lin and {Yang}, Yi and {Adams}, Scott M. and {Bufano}, Filomena and {Clubb}, Kelsey I. and {Foley}, Ryan J. and {Green}, Yoav and {Harmanen}, Jussi and {Ho}, Anna Y.~Q. and {Hook}, Isobel M. and {Hosseinzadeh}, Griffin and {Howell}, D. Andrew and {Kong}, Albert K.~H. and {Kotak}, Rubina and {Matheson}, Thomas and {McCully}, Curtis and {Milisavljevic}, Dan and {Pan}, Yen-Chen and {Poznanski}, Dovi and {Shivvers}, Isaac and {van Velzen}, Sjoert and {Verbeek}, Kars K.},
        title = "{The Palomar Transient Factory Core-collapse Supernova Host-galaxy Sample. I. Host-galaxy Distribution Functions and Environment Dependence of Core-collapse Supernovae}",
      journal = {\apjs},
         year = 2021,
        month = aug,
       volume = {255},
       number = {2},
          eid = {29},
        pages = {29},
          doi = {10.3847/1538-4365/abff5e},
archivePrefix = {arXiv},
       eprint = {2008.05988},
 primaryClass = {astro-ph.GA},
       adsurl = {https://ui.adsabs.harvard.edu/abs/2021ApJS..255...29S}
}

@ARTICLE{DESDR2,
       author = {{Abbott}, T.~M.~C. and {Adam{\'o}w}, M. and {Aguena}, M. and {Allam}, S. and {Amon}, A. and {Annis}, J. and {Avila}, S. and {Bacon}, D. and {Banerji}, M. and {Bechtol}, K. and {Becker}, M.~R. and {Bernstein}, G.~M. and {Bertin}, E. and {Bhargava}, S. and {Bridle}, S.~L. and {Brooks}, D. and {Burke}, D.~L. and {Carnero Rosell}, A. and {Carrasco Kind}, M. and {Carretero}, J. and {Castander}, F.~J. and {Cawthon}, R. and {Chang}, C. and {Choi}, A. and {Conselice}, C. and {Costanzi}, M. and {Crocce}, M. and {da Costa}, L.~N. and {Davis}, T.~M. and {De Vicente}, J. and {DeRose}, J. and {Desai}, S. and {Diehl}, H.~T. and {Dietrich}, J.~P. and {Drlica-Wagner}, A. and {Eckert}, K. and {Elvin-Poole}, J. and {Everett}, S. and {Evrard}, A.~E. and {Ferrero}, I. and {Fert{\'e}}, A. and {Flaugher}, B. and {Fosalba}, P. and {Friedel}, D. and {Frieman}, J. and {Garc{\'\i}a-Bellido}, J. and {Gaztanaga}, E. and {Gelman}, L. and {Gerdes}, D.~W. and {Giannantonio}, T. and {Gill}, M.~S.~S. and {Gruen}, D. and {Gruendl}, R.~A. and {Gschwend}, J. and {Gutierrez}, G. and {Hartley}, W.~G. and {Hinton}, S.~R. and {Hollowood}, D.~L. and {Honscheid}, K. and {Huterer}, D. and {James}, D.~J. and {Jeltema}, T. and {Johnson}, M.~D. and {Kent}, S. and {Kron}, R. and {Kuehn}, K. and {Kuropatkin}, N. and {Lahav}, O. and {Li}, T.~S. and {Lidman}, C. and {Lin}, H. and {MacCrann}, N. and {Maia}, M.~A.~G. and {Manning}, T.~A. and {Maloney}, J.~D. and {March}, M. and {Marshall}, J.~L. and {Martini}, P. and {Melchior}, P. and {Menanteau}, F. and {Miquel}, R. and {Morgan}, R. and {Myles}, J. and {Neilsen}, E. and {Ogando}, R.~L.~C. and {Palmese}, A. and {Paz-Chinch{\'o}n}, F. and {Petravick}, D. and {Pieres}, A. and {Plazas}, A.~A. and {Pond}, C. and {Rodriguez-Monroy}, M. and {Romer}, A.~K. and {Roodman}, A. and {Rykoff}, E.~S. and {Sako}, M. and {Sanchez}, E. and {Santiago}, B. and {Scarpine}, V. and {Serrano}, S. and {Sevilla-Noarbe}, I. and {Smith}, J. Allyn and {Smith}, M. and {Soares-Santos}, M. and {Suchyta}, E. and {Swanson}, M.~E.~C. and {Tarle}, G. and {Thomas}, D. and {To}, C. and {Tremblay}, P.~E. and {Troxel}, M.~A. and {Tucker}, D.~L. and {Turner}, D.~J. and {Varga}, T.~N. and {Walker}, A.~R. and {Wechsler}, R.~H. and {Weller}, J. and {Wester}, W. and {Wilkinson}, R.~D. and {Yanny}, B. and {Zhang}, Y. and {Nikutta}, R. and {Fitzpatrick}, M. and {Jacques}, A. and {Scott}, A. and {Olsen}, K. and {Huang}, L. and {Herrera}, D. and {Juneau}, S. and {Nidever}, D. and {Weaver}, B.~A. and {Adean}, C. and {Correia}, V. and {de Freitas}, M. and {Freitas}, F.~N. and {Singulani}, C. and {Vila-Verde}, G. and {Linea Science Server}},
        title = "{The Dark Energy Survey Data Release 2}",
      journal = {\apjs},
         year = 2021,
        month = aug,
       volume = {255},
       number = {2},
          eid = {20},
        pages = {20},
          doi = {10.3847/1538-4365/ac00b3},
archivePrefix = {arXiv},
       eprint = {2101.05765},
 primaryClass = {astro-ph.IM},
       adsurl = {https://ui.adsabs.harvard.edu/abs/2021ApJS..255...20A}
}

@ARTICLE{PS1DR2,
       author = {{Flewelling}, H.~A. and {Magnier}, E.~A. and {Chambers}, K.~C. and {Heasley}, J.~N. and {Holmberg}, C. and {Huber}, M.~E. and {Sweeney}, W. and {Waters}, C.~Z. and {Calamida}, A. and {Casertano}, S. and {Chen}, X. and {Farrow}, D. and {Hasinger}, G. and {Henderson}, R. and {Long}, K.~S. and {Metcalfe}, N. and {Narayan}, G. and {Nieto-Santisteban}, M.~A. and {Norberg}, P. and {Rest}, A. and {Saglia}, R.~P. and {Szalay}, A. and {Thakar}, A.~R. and {Tonry}, J.~L. and {Valenti}, J. and {Werner}, S. and {White}, R. and {Denneau}, L. and {Draper}, P.~W. and {Hodapp}, K.~W. and {Jedicke}, R. and {Kaiser}, N. and {Kudritzki}, R.~P. and {Price}, P.~A. and {Wainscoat}, R.~J. and {Chastel}, S. and {McLean}, B. and {Postman}, M. and {Shiao}, B.},
        title = "{The Pan-STARRS1 Database and Data Products}",
      journal = {\apjs},
         year = 2020,
        month = nov,
       volume = {251},
       number = {1},
          eid = {7},
        pages = {7},
          doi = {10.3847/1538-4365/abb82d},
archivePrefix = {arXiv},
       eprint = {1612.05243},
 primaryClass = {astro-ph.IM},
       adsurl = {https://ui.adsabs.harvard.edu/abs/2020ApJS..251....7F}
}

@ARTICLE{2MASS,
       author = {{Skrutskie}, M.~F. and {Cutri}, R.~M. and {Stiening}, R. and {Weinberg}, M.~D. and {Schneider}, S. and {Carpenter}, J.~M. and {Beichman}, C. and {Capps}, R. and {Chester}, T. and {Elias}, J. and {Huchra}, J. and {Liebert}, J. and {Lonsdale}, C. and {Monet}, D.~G. and {Price}, S. and {Seitzer}, P. and {Jarrett}, T. and {Kirkpatrick}, J.~D. and {Gizis}, J.~E. and {Howard}, E. and {Evans}, T. and {Fowler}, J. and {Fullmer}, L. and {Hurt}, R. and {Light}, R. and {Kopan}, E.~L. and {Marsh}, K.~A. and {McCallon}, H.~L. and {Tam}, R. and {Van Dyk}, S. and {Wheelock}, S.},
        title = "{The Two Micron All Sky Survey (2MASS)}",
      journal = {\aj},
         year = 2006,
        month = feb,
       volume = {131},
       number = {2},
        pages = {1163-1183},
          doi = {10.1086/498708},
       adsurl = {https://ui.adsabs.harvard.edu/abs/2006AJ....131.1163S}
}

@ARTICLE{WISE,
       author = {{Wright}, Edward L. and {Eisenhardt}, Peter R.~M. and {Mainzer}, Amy K. and {Ressler}, Michael E. and {Cutri}, Roc M. and {Jarrett}, Thomas and {Kirkpatrick}, J. Davy and {Padgett}, Deborah and {McMillan}, Robert S. and {Skrutskie}, Michael and {Stanford}, S.~A. and {Cohen}, Martin and {Walker}, Russell G. and {Mather}, John C. and {Leisawitz}, David and {Gautier}, Thomas N., III and {McLean}, Ian and {Benford}, Dominic and {Lonsdale}, Carol J. and {Blain}, Andrew and {Mendez}, Bryan and {Irace}, William R. and {Duval}, Valerie and {Liu}, Fengchuan and {Royer}, Don and {Heinrichsen}, Ingolf and {Howard}, Joan and {Shannon}, Mark and {Kendall}, Martha and {Walsh}, Amy L. and {Larsen}, Mark and {Cardon}, Joel G. and {Schick}, Scott and {Schwalm}, Mark and {Abid}, Mohamed and {Fabinsky}, Beth and {Naes}, Larry and {Tsai}, Chao-Wei},
        title = "{The Wide-field Infrared Survey Explorer (WISE): Mission Description and Initial On-orbit Performance}",
      journal = {\aj},
         year = 2010,
        month = dec,
       volume = {140},
       number = {6},
        pages = {1868-1881},
          doi = {10.1088/0004-6256/140/6/1868},
archivePrefix = {arXiv},
       eprint = {1008.0031},
 primaryClass = {astro-ph.IM},
       adsurl = {https://ui.adsabs.harvard.edu/abs/2010AJ....140.1868W}
}

@ARTICLE{Kron1980,
       author = {{Kron}, R.~G.},
        title = "{Photometry of a complete sample of faint galaxies.}",
      journal = {\apjs},
         year = 1980,
        month = jun,
       volume = {43},
        pages = {305-325},
          doi = {10.1086/190669},
       adsurl = {https://ui.adsabs.harvard.edu/abs/1980ApJS...43..305K}
}

@MISC{Prospector,
       author = {{Johnson}, Benjamin D. and {Leja}, Joel L. and {Conroy}, Charlie and {Speagle}, Joshua S.},
        title = "{Prospector: Stellar population inference from spectra and SEDs}",
         year = 2019,
        month = may,
          eid = {ascl:1905.025},
        pages = {ascl:1905.025},
archivePrefix = {ascl},
       eprint = {1905.025},
       adsurl = {https://ui.adsabs.harvard.edu/abs/2019ascl.soft05025J}
}

@ARTICLE{Conroy2009,
       author = {{Conroy}, Charlie and {Gunn}, James E. and {White}, Martin},
        title = "{The Propagation of Uncertainties in Stellar Population Synthesis Modeling. I. The Relevance of Uncertain Aspects of Stellar Evolution and the Initial Mass Function to the Derived Physical Properties of Galaxies}",
      journal = {\apj},
         year = 2009,
        month = jul,
       volume = {699},
       number = {1},
        pages = {486-506},
          doi = {10.1088/0004-637X/699/1/486},
archivePrefix = {arXiv},
       eprint = {0809.4261},
 primaryClass = {astro-ph},
       adsurl = {https://ui.adsabs.harvard.edu/abs/2009ApJ...699..486C}
}

@ARTICLE{Calzetti2000,
       author = {{Calzetti}, Daniela and {Armus}, Lee and {Bohlin}, Ralph C. and {Kinney}, Anne L. and {Koornneef}, Jan and {Storchi-Bergmann}, Thaisa},
        title = "{The Dust Content and Opacity of Actively Star-forming Galaxies}",
      journal = {\apj},
         year = 2000,
        month = apr,
       volume = {533},
       number = {2},
        pages = {682-695},
          doi = {10.1086/308692},
archivePrefix = {arXiv},
       eprint = {astro-ph/9911459},
 primaryClass = {astro-ph},
       adsurl = {https://ui.adsabs.harvard.edu/abs/2000ApJ...533..682C}
}

@ARTICLE{Miller2019,
       author = {{Miller}, Tim B. and {van Dokkum}, Pieter and {Mowla}, Lamiya and {van der Wel}, Arjen},
        title = "{A New View of the Size-Mass Distribution of Galaxies: Using r $_{20}$ and r $_{80}$ Instead of r $_{50}$}",
      journal = {\apjl},
         year = 2019,
        month = feb,
       volume = {872},
       number = {1},
          eid = {L14},
        pages = {L14},
          doi = {10.3847/2041-8213/ab0380},
archivePrefix = {arXiv},
       eprint = {1901.05017},
 primaryClass = {astro-ph.GA},
       adsurl = {https://ui.adsabs.harvard.edu/abs/2019ApJ...872L..14M}
}

@ARTICLE{Mowla2019,
       author = {{Mowla}, Lamiya and {van der Wel}, Arjen and {van Dokkum}, Pieter and {Miller}, Tim B.},
        title = "{A Mass-dependent Slope of the Galaxy Size-Mass Relation out to z {\ensuremath{\sim}} 3: Further Evidence for a Direct Relation between Median Galaxy Size and Median Halo Mass}",
      journal = {\apjl},
         year = 2019,
        month = feb,
       volume = {872},
       number = {1},
          eid = {L13},
        pages = {L13},
          doi = {10.3847/2041-8213/ab0379},
archivePrefix = {arXiv},
       eprint = {1901.05014},
 primaryClass = {astro-ph.GA},
       adsurl = {https://ui.adsabs.harvard.edu/abs/2019ApJ...872L..13M}
}

@ARTICLE{Jerkstrand2020,
       author = {{Jerkstrand}, Anders and {Maeda}, Keiichi and {Kawabata}, Koji S.},
        title = "{A type Ia supernova at the heart of superluminous transient SN 2006gy}",
      journal = {Science},
         year = 2020,
        month = jan,
       volume = {367},
       number = {6476},
        pages = {415-418},
          doi = {10.1126/science.aaw1469},
archivePrefix = {arXiv},
       eprint = {2002.10768},
 primaryClass = {astro-ph.SR},
       adsurl = {https://ui.adsabs.harvard.edu/abs/2020Sci...367..415J}
}

@ARTICLE{Ablimit2021,
       author = {{Ablimit}, Iminhaji},
        title = "{The CO White Dwarf + Intermediate-mass/Massive Star Binary Evolution: Possible Merger Origins for Peculiar Type Ia and II Supernovae}",
      journal = {\pasp},
         year = 2021,
        month = jul,
       volume = {133},
       number = {1025},
          eid = {074201},
        pages = {074201},
          doi = {10.1088/1538-3873/ac025c},
archivePrefix = {arXiv},
       eprint = {2101.03670},
 primaryClass = {astro-ph.SR},
       adsurl = {https://ui.adsabs.harvard.edu/abs/2021PASP..133g4201A}
}

@ARTICLE{Smith2007,
       author = {{Smith}, Nathan and {Li}, Weidong and {Foley}, Ryan J. and {Wheeler}, J. Craig and {Pooley}, David and {Chornock}, Ryan and {Filippenko}, Alexei V. and {Silverman}, Jeffrey M. and {Quimby}, Robert and {Bloom}, Joshua S. and {Hansen}, Charles},
        title = "{SN 2006gy: Discovery of the Most Luminous Supernova Ever Recorded, Powered by the Death of an Extremely Massive Star like {\ensuremath{\eta}} Carinae}",
      journal = {\apj},
         year = 2007,
        month = sep,
       volume = {666},
       number = {2},
        pages = {1116-1128},
          doi = {10.1086/519949},
archivePrefix = {arXiv},
       eprint = {astro-ph/0612617},
 primaryClass = {astro-ph},
       adsurl = {https://ui.adsabs.harvard.edu/abs/2007ApJ...666.1116S}
}

@ARTICLE{Smith2010,
       author = {{Smith}, Nathan and {Chornock}, Ryan and {Silverman}, Jeffrey M. and {Filippenko}, Alexei V. and {Foley}, Ryan J.},
        title = "{Spectral Evolution of the Extraordinary Type IIn Supernova 2006gy}",
      journal = {\apj},
         year = 2010,
        month = feb,
       volume = {709},
       number = {2},
        pages = {856-883},
          doi = {10.1088/0004-637X/709/2/856},
archivePrefix = {arXiv},
       eprint = {0906.2200},
 primaryClass = {astro-ph.HE},
       adsurl = {https://ui.adsabs.harvard.edu/abs/2010ApJ...709..856S}
}

@ARTICLE{Nomoto1984,
       author = {{Nomoto}, K. and {Thielemann}, F. -K. and {Yokoi}, K.},
        title = "{Accreting white dwarf models for type I supern. III. Carbon deflagration supernovae.}",
      journal = {\apj},
         year = 1984,
        month = nov,
       volume = {286},
        pages = {644-658},
          doi = {10.1086/162639},
       adsurl = {https://ui.adsabs.harvard.edu/abs/1984ApJ...286..644N}
}

@ARTICLE{Silverman2013_sample,
       author = {{Silverman}, Jeffrey M. and {Nugent}, Peter E. and {Gal-Yam}, Avishay and {Sullivan}, Mark and {Howell}, D. Andrew and {Filippenko}, Alexei V. and {Arcavi}, Iair and {Ben-Ami}, Sagi and {Bloom}, Joshua S. and {Cenko}, S. Bradley and {Cao}, Yi and {Chornock}, Ryan and {Clubb}, Kelsey I. and {Coil}, Alison L. and {Foley}, Ryan J. and {Graham}, Melissa L. and {Griffith}, Christopher V. and {Horesh}, Assaf and {Kasliwal}, Mansi M. and {Kulkarni}, Shrinivas R. and {Leonard}, Douglas C. and {Li}, Weidong and {Matheson}, Thomas and {Miller}, Adam A. and {Modjaz}, Maryam and {Ofek}, Eran O. and {Pan}, Yen-Chen and {Perley}, Daniel A. and {Poznanski}, Dovi and {Quimby}, Robert M. and {Steele}, Thea N. and {Sternberg}, Assaf and {Xu}, Dong and {Yaron}, Ofer},
        title = "{Type Ia Supernovae Strongly Interacting with Their Circumstellar Medium}",
      journal = {\apjs},
         year = 2013,
        month = jul,
       volume = {207},
       number = {1},
          eid = {3},
        pages = {3},
          doi = {10.1088/0067-0049/207/1/3},
archivePrefix = {arXiv},
       eprint = {1304.0763},
 primaryClass = {astro-ph.CO},
       adsurl = {https://ui.adsabs.harvard.edu/abs/2013ApJS..207....3S}
}

@ARTICLE{Fox2015,
       author = {{Fox}, Ori D. and {Silverman}, Jeffrey M. and {Filippenko}, Alexei V. and {Mauerhan}, Jon and {Becker}, Juliette and {Borish}, H. Jacob and {Cenko}, S. Bradley and {Clubb}, Kelsey I. and {Graham}, Melissa and {Hsiao}, Eric and {Kelly}, Patrick L. and {Lee}, William H. and {Marion}, G.~H. and {Milisavljevic}, Dan and {Parrent}, Jerod and {Shivvers}, Isaac and {Skrutskie}, Michael and {Smith}, Nathan and {Wilson}, John and {Zheng}, Weikang},
        title = "{On the nature of Type IIn/Ia-CSM supernovae: optical and near-infrared spectra of SN 2012ca and SN 2013dn}",
      journal = {\mnras},
         year = 2015,
        month = feb,
       volume = {447},
       number = {1},
        pages = {772-785},
          doi = {10.1093/mnras/stu2435},
archivePrefix = {arXiv},
       eprint = {1408.6239},
 primaryClass = {astro-ph.SR},
       adsurl = {https://ui.adsabs.harvard.edu/abs/2015MNRAS.447..772F}
}

@ARTICLE{Inserra2016,
       author = {{Inserra}, C. and {Fraser}, M. and {Smartt}, S.~J. and {Benetti}, S. and {Chen}, T. -W. and {Childress}, M. and {Gal-Yam}, A. and {Howell}, D.~A. and {Kangas}, T. and {Pignata}, G. and {Polshaw}, J. and {Sullivan}, M. and {Smith}, K.~W. and {Valenti}, S. and {Young}, D.~R. and {Parker}, S. and {Seccull}, T. and {McCrum}, M.},
        title = "{On Type IIn/Ia-CSM supernovae as exemplified by SN 2012ca*}",
      journal = {\mnras},
         year = 2016,
        month = jul,
       volume = {459},
       number = {3},
        pages = {2721-2740},
          doi = {10.1093/mnras/stw825},
archivePrefix = {arXiv},
       eprint = {1510.01109},
 primaryClass = {astro-ph.SR},
       adsurl = {https://ui.adsabs.harvard.edu/abs/2016MNRAS.459.2721I}
}

@ARTICLE{Perley2016,
       author = {{Perley}, D.~A. and {Quimby}, R.~M. and {Yan}, L. and {Vreeswijk}, P.~M. and {De Cia}, A. and {Lunnan}, R. and {Gal-Yam}, A. and {Yaron}, O. and {Filippenko}, A.~V. and {Graham}, M.~L. and {Laher}, R. and {Nugent}, P.~E.},
        title = "{Host-galaxy Properties of 32 Low-redshift Superluminous Supernovae from the Palomar Transient Factory}",
      journal = {\apj},
         year = 2016,
        month = oct,
       volume = {830},
       number = {1},
          eid = {13},
        pages = {13},
          doi = {10.3847/0004-637X/830/1/13},
archivePrefix = {arXiv},
       eprint = {1604.08207},
 primaryClass = {astro-ph.HE},
       adsurl = {https://ui.adsabs.harvard.edu/abs/2016ApJ...830...13P}
}

@ARTICLE{Kawabata2009,
       author = {{Kawabata}, Koji S. and {Tanaka}, Masaomi and {Maeda}, Keiichi and {Hattori}, Takashi and {Nomoto}, Ken'ichi and {Tominaga}, Nozomu and {Yamanaka}, Masayuki},
        title = "{Extremely Luminous Supernova 2006gy at Late Phase: Detection of Optical Emission from Supernova}",
      journal = {\apj},
         year = 2009,
        month = may,
       volume = {697},
       number = {1},
        pages = {747-757},
          doi = {10.1088/0004-637X/697/1/747},
archivePrefix = {arXiv},
       eprint = {0902.1440},
 primaryClass = {astro-ph.SR},
       adsurl = {https://ui.adsabs.harvard.edu/abs/2009ApJ...697..747K}
}

@ARTICLE{Benetti2014,
       author = {{Benetti}, S. and {Nicholl}, M. and {Cappellaro}, E. and {Pastorello}, A. and {Smartt}, S.~J. and {Elias-Rosa}, N. and {Drake}, A.~J. and {Tomasella}, L. and {Turatto}, M. and {Harutyunyan}, A. and {Taubenberger}, S. and {Hachinger}, S. and {Morales-Garoffolo}, A. and {Chen}, T. -W. and {Djorgovski}, S.~G. and {Fraser}, M. and {Gal-Yam}, A. and {Inserra}, C. and {Mazzali}, P. and {Pumo}, M.~L. and {Sollerman}, J. and {Valenti}, S. and {Young}, D.~R. and {Dennefeld}, M. and {Le Guillou}, L. and {Fleury}, M. and {L{\'e}get}, P. -F.},
        title = "{The supernova CSS121015:004244+132827: a clue for understanding superluminous supernovae}",
      journal = {\mnras},
         year = 2014,
        month = jun,
       volume = {441},
       number = {1},
        pages = {289-303},
          doi = {10.1093/mnras/stu538},
archivePrefix = {arXiv},
       eprint = {1310.1311},
 primaryClass = {astro-ph.SR},
       adsurl = {https://ui.adsabs.harvard.edu/abs/2014MNRAS.441..289B}
}

@ARTICLE{Prieto2007,
       author = {{Prieto}, J.~L. and {Garnavich}, P.~M. and {Phillips}, M.~M. and {DePoy}, D.~L. and {Parrent}, J. and {Pooley}, D. and {Dwarkadas}, V.~V. and {Baron}, E. and {Bassett}, B. and {Becker}, A. and {Cinabro}, D. and {DeJongh}, F. and {Dilday}, B. and {Doi}, M. and {Frieman}, J.~A. and {Hogan}, C.~J. and {Holtzman}, J. and {Jha}, S. and {Kessler}, R. and {Konishi}, K. and {Lampeitl}, H. and {Marriner}, J. and {Marshall}, J.~L. and {Miknaitis}, G. and {Nichol}, R.~C. and {Riess}, A.~G. and {Richmond}, M.~W. and {Romani}, R. and {Sako}, M. and {Schneider}, D.~P. and {Smith}, M. and {Takanashi}, N. and {Tokita}, K. and {van der Heyden}, K. and {Yasuda}, N. and {Zheng}, C. and {Wheeler}, J.~C. and {Barentine}, J. and {Dembicky}, J. and {Eastman}, J. and {Frank}, S. and {Ketzeback}, W. and {McMillan}, R.~J. and {Morrell}, N. and {Folatelli}, G. and {Contreras}, C. and {Burns}, C.~R. and {Freedman}, W.~L. and {Gonzalez}, S. and {Hamuy}, M. and {Krzeminski}, W. and {Madore}, B.~F. and {Murphy}, D. and {Persson}, S.~E. and {Roth}, M. and {Suntzeff}, N.~B.},
        title = "{A Study of the Type Ia/IIn Supernova 2005gj from X-ray to the Infrared: Paper I}",
      journal = {arXiv e-prints},
         year = 2007,
        month = jun,
          eid = {arXiv:0706.4088},
        pages = {arXiv:0706.4088},
archivePrefix = {arXiv},
       eprint = {0706.4088},
 primaryClass = {astro-ph},
       adsurl = {https://ui.adsabs.harvard.edu/abs/2007arXiv0706.4088P}
}

@ARTICLE{Fassia2001,
       author = {{Fassia}, A. and {Meikle}, W.~P.~S. and {Chugai}, N. and {Geballe}, T.~R. and {Lundqvist}, P. and {Walton}, N.~A. and {Pollacco}, D. and {Veilleux}, S. and {Wright}, G.~S. and {Pettini}, M. and {Kerr}, T. and {Puchnarewicz}, E. and {Puxley}, P. and {Irwin}, M. and {Packham}, C. and {Smartt}, S.~J. and {Harmer}, D.},
        title = "{Optical and infrared spectroscopy of the type IIn SN 1998S: days 3-127}",
      journal = {\mnras},
         year = 2001,
        month = aug,
       volume = {325},
       number = {3},
        pages = {907-930},
          doi = {10.1046/j.1365-8711.2001.04282.x},
archivePrefix = {arXiv},
       eprint = {astro-ph/0011340},
 primaryClass = {astro-ph},
       adsurl = {https://ui.adsabs.harvard.edu/abs/2001MNRAS.325..907F}
}

@ARTICLE{Germany2000,
       author = {{Germany}, Lisa M. and {Reiss}, David J. and {Sadler}, Elaine M. and {Schmidt}, Brian P. and {Stubbs}, C.~W.},
        title = "{SN 1997CY/GRB 970514: A New Piece in the Gamma-Ray Burst Puzzle?}",
      journal = {\apj},
         year = 2000,
        month = apr,
       volume = {533},
       number = {1},
        pages = {320-328},
          doi = {10.1086/308639},
archivePrefix = {arXiv},
       eprint = {astro-ph/9906096},
 primaryClass = {astro-ph},
       adsurl = {https://ui.adsabs.harvard.edu/abs/2000ApJ...533..320G}
}

@ARTICLE{Turatto2000,
       author = {{Turatto}, M. and {Suzuki}, T. and {Mazzali}, P.~A. and {Benetti}, S. and {Cappellaro}, E. and {Danziger}, I.~J. and {Nomoto}, K. and {Nakamura}, T. and {Young}, T.~R. and {Patat}, F.},
        title = "{The Properties of Supernova 1997CY Associated with GRB 970514}",
      journal = {\apjl},
         year = 2000,
        month = may,
       volume = {534},
       number = {1},
        pages = {L57-L61},
          doi = {10.1086/312653},
archivePrefix = {arXiv},
       eprint = {astro-ph/9910324},
 primaryClass = {astro-ph},
       adsurl = {https://ui.adsabs.harvard.edu/abs/2000ApJ...534L..57T}
}

@ARTICLE{Rigon2003,
       author = {{Rigon}, L. and {Turatto}, M. and {Benetti}, S. and {Pastorello}, A. and {Cappellaro}, E. and {Aretxaga}, I. and {Vega}, O. and {Chavushyan}, V. and {Patat}, F. and {Danziger}, I.~J. and {Salvo}, M.},
        title = "{SN 1999E: another piece in the supernova-gamma-ray burst connection puzzle}",
      journal = {\mnras},
         year = 2003,
        month = mar,
       volume = {340},
       number = {1},
        pages = {191-196},
          doi = {10.1046/j.1365-8711.2003.06282.x},
archivePrefix = {arXiv},
       eprint = {astro-ph/0211432},
 primaryClass = {astro-ph},
       adsurl = {https://ui.adsabs.harvard.edu/abs/2003MNRAS.340..191R}
}

@ARTICLE{Hodgkin2021,
       author = {{Hodgkin}, S.~T. and {Harrison}, D.~L. and {Breedt}, E. and {Wevers}, T. and {Rixon}, G. and {Delgado}, A. and {Yoldas}, A. and {Kostrzewa-Rutkowska}, Z. and {Wyrzykowski}, {\L}. and {van Leeuwen}, M. and {Blagorodnova}, N. and {Campbell}, H. and {Eappachen}, D. and {Fraser}, M. and {Ihanec}, N. and {Koposov}, S.~E. and {Kruszy{\'n}ska}, K. and {Marton}, G. and {Rybicki}, K.~A. and {Brown}, A.~G.~A. and {Burgess}, P.~W. and {Busso}, G. and {Cowell}, S. and {De Angeli}, F. and {Diener}, C. and {Evans}, D.~W. and {Gilmore}, G. and {Holland}, G. and {Jonker}, P.~G. and {van Leeuwen}, F. and {Mignard}, F. and {Osborne}, P.~J. and {Portell}, J. and {Prusti}, T. and {Richards}, P.~J. and {Riello}, M. and {Seabroke}, G.~M. and {Walton}, N.~A. and {{\'A}brah{\'a}m}, P. and {Altavilla}, G. and {Baker}, S.~G. and {Bastian}, U. and {O'Brien}, P. and {de Bruijne}, J. and {Butterley}, T. and {Carrasco}, J.~M. and {Casta{\~n}eda}, J. and {Clark}, J.~S. and {Clementini}, G. and {Copperwheat}, C.~M. and {Cropper}, M. and {Damljanovic}, G. and {Davidson}, M. and {Davis}, C.~J. and {Dennefeld}, M. and {Dhillon}, V.~S. and {Dolding}, C. and {Dominik}, M. and {Esquej}, P. and {Eyer}, L. and {Fabricius}, C. and {Fridman}, M. and {Froebrich}, D. and {Garralda}, N. and {Gomboc}, A. and {Gonz{\'a}lez-Vidal}, J.~J. and {Guerra}, R. and {Hambly}, N.~C. and {Hardy}, L.~K. and {Holl}, B. and {Hourihane}, A. and {Japelj}, J. and {Kann}, D.~A. and {Kiss}, C. and {Knigge}, C. and {Kolb}, U. and {Komossa}, S. and {K{\'o}sp{\'a}l}, {\'A}. and {Kov{\'a}cs}, G. and {Kun}, M. and {Leto}, G. and {Lewis}, F. and {Littlefair}, S.~P. and {Mahabal}, A.~A. and {Mundell}, C.~G. and {Nagy}, Z. and {Padeletti}, D. and {Palaversa}, L. and {Pigulski}, A. and {Pretorius}, M.~L. and {van Reeven}, W. and {Ribeiro}, V.~A.~R.~M. and {Roelens}, M. and {Rowell}, N. and {Schartel}, N. and {Scholz}, A. and {Schwope}, A. and {Sip{\H{o}}cz}, B.~M. and {Smartt}, S.~J. and {Smith}, M.~D. and {Serraller}, I. and {Steeghs}, D. and {Sullivan}, M. and {Szabados}, L. and {Szegedi-Elek}, E. and {Tisserand}, P. and {Tomasella}, L. and {van Velzen}, S. and {Whitelock}, P.~A. and {Wilson}, R.~W. and {Young}, D.~R.},
        title = "{Gaia Early Data Release 3. Gaia photometric science alerts}",
      journal = {\aap},
         year = 2021,
        month = aug,
       volume = {652},
          eid = {A76},
        pages = {A76},
          doi = {10.1051/0004-6361/202140735},
archivePrefix = {arXiv},
       eprint = {2106.01394},
 primaryClass = {astro-ph.IM},
       adsurl = {https://ui.adsabs.harvard.edu/abs/2021A&A...652A..76H}
}

@ARTICLE{Dilday2012,
       author = {{Dilday}, B. and {Howell}, D.~A. and {Cenko}, S.~B. and {Silverman}, J.~M. and {Nugent}, P.~E. and {Sullivan}, M. and {Ben-Ami}, S. and {Bildsten}, L. and {Bolte}, M. and {Endl}, M. and {Filippenko}, A.~V. and {Gnat}, O. and {Horesh}, A. and {Hsiao}, E. and {Kasliwal}, M.~M. and {Kirkman}, D. and {Maguire}, K. and {Marcy}, G.~W. and {Moore}, K. and {Pan}, Y. and {Parrent}, J.~T. and {Podsiadlowski}, P. and {Quimby}, R.~M. and {Sternberg}, A. and {Suzuki}, N. and {Tytler}, D.~R. and {Xu}, D. and {Bloom}, J.~S. and {Gal-Yam}, A. and {Hook}, I.~M. and {Kulkarni}, S.~R. and {Law}, N.~M. and {Ofek}, E.~O. and {Polishook}, D. and {Poznanski}, D.},
        title = "{PTF 11kx: A Type Ia Supernova with a Symbiotic Nova Progenitor}",
      journal = {Science},
         year = 2012,
        month = aug,
       volume = {337},
       number = {6097},
        pages = {942},
          doi = {10.1126/science.1219164},
archivePrefix = {arXiv},
       eprint = {1207.1306},
 primaryClass = {astro-ph.CO},
       adsurl = {https://ui.adsabs.harvard.edu/abs/2012Sci...337..942D}
}

@ARTICLE{Silverman2013,
       author = {{Silverman}, Jeffrey M. and {Nugent}, Peter E. and {Gal-Yam}, Avishay and {Sullivan}, Mark and {Howell}, D. Andrew and {Filippenko}, Alexei V. and {Pan}, Yen-Chen and {Cenko}, S. Bradley and {Hook}, Isobel M.},
        title = "{Late-time Spectral Observations of the Strongly Interacting Type Ia Supernova PTF11kx}",
      journal = {\apj},
         year = 2013,
        month = aug,
       volume = {772},
       number = {2},
          eid = {125},
        pages = {125},
          doi = {10.1088/0004-637X/772/2/125},
archivePrefix = {arXiv},
       eprint = {1303.7234},
 primaryClass = {astro-ph.CO},
       adsurl = {https://ui.adsabs.harvard.edu/abs/2013ApJ...772..125S}
}

@ARTICLE{PS1,
       author = {{Chambers}, K.~C. and {Magnier}, E.~A. and {Metcalfe}, N. and {Flewelling}, H.~A. and {Huber}, M.~E. and {Waters}, C.~Z. and {Denneau}, L. and {Draper}, P.~W. and {Farrow}, D. and {Finkbeiner}, D.~P. and {Holmberg}, C. and {Koppenhoefer}, J. and {Price}, P.~A. and {Rest}, A. and {Saglia}, R.~P. and {Schlafly}, E.~F. and {Smartt}, S.~J. and {Sweeney}, W. and {Wainscoat}, R.~J. and {Burgett}, W.~S. and {Chastel}, S. and {Grav}, T. and {Heasley}, J.~N. and {Hodapp}, K.~W. and {Jedicke}, R. and {Kaiser}, N. and {Kudritzki}, R. -P. and {Luppino}, G.~A. and {Lupton}, R.~H. and {Monet}, D.~G. and {Morgan}, J.~S. and {Onaka}, P.~M. and {Shiao}, B. and {Stubbs}, C.~W. and {Tonry}, J.~L. and {White}, R. and {Ba{\~n}ados}, E. and {Bell}, E.~F. and {Bender}, R. and {Bernard}, E.~J. and {Boegner}, M. and {Boffi}, F. and {Botticella}, M.~T. and {Calamida}, A. and {Casertano}, S. and {Chen}, W. -P. and {Chen}, X. and {Cole}, S. and {Deacon}, N. and {Frenk}, C. and {Fitzsimmons}, A. and {Gezari}, S. and {Gibbs}, V. and {Goessl}, C. and {Goggia}, T. and {Gourgue}, R. and {Goldman}, B. and {Grant}, P. and {Grebel}, E.~K. and {Hambly}, N.~C. and {Hasinger}, G. and {Heavens}, A.~F. and {Heckman}, T.~M. and {Henderson}, R. and {Henning}, T. and {Holman}, M. and {Hopp}, U. and {Ip}, W. -H. and {Isani}, S. and {Jackson}, M. and {Keyes}, C.~D. and {Koekemoer}, A.~M. and {Kotak}, R. and {Le}, D. and {Liska}, D. and {Long}, K.~S. and {Lucey}, J.~R. and {Liu}, M. and {Martin}, N.~F. and {Masci}, G. and {McLean}, B. and {Mindel}, E. and {Misra}, P. and {Morganson}, E. and {Murphy}, D.~N.~A. and {Obaika}, A. and {Narayan}, G. and {Nieto-Santisteban}, M.~A. and {Norberg}, P. and {Peacock}, J.~A. and {Pier}, E.~A. and {Postman}, M. and {Primak}, N. and {Rae}, C. and {Rai}, A. and {Riess}, A. and {Riffeser}, A. and {Rix}, H.~W. and {R{\"o}ser}, S. and {Russel}, R. and {Rutz}, L. and {Schilbach}, E. and {Schultz}, A.~S.~B. and {Scolnic}, D. and {Strolger}, L. and {Szalay}, A. and {Seitz}, S. and {Small}, E. and {Smith}, K.~W. and {Soderblom}, D.~R. and {Taylor}, P. and {Thomson}, R. and {Taylor}, A.~N. and {Thakar}, A.~R. and {Thiel}, J. and {Thilker}, D. and {Unger}, D. and {Urata}, Y. and {Valenti}, J. and {Wagner}, J. and {Walder}, T. and {Walter}, F. and {Watters}, S.~P. and {Werner}, S. and {Wood-Vasey}, W.~M. and {Wyse}, R.},
        title = "{The Pan-STARRS1 Surveys}",
      journal = {arXiv e-prints},
         year = 2016,
        month = dec,
          eid = {arXiv:1612.05560},
        pages = {arXiv:1612.05560},
archivePrefix = {arXiv},
       eprint = {1612.05560},
 primaryClass = {astro-ph.IM},
       adsurl = {https://ui.adsabs.harvard.edu/abs/2016arXiv161205560C}
}

@ARTICLE{DES,
       author = {{Dark Energy Survey Collaboration} and {Abbott}, T. and {Abdalla}, F.~B. and {Aleksi{\'c}}, J. and {Allam}, S. and {Amara}, A. and {Bacon}, D. and {Balbinot}, E. and {Banerji}, M. and {Bechtol}, K. and {Benoit-L{\'e}vy}, A. and {Bernstein}, G.~M. and {Bertin}, E. and {Blazek}, J. and {Bonnett}, C. and {Bridle}, S. and {Brooks}, D. and {Brunner}, R.~J. and {Buckley-Geer}, E. and {Burke}, D.~L. and {Caminha}, G.~B. and {Capozzi}, D. and {Carlsen}, J. and {Carnero-Rosell}, A. and {Carollo}, M. and {Carrasco-Kind}, M. and {Carretero}, J. and {Castander}, F.~J. and {Clerkin}, L. and {Collett}, T. and {Conselice}, C. and {Crocce}, M. and {Cunha}, C.~E. and {D'Andrea}, C.~B. and {da Costa}, L.~N. and {Davis}, T.~M. and {Desai}, S. and {Diehl}, H.~T. and {Dietrich}, J.~P. and {Dodelson}, S. and {Doel}, P. and {Drlica-Wagner}, A. and {Estrada}, J. and {Etherington}, J. and {Evrard}, A.~E. and {Fabbri}, J. and {Finley}, D.~A. and {Flaugher}, B. and {Foley}, R.~J. and {Fosalba}, P. and {Frieman}, J. and {Garc{\'\i}a-Bellido}, J. and {Gaztanaga}, E. and {Gerdes}, D.~W. and {Giannantonio}, T. and {Goldstein}, D.~A. and {Gruen}, D. and {Gruendl}, R.~A. and {Guarnieri}, P. and {Gutierrez}, G. and {Hartley}, W. and {Honscheid}, K. and {Jain}, B. and {James}, D.~J. and {Jeltema}, T. and {Jouvel}, S. and {Kessler}, R. and {King}, A. and {Kirk}, D. and {Kron}, R. and {Kuehn}, K. and {Kuropatkin}, N. and {Lahav}, O. and {Li}, T.~S. and {Lima}, M. and {Lin}, H. and {Maia}, M.~A.~G. and {Makler}, M. and {Manera}, M. and {Maraston}, C. and {Marshall}, J.~L. and {Martini}, P. and {McMahon}, R.~G. and {Melchior}, P. and {Merson}, A. and {Miller}, C.~J. and {Miquel}, R. and {Mohr}, J.~J. and {Morice-Atkinson}, X. and {Naidoo}, K. and {Neilsen}, E. and {Nichol}, R.~C. and {Nord}, B. and {Ogando}, R. and {Ostrovski}, F. and {Palmese}, A. and {Papadopoulos}, A. and {Peiris}, H.~V. and {Peoples}, J. and {Percival}, W.~J. and {Plazas}, A.~A. and {Reed}, S.~L. and {Refregier}, A. and {Romer}, A.~K. and {Roodman}, A. and {Ross}, A. and {Rozo}, E. and {Rykoff}, E.~S. and {Sadeh}, I. and {Sako}, M. and {S{\'a}nchez}, C. and {Sanchez}, E. and {Santiago}, B. and {Scarpine}, V. and {Schubnell}, M. and {Sevilla-Noarbe}, I. and {Sheldon}, E. and {Smith}, M. and {Smith}, R.~C. and {Soares-Santos}, M. and {Sobreira}, F. and {Soumagnac}, M. and {Suchyta}, E. and {Sullivan}, M. and {Swanson}, M. and {Tarle}, G. and {Thaler}, J. and {Thomas}, D. and {Thomas}, R.~C. and {Tucker}, D. and {Vieira}, J.~D. and {Vikram}, V. and {Walker}, A.~R. and {Wechsler}, R.~H. and {Weller}, J. and {Wester}, W. and {Whiteway}, L. and {Wilcox}, H. and {Yanny}, B. and {Zhang}, Y. and {Zuntz}, J.},
        title = "{The Dark Energy Survey: more than dark energy - an overview}",
      journal = {\mnras},
         year = 2016,
        month = aug,
       volume = {460},
       number = {2},
        pages = {1270-1299},
          doi = {10.1093/mnras/stw641},
archivePrefix = {arXiv},
       eprint = {1601.00329},
 primaryClass = {astro-ph.CO},
       adsurl = {https://ui.adsabs.harvard.edu/abs/2016MNRAS.460.1270D}
}

@ARTICLE{DESILIS,
       author = {{Dey}, Arjun and {Schlegel}, David J. and {Lang}, Dustin and {Blum}, Robert and {Burleigh}, Kaylan and {Fan}, Xiaohui and {Findlay}, Joseph R. and {Finkbeiner}, Doug and {Herrera}, David and {Juneau}, St{\'e}phanie and {Landriau}, Martin and {Levi}, Michael and {McGreer}, Ian and {Meisner}, Aaron and {Myers}, Adam D. and {Moustakas}, John and {Nugent}, Peter and {Patej}, Anna and {Schlafly}, Edward F. and {Walker}, Alistair R. and {Valdes}, Francisco and {Weaver}, Benjamin A. and {Y{\`e}che}, Christophe and {Zou}, Hu and {Zhou}, Xu and {Abareshi}, Behzad and {Abbott}, T.~M.~C. and {Abolfathi}, Bela and {Aguilera}, C. and {Alam}, Shadab and {Allen}, Lori and {Alvarez}, A. and {Annis}, James and {Ansarinejad}, Behzad and {Aubert}, Marie and {Beechert}, Jacqueline and {Bell}, Eric F. and {BenZvi}, Segev Y. and {Beutler}, Florian and {Bielby}, Richard M. and {Bolton}, Adam S. and {Brice{\~n}o}, C{\'e}sar and {Buckley-Geer}, Elizabeth J. and {Butler}, Karen and {Calamida}, Annalisa and {Carlberg}, Raymond G. and {Carter}, Paul and {Casas}, Ricard and {Castander}, Francisco J. and {Choi}, Yumi and {Comparat}, Johan and {Cukanovaite}, Elena and {Delubac}, Timoth{\'e}e and {DeVries}, Kaitlin and {Dey}, Sharmila and {Dhungana}, Govinda and {Dickinson}, Mark and {Ding}, Zhejie and {Donaldson}, John B. and {Duan}, Yutong and {Duckworth}, Christopher J. and {Eftekharzadeh}, Sarah and {Eisenstein}, Daniel J. and {Etourneau}, Thomas and {Fagrelius}, Parker A. and {Farihi}, Jay and {Fitzpatrick}, Mike and {Font-Ribera}, Andreu and {Fulmer}, Leah and {G{\"a}nsicke}, Boris T. and {Gaztanaga}, Enrique and {George}, Koshy and {Gerdes}, David W. and {Gontcho}, Satya Gontcho A. and {Gorgoni}, Claudio and {Green}, Gregory and {Guy}, Julien and {Harmer}, Diane and {Hernandez}, M. and {Honscheid}, Klaus and {Huang}, Lijuan Wendy and {James}, David J. and {Jannuzi}, Buell T. and {Jiang}, Linhua and {Joyce}, Richard and {Karcher}, Armin and {Karkar}, Sonia and {Kehoe}, Robert and {Kneib}, Jean-Paul and {Kueter-Young}, Andrea and {Lan}, Ting-Wen and {Lauer}, Tod R. and {Le Guillou}, Laurent and {Le Van Suu}, Auguste and {Lee}, Jae Hyeon and {Lesser}, Michael and {Perreault Levasseur}, Laurence and {Li}, Ting S. and {Mann}, Justin L. and {Marshall}, Robert and {Mart{\'\i}nez-V{\'a}zquez}, C.~E. and {Martini}, Paul and {du Mas des Bourboux}, H{\'e}lion and {McManus}, Sean and {Meier}, Tobias Gabriel and {M{\'e}nard}, Brice and {Metcalfe}, Nigel and {Mu{\~n}oz-Guti{\'e}rrez}, Andrea and {Najita}, Joan and {Napier}, Kevin and {Narayan}, Gautham and {Newman}, Jeffrey A. and {Nie}, Jundan and {Nord}, Brian and {Norman}, Dara J. and {Olsen}, Knut A.~G. and {Paat}, Anthony and {Palanque-Delabrouille}, Nathalie and {Peng}, Xiyan and {Poppett}, Claire L. and {Poremba}, Megan R. and {Prakash}, Abhishek and {Rabinowitz}, David and {Raichoor}, Anand and {Rezaie}, Mehdi and {Robertson}, A.~N. and {Roe}, Natalie A. and {Ross}, Ashley J. and {Ross}, Nicholas P. and {Rudnick}, Gregory and {Safonova}, Sasha and {Saha}, Abhijit and {S{\'a}nchez}, F. Javier and {Savary}, Elodie and {Schweiker}, Heidi and {Scott}, Adam and {Seo}, Hee-Jong and {Shan}, Huanyuan and {Silva}, David R. and {Slepian}, Zachary and {Soto}, Christian and {Sprayberry}, David and {Staten}, Ryan and {Stillman}, Coley M. and {Stupak}, Robert J. and {Summers}, David L. and {Sien Tie}, Suk and {Tirado}, H. and {Vargas-Maga{\~n}a}, Mariana and {Vivas}, A. Katherina and {Wechsler}, Risa H. and {Williams}, Doug and {Yang}, Jinyi and {Yang}, Qian and {Yapici}, Tolga and {Zaritsky}, Dennis and {Zenteno}, A. and {Zhang}, Kai and {Zhang}, Tianmeng and {Zhou}, Rongpu and {Zhou}, Zhimin},
        title = "{Overview of the DESI Legacy Imaging Surveys}",
      journal = {\aj},
         year = 2019,
        month = may,
       volume = {157},
       number = {5},
          eid = {168},
        pages = {168},
          doi = {10.3847/1538-3881/ab089d},
archivePrefix = {arXiv},
       eprint = {1804.08657},
 primaryClass = {astro-ph.IM},
       adsurl = {https://ui.adsabs.harvard.edu/abs/2019AJ....157..168D}
}

@ARTICLE{emcee2013,
       author = {{Foreman-Mackey}, Daniel and {Hogg}, David W. and {Lang}, Dustin and {Goodman}, Jonathan},
        title = "{emcee: The MCMC Hammer}",
      journal = {\pasp},
         year = 2013,
        month = mar,
       volume = {125},
       number = {925},
        pages = {306},
          doi = {10.1086/670067},
archivePrefix = {arXiv},
       eprint = {1202.3665},
 primaryClass = {astro-ph.IM},
       adsurl = {https://ui.adsabs.harvard.edu/abs/2013PASP..125..306F}
}

@ARTICLE{Harutyunyan2008,
       author = {{Harutyunyan}, A.~H. and {Pfahler}, P. and {Pastorello}, A. and {Taubenberger}, S. and {Turatto}, M. and {Cappellaro}, E. and {Benetti}, S. and {Elias-Rosa}, N. and {Navasardyan}, H. and {Valenti}, S. and {Stanishev}, V. and {Patat}, F. and {Riello}, M. and {Pignata}, G. and {Hillebrandt}, W.},
        title = "{ESC supernova spectroscopy of non-ESC targets}",
      journal = {\aap},
         year = 2008,
        month = sep,
       volume = {488},
       number = {1},
        pages = {383-399},
          doi = {10.1051/0004-6361:20078859},
archivePrefix = {arXiv},
       eprint = {0804.1939},
 primaryClass = {astro-ph},
       adsurl = {https://ui.adsabs.harvard.edu/abs/2008A&A...488..383H}
}

@ARTICLE{Johnson2021,
       author = {{Johnson}, Benjamin D. and {Leja}, Joel and {Conroy}, Charlie and {Speagle}, Joshua S.},
        title = "{Stellar Population Inference with Prospector}",
      journal = {\apjs},
         year = 2021,
        month = jun,
       volume = {254},
       number = {2},
          eid = {22},
        pages = {22},
          doi = {10.3847/1538-4365/abef67},
archivePrefix = {arXiv},
       eprint = {2012.01426},
 primaryClass = {astro-ph.GA},
       adsurl = {https://ui.adsabs.harvard.edu/abs/2021ApJS..254...22J}
}

@software{sedpy,
  author       = {{Johnson}, Benjamin D.},
  title        = {bd-j/sedpy: sedpy v0.2.0},
  month        = mar,
  year         = 2021,
  publisher    = {Zenodo},
  version      = {v0.2.0},
  doi          = {10.5281/zenodo.4582723},
  url          = {https://doi.org/10.5281/zenodo.4582723}
}

@ARTICLE{Conroy2010,
       author = {{Conroy}, Charlie and {Gunn}, James E.},
        title = "{The Propagation of Uncertainties in Stellar Population Synthesis Modeling. III. Model Calibration, Comparison, and Evaluation}",
      journal = {\apj},
         year = 2010,
        month = apr,
       volume = {712},
       number = {2},
        pages = {833-857},
          doi = {10.1088/0004-637X/712/2/833},
archivePrefix = {arXiv},
       eprint = {0911.3151},
 primaryClass = {astro-ph.CO},
       adsurl = {https://ui.adsabs.harvard.edu/abs/2010ApJ...712..833C}
}

@INPROCEEDINGS{Crawford2010,
author = {{Crawford}, S.~M. and {Still}, M. and {Schellart}, P. and {Balona}, L. and
     {Buckley}, D.~A.~H. and {Dugmore}, G. and {Gulbis}, A.~A.~S. and
     {Kniazev}, A. and {Kotze}, M. and {Loaring}, N. and {Nordsieck}, K.~H. and
     {Pickering}, T.~E. and {Potter}, S. and {Romero Colmenero}, E. and
     {Vaisanen}, P. and {Williams}, T. and {Zietsman}, E.},
title = "{PySALT: the SALT science pipeline}",
booktitle = {Society of Photo-Optical Instrumentation Engineers (SPIE) Conference Series},
 year = 2010,
series = {Society of Photo-Optical Instrumentation Engineers (SPIE) Conference Series},
volume = 7737,
month = jul,
  eid = {773725},
pages = {25},
  doi = {10.1117/12.857000},
adsurl = {http://adsabs.harvard.edu/abs/2010SPIE.7737E..25C}
}

@INPROCEEDINGS{Smith2006,
author = {{Smith}, M.~P. and {Nordsieck}, K.~H. and {Burgh}, E.~B. and
     {Percival}, J.~W. and {Williams}, T.~B. and {O'Donohue}, D. and
     {O'Connor}, J. and {Schier}, J.~A.},
 title = "{The prime focus imaging spectrograph for the Southern African Large Telescope: structural and mechanical design and commissioning}",
booktitle = {Society of Photo-Optical Instrumentation Engineers (SPIE) Conference Series},
  year = 2006,
series = {\procspie},
volume = 6269,
 month = jun,
   eid = {62692A},
 pages = {62692A},
   doi = {10.1117/12.672415},
adsurl = {http://adsabs.harvard.edu/abs/2006SPIE.6269E..2AS}
}

@ARTICLE{Moriya2013,
       author = {{Moriya}, Takashi J. and {Blinnikov}, Sergei I. and {Tominaga}, Nozomu and {Yoshida}, Naoki and {Tanaka}, Masaomi and {Maeda}, Keiichi and {Nomoto}, Ken'ichi},
        title = "{Light-curve modelling of superluminous supernova 2006gy: collision between supernova ejecta and a dense circumstellar medium}",
      journal = {\mnras},
         year = 2013,
        month = jan,
       volume = {428},
       number = {2},
        pages = {1020-1035},
          doi = {10.1093/mnras/sts075},
archivePrefix = {arXiv},
       eprint = {1204.6109},
 primaryClass = {astro-ph.HE},
       adsurl = {https://ui.adsabs.harvard.edu/abs/2013MNRAS.428.1020M}
}

@INPROCEEDINGS{Howell2017GSP,
       author = {{Howell}, Dale Andrew and {Global Supernova Project}},
        title = "{The Global Supernova Project}",
    booktitle = {American Astronomical Society Meeting Abstracts \#230},
         year = 2017,
       series = {American Astronomical Society Meeting Abstracts},
       volume = {230},
        month = jun,
          eid = {318.03},
        pages = {318.03},
       adsurl = {https://ui.adsabs.harvard.edu/abs/2017AAS...23031803H}
}

@Inbook{Howell2017SLSN,
author="Howell, D. Andrew",
editor="Alsabti, Athem W.
and Murdin, Paul",
title="Superluminous Supernovae",
bookTitle="Handbook of Supernovae",
year="2017",
publisher="Springer International Publishing",
address="Cham",
pages="1--29",
isbn="978-3-319-20794-0",
doi="10.1007/978-3-319-20794-0_41-1",
url="https://doi.org/10.1007/978-3-319-20794-0_41-1"
}
